\pdfoutput=1
\documentclass[mnsc, sglanonrev]{informs4_WP}
\RequirePackage{fix-cm}
\RequirePackage{tgtermes}
\RequirePackage{newtxtext}
\RequirePackage{newtxmath}

\RequirePackage{bm}
\RequirePackage{endnotes}

\usepackage{enumitem}
\setlist[itemize]{label=\textbullet}
\usepackage{booktabs}
\usepackage{multirow}
\usepackage{threeparttable}
\usepackage{graphicx}
\usepackage[position=b]{subfig}
\usepackage{placeins}

\usepackage{tikz}

\usepackage{hyperref}
\usepackage{float}
\usepackage{authblk}
\usepackage{comment}
\hypersetup{%
  colorlinks=true,
  citecolor=blue,
  linkcolor=blue,
  urlcolor=blue
}

\usepackage{xcolor}
\usepackage{colortbl}
\usepackage{amssymb}

\usepackage{bbm}
\usepackage[normalem]{ulem}
\useunder{\uline}{\ul}{}

\OneAndAHalfSpacedXII

\usepackage{changepage}
\usepackage{url}
\usepackage{sansmath}

\newcommand{\proofheading}[1]{%
  \par\addvspace{0.75\baselineskip}%
  \noindent{\normalfont\bfseries #1}\par\nobreak
  \smallskip\noindent\ignorespaces
}
\newcommand{\proofstep}[1]{%
  \par\addvspace{0.4\baselineskip}%
  \noindent{\normalfont\itshape #1}\quad\ignorespaces
}
\newcommand{\proofend}{%
  \par\nobreak\noindent\hfill\ensuremath{\square}\par
  \addvspace{0.6\baselineskip}%
}

\newcolumntype{L}[1]{>{\raggedright\arraybackslash}b{#1}}
\newcolumntype{C}[1]{>{\centering\arraybackslash}b{#1}}
\newcolumntype{R}[1]{>{\raggedleft\arraybackslash}b{#1}}
\usepackage{natbib}
\usepackage{bibunits}
 \bibpunct[]{(}{)}{,}{a}{}{,}%
 \def\bibfont{\footnotesize}%

\usepackage{import}

\usepackage{array}
\usepackage{verbatim}
\usepackage{amsfonts}
\usepackage{dsfont}
\usepackage{accents}
\usepackage{mathtools}
\usepackage[framemethod=TikZ]{mdframed}
\usepackage{pdflscape}

\usepackage{etoolbox}

\usepackage{framed}

\theoremstyle{plain}
\newtheorem{theorem}{Theorem}
\newtheorem{proposition}{Proposition}
\newtheorem{lemma}{Lemma}
\newtheorem{corollary}{Corollary}

\ifdefined\argmax\else\DeclareMathOperator*{\argmax}{arg\,max}\fi

\ifdefined\theorem\else\TheoremsNumberedThrough\fi
\ECRepeatTheorems

\EquationsNumberedThrough

\MANUSCRIPTNO{}

\begin{document}

\RUNAUTHOR{Sinchaisri and Wang}
\RUNTITLE{Designing AI-Mediated Communication in Teams}

\TITLE{Designing AI-Mediated Communication in Teams}

\ARTICLEAUTHORS{%
\AUTHOR{Wichinpong Park Sinchaisri} \AFF{Haas School of Business, University of California, Berkeley, \EMAIL{parksinchaisri@berkeley.edu}}

\AUTHOR{Zhiwei Wang} \AFF{Department of Economics, University of California, Berkeley, \EMAIL{zhiwei.wang@berkeley.edu}}
}

\ABSTRACT{
AI assistants increasingly participate in work that requires employees to exchange information and coordinate decisions. We study how a firm should design this communication when workers' interests conflict with each other's and the firm's. Two workers privately know local conditions and their preferred responses, while the assistant knows aggregate project conditions. The firm discloses project information before workers report and choose their own actions. Transparency improves adaptation but can undermine truthful preference reporting. When workers agree with the firm on the productive direction, coarsening project information under weak coordination can preserve useful preference reporting at a cost in forecast precision. Under strong coordination, the firm instead prefers full project disclosure and factual reports, because preference information induces costly accommodation of private objectives. Under weak coordination, a firm-optimal policy can leave workers worse off than direct exchange; requiring their acceptance can call for further coarsening. Releasing project context after reports can sustain factual communication that early release would prevent. With separate delivery decisions that cannot be revised jointly, endogenous forwarding weakly improves firm-preferred performance over unchanged forwarding and strictly improves it in some environments. Numerical comparisons show that full initial disclosure captures most gains in the aligned environments considered, while further design has greater value when workers can favor opposing directions. The results distinguish the credibility of employees' reports from their organizational value and show how disclosure timing affects both.
}


\begin{bibunit}[informs2014]
\maketitle

\section{Introduction}\label{sec:exchange-introduction}

Coordinating a team requires information that its members may have reasons to withhold or distort. Employees know local operating conditions, but their judgments about how to respond also reflect personal objectives. A regional manager may favor a more aggressive adjustment than headquarters would choose, or may place greater weight on the costs borne by their own unit. Coworkers need to anticipate these preferences because their decisions are interdependent. The firm, however, need not benefit from every piece of information that helps employees predict one another. A communication system must therefore address two questions: what employees can be induced to reveal, and which of that information improves organizational performance.

AI assistants make these questions relevant to the design of work processes. Research on generative AI in teams examines how access to an assistant changes collaboration and the use of expertise \citep{dellacqua2025the}, while work on decision-oriented dialogue studies how an assistant can help elicit and communicate information needed for a joint decision \citep{lin2024decision}. An assistant can provide project context before asking employees for their assessments, then determine what to convey to each participant. These choices affect the information available for operating decisions. They also affect the incentives behind the assessments themselves. Even accurate processing of a worker's message cannot recover private information that the worker chose not to communicate.

Consider two regional managers adjusting their operating plans. Each knows whether local conditions call for an increase or a decrease in activity. Headquarters has an aggregate forecast that determines the appropriate scale of the adjustment. The managers must reconcile local needs with the cost of inconsistent plans, and each privately favors a stronger or weaker response than headquarters does. An assistant can make the forecast public and relay the managers' reports before they act. Revealing a manager's preferred intensity helps the counterpart anticipate the manager's decision, but can also induce that counterpart to accommodate an objective the firm does not share. Moreover, the forecast changes how attractive it is to exaggerate a preferred intensity. Thus, providing a more precise forecast can improve adaptation while impairing the communication on which coordination depends.

We develop a model of two workers who retain authority over their operating actions. Each privately observes a local condition and a preference type. The local condition determines the productive direction of the worker's action; the preference type shifts the worker's desired intensity relative to the firm's target. Workers incur losses from departing from their private targets and from failing to coordinate with each other. The firm observes aggregate project conditions and commits to a public disclosure policy. After receiving the project signal, workers send costless, unverifiable reports and choose their actions. The baseline assistant forwards reports unchanged. Because the reporting and operating decisions are voluntary, the firm must anticipate how its project signal changes the information workers will supply.

The central mechanism is a conflict between the informational value of project transparency and its effect on preference reporting. On an aligned support, where both preference types agree with the firm on the productive direction, workers can credibly communicate local conditions without revealing preferences. Full preference reporting requires an additional incentive constraint. When the expected aggregate state is sufficiently high, a worker with a moderate target benefits from claiming a more intense preference to move the coworker's action. Preference communication then becomes less informative even though factual communication remains feasible. At low coordination weights, the firm can benefit from information coarsening that preserves preference communication. Its optimal signal can pool project states around the reporting threshold, sacrificing some forecast precision to sustain more informative worker reports. We characterize the disclosure problem and derive the optimal interval when it is interior to a uniform aggregate support. Workers value preference accommodation more than the firm does. Relative to direct exchange without project information, a firm-optimal disclosure policy can therefore make both workers worse off ex ante.

The value of preference communication changes with the strength of coordination. Better knowledge of a coworker's objective improves prediction of the coworker's behavior, but also changes actions toward that objective. When coordination is sufficiently strong, this accommodation makes additional preference information costly to the firm. The firm then prefers full project disclosure accompanied by factual reports, even in states where workers could credibly reveal their preferences. This reversal distinguishes the credibility of a report from its organizational value.

This conflict matters for adoption. If employees must agree to use a communication assistant, the firm's preferred information policy may be infeasible even when it improves organizational performance. We impose ex-ante worker acceptance relative to direct exchange without project information and solve the firm's problem within a single-interval disclosure family. When acceptance binds, the firm widens the pool beyond its unconstrained choice. The additional pooling preserves preference communication in more project states, compensating workers through the information policy itself. A nearby adjustment of the pool also gives workers a strict incentive to report their preferences truthfully at the pooled signal. These results identify two distinct costs of implementation: protecting workers' outside options and providing a positive reporting incentive.

Three extensions follow. With endogenous forwarding, two separate processors each observe the aggregate state and one worker's report and choose what to deliver. Each processor controls only its own delivery, and neither can revise both messages jointly. Under this restriction, firm-preferred forwarding weakly improves on unchanged forwarding, strictly for some parameters. In continuing teams, a preference report also affects accommodation in future periods. Sufficiently low expected future project conditions make exaggeration costly and can sustain early truthful reporting with full project disclosure; high expected future conditions have the opposite effect. Finally, releasing project context after workers report can sustain factual communication that releasing the same context beforehand would prevent.

The paper contributes a theory of communication design in which project disclosure changes both the supply and the organizational value of employees' private information. The distinction between productive facts and personal objectives explains why preserving more communication can be desirable at low coordination weights and undesirable at high weights. It also links the firm's disclosure decision to worker acceptance and to the governance of delivery. The main disclosure problem fixes project communication before reports; delayed disclosure is a separate extension. A human intermediary with the same information and commitment capabilities could implement these policies. Indeed, direct exchange with the same public signal and reporting opportunities is equivalent to unchanged forwarding. The baseline benefit is therefore attributable to project information and its design. Throughout, we select the reporting equilibrium with the lowest expected firm loss.

\subsection{Related literature}\label{subsec:exchange-literature}

The interaction between public disclosure and private communication is central to our contribution. In cheap talk, incentive compatibility limits the information an informed party can convey \citep{crawford1982strategic}; in persuasion, commitment makes the information structure a choice variable \citep{kamenica2011bayesian,bergemann2019information}. \citet{chen2012value} studies public information arriving before or after a sender reports and shows that greater signal precision can reduce communication and receiver welfare. Informational control can also improve communication by changing what a sender learns before reporting \citep{ivanov2010informational,arieli2024optimal}. \citet{kreutzkamp2025persuasion} characterize signal choice when the sender can commit to an information structure but cannot commit to truthful reporting afterward. \citet{bonatti2026information} characterize implementable elicitation and action-recommendation mechanisms in linear--quadratic Gaussian games, with and without transfers. Our workers exchange unverifiable reports before choosing their own actions. The firm designs project context to influence this communication, and the organizational value of preference information can differ from its value to the workers. Our firm discloses an aggregate state that is independent of the information workers already hold. This state changes their productive targets and hence their incentives to reveal a different object, their private preferences. The additional distinction is between the credibility of preference communication and its value to a firm whose objectives differ from those of both decision makers. The optimal signal follows from this endogenous continuation value, using supporting bounds in the spirit of \citet{dworczak2019the}.

Organizational coordination models explain why communication and authority must be considered jointly. \citet{dessein2002authority} compares strategic communication with delegation of authority. \citet{alonso2008when} and \citet{rantakari2008governing} examine how the need for adaptation and coordination affects the allocation of decision rights when communication is strategic. Communication networks and the costs of influencing other decision makers provide further organizational instruments \citep{hagenbach2010strategic,galeotti2013strategic,calvo2015communication}. We retain decentralized authority and a fixed communication network, allowing the firm to change the project context workers receive before they communicate. This instrument is related to the design of public information in coordination games \citep{morris2002social}. Most closely, \citet{tamura2025information} characterizes organizational disclosure when agents adapt to local states and coordinate over a network. In our model, a public signal also changes the private information exchanged afterward. The firm's disclosure choice must account for the resulting change in worker communication, including the possibility that additional preference information worsens its objective.

The distinction between factual and preference information also connects to research on the disclosure of conflicting interests. \citet{li2008when} show that revealing an adviser's bias can impair communication. \citet{heller2024communication} examine coordination with private values and distinguish the roles of preference rankings and intensities. Our aligned environment allows productive direction to be communicated even when preference intensity cannot be revealed. Holding that factual information fixed, we identify a coordination threshold at which the firm's value of additional preference information changes sign. Workers continue to value preference accommodation more than the firm does. This welfare difference motivates a disclosure constraint based on the workers' outside options, rather than an assumption that the organization's preferred communication policy is also preferred by its members.

The delivery extension draws on the economics of mediation. Noise and filtering can improve strategic communication \citep{blume2007noisy,goltsman2009mediation,arieli2023mediated}, while intermediaries with their own objectives face incentives to distort the messages they transmit \citep{ivanov2010communication,ambrus2013hierarchical}. \citet{corrao2023the} distinguish the value of commitment from the limits imposed by unverifiable reports in sender-aligned mediation. \citet{kamenica2024commitment} characterize when commitment is valuable in finite sender--receiver games. Our intermediary has an additional source of information, the aggregate project state, and communicates with two workers who both make operating decisions. The forwarding extension studies separate processors with unilateral delivery choices.

Persistent preferences introduce a dynamic communication incentive. Studies of repeated communication and learning show that future interactions can affect the information transmitted today \citep{ivanov2016dynamic,atakan2020starting}. Here a preference claim changes how a coworker responds to fresh local conditions in later periods. Future accommodation can either deter or encourage exaggeration, depending on the expected aggregate environment. This mechanism yields a sufficient region in which full project disclosure and early preference learning replace the static pooling policy. The absence of intermediate action and payoff feedback isolates learning through communication; additional performance feedback would create a separate source of information about preferences.

Our operational setting complements research on how to incorporate human information into algorithmic decisions. \citet{ibrahim2021eliciting} compare methods for eliciting human judgment that differ in how directly they solicit private information. \citet{boyaci2024human} study the effect of machine input under cognitive limitations, and \citet{balakrishnan2026human} examine advice weighting when people possess information unavailable to an algorithm. Advice can also improve sequential decisions \citep{bastani2026improving}, while generative AI can change workers' task choices \citep{keppler2026teacher}. In our model, the employee's contribution is a strategic report to a coworker, and its value depends on the coworker's subsequent action. Coarsening affects incentives to supply that report, in contrast to its role in decisions with errors in probabilistic reasoning \citep{dreyfuss2026calibrated}. The setting also differs from advice designed in anticipation of a recipient's consultation with personal AI \citep{liu2026strategic}: the firm here influences communication among several human decision makers.

Incentive-based models of human--AI work design examine authority and information acquisition \citep{athey2020the}, compensation for oversight \citep{bastani2026the}, and the interaction between system configuration and hidden effort \citep{dai2026designing}. These studies make human responses to system design endogenous. Our corresponding margin is communication: a more informative project signal can reduce the information workers contribute, and a performance-improving policy can require an adjustment to secure their acceptance. The distinction between firm and worker objectives also relates to credible information provision by platforms \citep{avinadav2025information} and to trust under information sharing, advice, and delegation \citep{ozer2018information}. We obtain the adoption conflict with Bayesian workers and fixed preferences. Worker acceptance restricts the feasible disclosure policy even when cognitive error and distrust are absent.

\section{Model}\label{sec:exchange-setup}

We consider a firm that uses AI as an information intermediary on a single project. AI knows the aggregate state, while each worker knows their local condition and private preference type. In the baseline, workers communicate only through AI. AI first sends a public signal about the aggregate state. Each worker then sends a reply, and AI forwards both replies unchanged to both workers. Workers finally choose their own operating actions.

\subsection{Setup}

There are two workers, indexed by $i,j\in\{1,2\}$, with $j\ne i$ when referring to a coworker. Worker $i$ privately observes their preference type $T_i$ and local condition $X_i$. The preference type determines the worker's private bias relative to the firm's target, while the local condition determines which action is productive. Types are independent across workers, and local conditions are independent of types and of one another:
\begin{equation}
 T_i\in\{-1,1\},\qquad \Pr(T_i=1)=q_i,\qquad
 X_i\in\{-1,1\},\qquad \Pr(X_i=1)=\frac12
 \label{eq:exchange-types}
\end{equation}
We allow arbitrary given priors $q_i\in[0,1]$; balanced beliefs correspond to $q_1=q_2=1/2$. We call $T_i=1$ the high type and $T_i=-1$ the low type. These labels refer to the two private biases in the preferred-action function defined below. The project is affected by a common aggregate state $\theta$, observed only by AI initially. We assume $\theta\sim U[\theta_L,\theta_H]$, with $0\leq\theta_L<\theta_H<\infty$, and denote this uniform prior by $F$. The aggregate state is independent of worker information.

Let $H_i$ denote the firm's target for worker $i$. The worker has a private bias $B_i$ relative to this target, so the worker's preferred action is $D_i=H_i+B_i$. Specifically,
\begin{equation}
 H_i=\theta hX_i,\qquad B_i=\beta T_iX_i,\qquad
 D_i=H_i+B_i=(\theta h+\beta T_i)X_i,\qquad h,\beta>0
 \label{eq:exchange-targets}
\end{equation}
The parameter $h$ determines how strongly the firm's preferred action responds to the aggregate state, and $\beta$ measures the magnitude of the worker's private bias. For a positive aggregate state, the high type's preferred action has the same sign as the firm's target but a larger magnitude. If $\theta h>\beta$, both worker types are aligned with the firm in direction: their preferred actions are positively correlated with the firm's target as the local condition varies. If $0<\theta h<\beta$, the low type is misaligned: its preferred action has the opposite sign from the firm's target. These preferences can represent a manager whose local incentives favor expansion or contraction more strongly than headquarters does. The model takes this preference difference as given.

After the information exchange, worker $i$ chooses $a_i\in\mathbb R$. The worker's loss and the firm's loss per worker are
\begin{align}
 \ell_i&=(1-\gamma)(a_i-D_i)^2+\gamma(a_i-a_j)^2,\qquad 0<\gamma<1
 \label{eq:exchange-workerloss}\\
 L_O&=\frac{1-\gamma}{2}\sum_{i=1}^2(a_i-H_i)^2
                              +\gamma(a_1-a_2)^2
 \label{eq:exchange-firmloss}
\end{align}
The first squared term penalizes departures from the decision maker's preferred action: $D_i$ for worker $i$ and $H_i$ for the firm. The coordination term penalizes differences between the workers' actions. The parameter $\gamma$ determines the weight on coordination relative to matching the preferred action.

\subsection{AI disclosure and worker replies}\label{subsec:exchange-policies}

The firm commits to AI's signal policy before information is realized. AI shares the firm's objective. Write the common prior beliefs about worker types as
\begin{equation}
 q_i=\Pr(T_i=1),\qquad q=(q_1,q_2)
 \label{eq:exchange-prior}
\end{equation}
Given the aggregate state, AI's initial public signal satisfies
\begin{equation}
 s\sim\sigma(\cdot\mid\theta,q)
 \label{eq:exchange-signal}
\end{equation}
The policy is common knowledge, and the same realized signal is observed by both workers. It can reveal the state, conceal it, or disclose it partially. Workers update the aggregate-state distribution using this policy. Write its first two posterior moments as
\begin{equation}
 \mu=\mathbb E[\theta\mid q,s,\sigma],\qquad
 \eta=\mathbb E[\theta^2\mid q,s,\sigma]
 \label{eq:exchange-moments}
\end{equation}

After observing $s$, each worker sends one report from the message set $\mathcal R=\{1,2,3,4\}$. We use pure reporting strategies:
\begin{equation}
 r_i=\lambda_i(T_i,X_i,q,s)\in\mathcal R
 \label{eq:exchange-reply}
\end{equation}
Conditional on $(q,s)$, the private state $(T_i,X_i)$ has at most four possible values. Four messages therefore suffice to represent every deterministic reporting rule, including full revelation and any pooling of private states. The initial public signal may be randomized.

AI forwards the pair $r=(r_1,r_2)$ unchanged to both workers and releases no further project information in the baseline. The coworker interprets each report using the equilibrium reporting strategy. Worker $i$ then chooses an operating action based on $(T_i,X_i,q,s,r)$. Section~\ref{sec:exchange-relay-extension} allows separate processors to choose deliveries after receiving reports; Section~\ref{subsec:exchange-timing-implications} studies a committed project update after reporting.

Workers choose operating actions simultaneously, without observing their coworker's actual action. The firm designs $\sigma$, while the workers' replies and actions must satisfy sequential Bayesian optimality.

\subsection{Timing}

\begin{figure}[htbp]
\centering
\includegraphics[width=\textwidth]{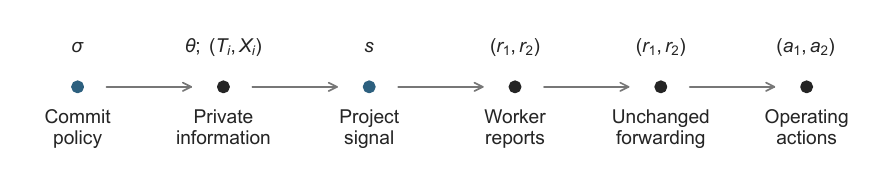}
\caption{Information and decisions in the baseline model}\label{fig:exchange-model-timing}
\par\vspace{8pt}\noindent\begin{minipage}{\textwidth}\footnotesize
The firm chooses the public disclosure policy $\sigma$. Workers choose their own reports and actions; neither action is observed by the coworker before both have been chosen. The baseline forwards reports unchanged. State-dependent delivery is introduced in Section~\ref{sec:exchange-relay-extension}.
\end{minipage}
\end{figure}

The firm commits before private information is realized. AI then observes $\theta$, and each worker observes their own $(T_i,X_i)$. Figure~\ref{fig:exchange-model-timing} shows the subsequent sequence. Workers report simultaneously and, after receiving both unchanged replies, choose their operating actions simultaneously.

\section{Reporting and operating decisions}\label{sec:exchange-terminal}

We first solve the workers' reporting and operating decisions after an AI signal. Fix the independent type beliefs $q=(q_1,q_2)$, signal policy $\sigma$, realized public signal $s$, and proposed reporting strategies. The signal induces the aggregate-state posterior moments $(\mu,\eta)$ defined in~\eqref{eq:exchange-moments}.

Let $p_i(T,x\mid q,s,r_i)$ denote the Bayesian posterior induced by worker $i$'s reply. Workers' private information is independent across workers and of the aggregate state. Consequently, their replies provide no additional information about $\theta$ beyond $s$, and the two worker posteriors factor. Appendix~\ref{subsec:exchange-posteriors} gives the reply probabilities and updating formulas, including beliefs at unused replies.

\subsection{Operating actions}\label{subsec:exchange-actions}

The worker's expected preferred target and the coworker's prediction of it are, respectively,
\begin{equation}
 d_i=(h\mu+\beta T_i)X_i,\qquad
 \widehat D_i(r_i)=\sum_{T,x}(h\mu+\beta T)x\,p_i(T,x\mid q,s,r_i)
 \label{eq:exchange-prediction}
\end{equation}
Here $d_i$ is a numerical target conditional on the worker's private information and the initial signal; $\widehat D_i$ averages that target under the coworker's reply-induced posterior.

\begin{lemma}[Operating actions]\label{lem:exchange-actions}
For any proposed reply profile and its Bayesian posteriors, the unique bounded Bayesian Nash operating actions are
\begin{equation}
 a_i=(1-\gamma)d_i+
        \frac{\gamma^2\widehat D_i+\gamma\widehat D_j}{1+\gamma}
 \label{eq:exchange-actions}
\end{equation}
\end{lemma}

Each worker combines their own target with the inferred targets of both workers. Reporting matters because it changes the coworker's prediction and hence their action. Appendix~\ref{subsec:proof-main-actions} proves uniqueness.

\subsection{The worker's reply incentive}\label{subsec:exchange-reply-ic}

A worker chooses their report anticipating both the coworker's response and their own subsequent action adjustment. Substituting the operating actions into the worker's loss and averaging over the coworker's independent report gives the following report-dependent term. Appendix~\ref{app:exchange-reply-derivation} gives the full loss calculation.

The reply-dependent part of worker $i$'s loss is
\begin{equation}
 \gamma(1-\gamma)\left\{d_i-\frac{\gamma}{1+\gamma}\widehat D_i(r_i)\right\}^2
 \label{eq:exchange-message-cost}
\end{equation}
For the reporting strategies to form an equilibrium, they must satisfy the incentive condition at every private state with positive probability:
\begin{equation}
 \lambda_i(T_i,X_i,q,s)\in\operatorname*{arg\,min}_{\widetilde r_i\in\mathcal R}
                \left\{d_i-\frac{\gamma}{1+\gamma}\widehat D_i(\widetilde r_i)\right\}^2
 \label{eq:exchange-terminal-ic}
\end{equation}

\subsection{When replies reveal direction and intensity}\label{subsec:exchange-communication}

Fix the type beliefs $q=(q_1,q_2)$ and the aggregate-state posterior mean $\mu>0$. Worker $i$ knows their expected preferred target $d_i=(h\mu+\beta T_i)X_i$, while their coworker must infer this target from the reply. Define the upper and lower target magnitudes as
\begin{equation}
 v_u=h\mu+\beta,\qquad v_l=|h\mu-\beta|
 \label{eq:exchange-target-magnitudes}
\end{equation}
The high type's two local conditions generate targets $\pm v_u$, each with probability $q_i/2$. The low type generates $\pm v_l$, each with probability $(1-q_i)/2$. When $0<q_i<1$ and $h\mu\ne\beta$, the four possible targets are ranked as
\[
 -v_u<-v_l<v_l<v_u
\]
The following three reporting outcomes organize the analysis. Under the alignment assumption $h\theta_L\geq\beta$ in Section~\ref{subsec:exchange-aligned}, the firm's preferred outcome is attained by the factual or full-reporting rules characterized there. Other reporting equilibria can exist; Appendix~\ref{app:exchange-reporting} characterizes all pure reporting equilibria. 

\paragraph{Full revelation.}
Each private state $(T_i,X_i)$ with positive probability sends a different reply. The coworker then knows both variables and hence $d_i$, so $\widehat D_i(r_i)=d_i$. If $0<q_i<1$, there are four possible private states and four distinct replies are needed.

\paragraph{Direction-only reporting.}
All states with $d_i>0$ send the same reply, while all states with $d_i<0$ send another. For example, use labels 1 and 2 from $\mathcal R$:
\[
 \lambda_i(T_i,X_i,q,s)=
 \begin{cases}
  1,&d_i>0\\
  2,&d_i<0
 \end{cases}
\]
We call reply 1 positive because it tells the coworker that $d_i>0$, and reply 2 negative because it tells the coworker that $d_i<0$. When $0<q_i<1$, the coworker learns which direction the worker prefers, but not whether the target has magnitude $v_u$ or $v_l$. The coworker therefore predicts
\[
 \widehat D_i(1)=q_i v_u+(1-q_i)v_l,\qquad
 \widehat D_i(2)=-\widehat D_i(1)
\]
We call the resulting equilibrium direction-only when this reporting rule is optimal for both workers and their operating actions satisfy~\eqref{eq:exchange-actions}.

What a direction-only reply reveals about the local condition depends on the aggregate-state posterior. If $h\mu>\beta$, both types' targets have the same sign as $X_i$. Reply 1 then reveals $X_i=1$ and reply 2 reveals $X_i=-1$, but neither reveals $T_i$. If $h\mu<\beta$, the low type's target has the opposite sign from $X_i$, so the reply reveals $T_iX_i$ instead. A positive reply can then come from either a high type with $X_i=1$ or a low type with $X_i=-1$. 

\paragraph{Babbling.}
Every private state sends the same reply, so the coworker learns nothing about either $T_i$ or $X_i$. The coworker's predicted target remains $\widehat D_i=\mathbb E[d_i\mid q,s]=0$.

\section{Project disclosure and worker communication}\label{sec:exchange-design}

The firm chooses the initial public signal anticipating the reporting and action game. For each signal, we select the worker equilibrium giving the firm its lowest expected loss. This is the firm-preferred equilibrium.

The firm's problem is
\begin{equation}
 \inf_{\sigma}\ \inf_{(\lambda,p,a)\in\mathcal E(q,F,\sigma)}
                 \mathbb E_O[L_O\mid q,F,\sigma,\lambda,p,a]
 \label{eq:exchange-design-problem}
\end{equation}
Here $\mathcal E(q,F,\sigma)$ contains reporting strategies $\lambda$, Bayesian posteriors $p$, and operating actions $a$. The posteriors satisfy Bayes' rule~\eqref{eq:exchange-posterior}, the actions satisfy~\eqref{eq:exchange-actions}, and the replies satisfy~\eqref{eq:exchange-terminal-ic}. Beliefs at unused replies are included in $p$. Thus the strategies and posteriors must be consistent: changing a reporting rule changes how the coworker interprets the worker's replies. The expectation averages the aggregate state, both worker types, both local conditions, and the initial AI signal under these same strategies and posteriors.

To evaluate a reporting equilibrium after a given signal, substitute the Nash operating actions into the firm's loss. Its conditional expected loss is
\begin{equation}
 \begin{aligned}
 \mathbb E_O[L_O\mid q,s]
 &=(1-\gamma)h^2\eta-(1-\gamma)^3h^2\mu^2+(1-\gamma)^2(1+\gamma)\beta^2\\
 &\quad+\gamma(1-\gamma)^2\beta h\mu\sum_i(2q_i-1)-\sum_i\Delta_i(q,s)
 \end{aligned}
 \label{eq:exchange-expected-firm-loss}
\end{equation}
where the contribution of worker $i$'s reply is
\begin{equation}
 \Delta_i(q,s)
 =\frac{(1-\gamma)\gamma^2}{1+\gamma}h\mu\,
       \mathbb E_O[X_i\widehat D_i\mid q,s]
      +b_\gamma\mathbb E_O[\widehat D_i^2\mid q,s]
 \label{eq:exchange-firm-decomposition}
\end{equation}
and
\begin{equation}
 b_\gamma=\frac{(1-\gamma)\gamma^2(1-2\gamma-\gamma^2)}{2(1+\gamma)^2}
 \label{eq:exchange-firm-coefficients}
\end{equation}
These expectations average worker types, local conditions, and their replies at the given signal; the remaining aggregate-state uncertainty enters through $\mu$ and $\eta$. Appendix~\ref{app:exchange-firm} derives the formula. The first term in $\Delta_i$ measures how the coworker's inferred target tracks the productive condition. The second depends on the precision of that target prediction. Its coefficient $b_\gamma$ is positive when $\gamma<\sqrt2-1$ and negative when $\gamma>\sqrt2-1$, so additional information about private objectives can help or hurt the firm.

We begin with the case in which both types favor the firm's productive direction. We first compare disclosure of local conditions with additional disclosure of private objectives, and then derive the optimal AI signal when its pooling interval is interior. We then briefly contrast the case in which the low type favors the opposite direction.

\subsection{Factual and preference disclosure}\label{subsec:exchange-aligned}

We assume $h\theta_L\geq\beta$, so every aggregate-state posterior satisfies $h\mu\geq\beta$. When the inequality is strict, both types prefer the same direction as the firm, although they differ in how strongly they want to respond to the local condition. At equality, the low type's preferred target is zero. Reporting $X_i$ tells the coworker what the productive condition is. Reporting $(T_i,X_i)$ provides the same factual information and additionally reveals the worker's private objective. This case isolates the value of preference disclosure without changing how much the coworker learns about $X_i$. The main analysis focuses on this region; Appendix~\ref{app:exchange-general} gives the reporting and AI-disclosure results without the alignment condition.

The aggregate-state posterior mean determines whether the worker agrees with the firm's direction and whether full reporting is incentive compatible. The corresponding thresholds are
\begin{equation}
 \mu_0=\frac{\beta}{h},\qquad \mu_c=\frac{(1+\gamma)\beta}{h}
 \label{eq:exchange-aligned-thresholds}
\end{equation}
The threshold $\mu_0$ separates aligned and opposite-direction targets. The threshold $\mu_c$ is the highest posterior mean at which full worker revelation is incentive compatible in the aligned region.

\begin{proposition}[Factual information and additional preference disclosure]\label{prop:exchange-aligned-reporting}
Suppose $0<\gamma<1$, $h\mu\geq\beta$, and the independent type beliefs are $q\in[0,1]^2$. Reporting $X_i$ without revealing type is an equilibrium. For a worker whose type is uncertain, full revelation is an equilibrium if and only if $\mu\leq\mu_c$. At a posterior where both are feasible, their contributions to reducing firm loss satisfy
\begin{equation}
 \Delta_i^{\mathrm{full}}-\Delta_i^{\mathrm{dir}}
 =4b_\gamma\beta^2q_i(1-q_i)
 \label{eq:exchange-preference-disclosure-value}
\end{equation}
Here $\Delta_i^{\mathrm{full}}$ and $\Delta_i^{\mathrm{dir}}$ denote the contributions in~\eqref{eq:exchange-firm-decomposition} under full revelation and reporting $X_i$, respectively.

Among all reporting equilibria, the firm prefers full worker revelation when $\gamma<\sqrt2-1$ and $\mu\leq\mu_c$, and reporting $X_i$ without revealing type otherwise. If $q_i\in\{0,1\}$, the two rules reveal the same supported private information and give the same outcome.
\end{proposition}

The proof is in Appendix~\ref{subsec:proof-main-aligned-reporting}.

Additional preference information changes the coworker's response intensity, not their knowledge of the productive condition. When coordination incentives are relatively weak, the firm benefits from this more precise response. When they are strong, disclosing private objectives can move the workers' coordinated actions away from the firm's target. The factor $q_i(1-q_i)$ measures how much preference uncertainty remains: the distinction disappears when type is already known.

For the signal-design problem, denote the lowest conditional expected firm loss by $L(\mu,\eta,q)$. Equation~\eqref{eq:exchange-expected-firm-loss} gives
\begin{equation}
 \begin{aligned}
 L(\mu,\eta,q)
 &=(1-\gamma)h^2\eta+(1-\gamma)^2(1+\gamma)\beta^2\\
 &\quad+\gamma(1-\gamma)^2\beta h\mu\sum_i(2q_i-1)-G(\mu,q)
 \end{aligned}
 \label{eq:exchange-frontier}
\end{equation}
where the benefit from the posterior and the resulting worker replies is
\begin{equation}
 G(\mu,q)=(1-\gamma)^3h^2\mu^2+\sum_iG_i(\mu,q_i)
 \label{eq:exchange-reward}
\end{equation}
Here $G_i$ is the largest attainable contribution from worker $i$'s reply. On the aligned region, Proposition~\ref{prop:exchange-aligned-reporting} gives
\begin{equation}
 G_i(\mu,q_i)=\Delta_i^{\mathrm{dir}}(\mu,q_i)
 +4\beta^2q_i(1-q_i)\max\{b_\gamma,0\}\mathbf1\{\mu\leq\mu_c\}
 \label{eq:exchange-aligned-reply-gain}
\end{equation}
The term $\Delta_i^{\mathrm{dir}}$ is obtained by substituting $\widehat D_i=[h\mu+\beta(2q_i-1)]X_i$ into~\eqref{eq:exchange-firm-decomposition}.

\subsection{Optimal AI disclosure when targets are aligned}\label{subsec:exchange-aligned-policy}

Suppose $\theta_L\geq\mu_0$, so every aggregate-state posterior is in the aligned region. Since $\mathbb E[\eta]=\mathbb E[\theta^2]$ and $\mathbb E[\mu]=\mathbb E[\theta]$, all terms in~\eqref{eq:exchange-frontier} except $-G(\mu,q)$ have the same expectation under every AI signal policy. The firm therefore maximizes $\mathbb E[G(\mu,q)]$. Substituting~\eqref{eq:exchange-aligned-reply-gain}, the part of $G$ that is not constant or linear in $\mu$ is
\begin{equation}
 A_\gamma h^2\mu^2+
 4\beta^2\max\{b_\gamma,0\}\sum_iq_i(1-q_i)\mathbf1\{\mu\leq\mu_c\},
 \qquad A_\gamma=\frac{(1-\gamma)(1+\gamma^2)}{(1+\gamma)^2}>0
 \label{eq:exchange-aggregate-coefficient}
\end{equation}
The quadratic term favors more aggregate-state information. The second term is the additional benefit of preference disclosure, which is available only while the posterior permits full worker revelation.

\begin{proposition}[Optimal AI disclosure on an aligned support]\label{prop:exchange-aligned-signal}
Suppose $\theta\sim U[\theta_L,\theta_H]$, $\theta_L\geq\mu_0$, $0<\gamma<1$, and $q\in[0,1]^2$ consists of independent type beliefs. If both workers' types are already known, full aggregate-state revelation is optimal and workers report $X_i$, revealing all their remaining private information. Otherwise, the firm's optimal disclosure and preferred worker reporting equilibria are as follows.
\begin{itemize}
 \item \textbf{High coordination: $\sqrt2-1\leq\gamma<1$.} Full aggregate-state revelation is optimal. A firm-preferred worker equilibrium reports $X_i$ without revealing $T_i$ after every signal. At $\gamma=\sqrt2-1$, the firm can also be indifferent to additional type disclosure when it is incentive compatible.

 \item \textbf{Low coordination with $\mu_c$ outside the support: $0<\gamma<\sqrt2-1$ and $\mu_c\notin(\theta_L,\theta_H)$.} Full aggregate-state revelation is optimal. The firm prefers full worker reporting of $(T_i,X_i)$ at revealed states $\theta\leq\mu_c$, and reporting $X_i$ without revealing type at $\theta>\mu_c$. Thus full worker revelation occurs throughout the support if $\mu_c\geq\theta_H$, while types are concealed at almost every state if $\mu_c\leq\theta_L$.

 \item \textbf{Low coordination with $\mu_c$ inside the support: $0<\gamma<\sqrt2-1$ and $\theta_L<\mu_c<\theta_H$.} Pooling a sufficiently small interval centered at $\mu_c$ strictly improves on full aggregate-state revelation. Define
\begin{equation}
 w_{\mu_c}=\frac{\beta}{h}
           \sqrt{\frac{2b_\gamma}{A_\gamma}\sum_iq_i(1-q_i)}
 \label{eq:exchange-pooling-width}
\end{equation}
If $[\mu_c-w_{\mu_c},\mu_c+w_{\mu_c}]\subseteq[\theta_L,\theta_H]$, an optimal policy is
\begin{equation}
 s(\theta)=
 \begin{cases}
  \mathsf P_{\mu_c},&\theta\in[\mu_c-w_{\mu_c},\mu_c+w_{\mu_c}]\\
  \theta,&\text{otherwise}
 \end{cases}
 \label{eq:exchange-aligned-policy}
\end{equation}
The firm prefers full worker reporting of $(T_i,X_i)$ inside the pool, where the posterior mean is $\mu_c$, and at revealed states below the pool. At revealed states above the pool, it prefers reporting $X_i$ without revealing type. The policy is optimal among all initial public AI signal policies in the baseline, including randomized policies and policies pooling disconnected sets. Its expected reduction in firm loss relative to full aggregate-state revelation is
\begin{equation}
 \frac{8b_\gamma\beta^2w_{\mu_c}}{3(\theta_H-\theta_L)}\sum_iq_i(1-q_i)>0
 \label{eq:exchange-strict-gain}
\end{equation}
\end{itemize}
\end{proposition}

The proof is in Appendix~\ref{subsec:proof-main-aligned-signal}.

AI coarsens aggregate-state information to obtain more information from the workers. States immediately above $\mu_c$ would otherwise make full preference disclosure unprofitable for a low type. Pooling them with nearby lower states keeps that disclosure incentive compatible. A wider pool obtains the reporting benefit at more states but sacrifices more adaptation to the aggregate state. Equation~\eqref{eq:exchange-pooling-width} balances these effects. The pool is wider when preference uncertainty $\sum_iq_i(1-q_i)$ is larger.

\begin{figure}[htbp]
\centering
\subfloat[Reporting at a given posterior]{\includegraphics[width=0.485\textwidth]{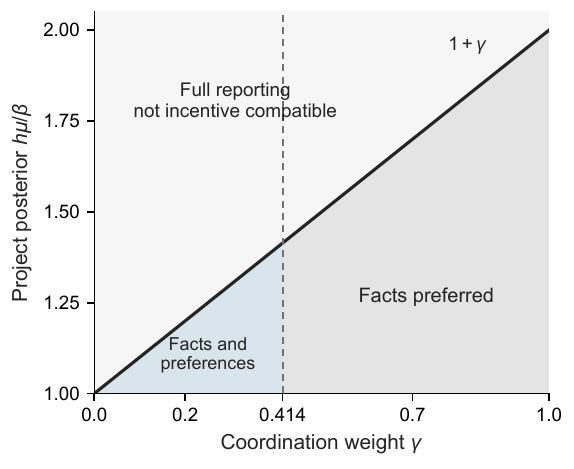}}\hfill
\subfloat[Optimal initial disclosure]{\includegraphics[width=0.485\textwidth]{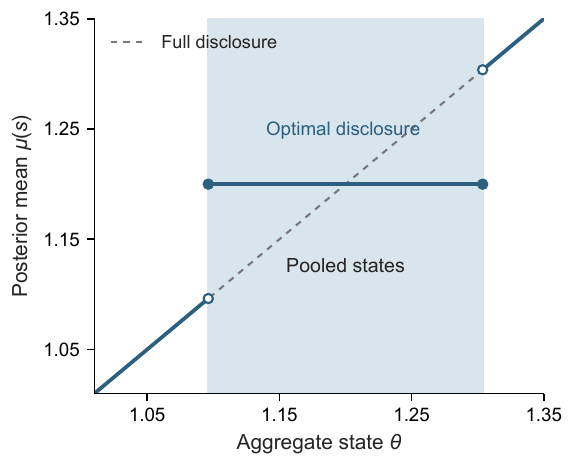}}
\caption{Reporting content and project disclosure}\label{fig:exchange-policy-map}
\par\vspace{8pt}\noindent
\begin{minipage}{\textwidth}\footnotesize
Panel (a) shows a firm-preferred reporting outcome under alignment and uncertain types. The sloped line is the full-reporting incentive boundary; the vertical line is the firm's indifference threshold for additional preference information. The region below $h\mu/\beta=1$ is excluded. Panel (b) uses $h=\beta=1$, $\gamma=0.2$, $q_1=q_2=0.5$, and $\theta\sim U[1.01,1.35]$. The shaded interval is pooled to $\mu_c=1.2$, with half-width approximately $0.104$; states outside it are revealed.
\end{minipage}
\end{figure}

Figure~\ref{fig:exchange-policy-map} separates reporting incentives from the firm's value of preference information. Full reporting can be feasible yet undesirable to the firm. When coordination is weak, pooling states around the reporting threshold preserves preference communication while limiting the loss of project precision.

\subsection{When a type favors the opposite direction}\label{subsec:exchange-opposite-contrast}

Below $\mu_0$, the low type prefers the opposite direction from the firm. A reply revealing the worker's preferred direction then identifies $T_iX_i$, not $X_i$. With balanced type beliefs, it conveys no information about the productive condition at all. Thus factual disclosure and disclosure of intended behavior are no longer interchangeable. High-type revelation, in which only the high type's replies distinguish its local condition and the low type's replies are pooled, can instead identify the high type's condition while leaving the low type's condition unknown.

Pooling can therefore change whether workers communicate productive facts or only their preferred response direction, rather than just how much type information they reveal. Appendix~\ref{app:exchange-general} characterizes the additional reporting thresholds and the conditions under which pooling near them helps.

The public signal can be accompanied by a recommendation of the incentive-compatible reporting rule. Appending a rule that is already a known function of the signal leaves information and equilibrium payoffs unchanged. Appendix~\ref{subsec:exchange-report-recommendations} formalizes this interpretation.

\subsection{Comparison with direct worker communication}\label{subsec:exchange-direct-comparison}

Under direct communication, workers exchange replies without an initial AI signal. We keep the same private information, reply space, simultaneous timing, and firm-preferred equilibrium selection. Reports remain strategic and are interpreted using Bayes' rule. Neither worker observes the aggregate state, so their posterior remains the prior. The corresponding firm loss and the loss under the firm's optimal initial AI signal are
\begin{align}
 L^{\mathrm{direct}}(q)&=L(\mathbb E[\theta],\mathbb E[\theta^2],q)
 \label{eq:exchange-direct-loss}\\
 L^{\mathrm{AI},*}(q)&=\inf_\sigma\mathbb E_s[L(\mu(s),\eta(s),q)]
                               \leq L^{\mathrm{direct}}(q)
 \label{eq:exchange-direct-ai-ordering}
\end{align}
The inequality follows because a constant initial signal reproduces direct communication. Let $\ell_i^{\mathrm{direct}}$ and $\ell_i^{\mathrm{AI},*}$ be each worker's ex-ante loss under these firm-selected outcomes, and write $\overline \ell=(\ell_1+\ell_2)/2$. These losses average both worker types, both local conditions, the aggregate state, and the initial signal.

\begin{proposition}[Firm and worker gains from AI mediation]\label{prop:exchange-direct-worker-loss}
Suppose $h\theta_L\geq\beta$ and the type beliefs $q=(q_1,q_2)$ are independent. Let $\sigma^*$ be a firm-optimal initial signal, and let $\overline\theta=\mathbb E[\theta]$.
\begin{enumerate}
 \item If $\sqrt2-1\leq\gamma<1$, full aggregate-state revelation is optimal, with local-condition reporting under both AI mediation and direct communication. Under this selection,
 \begin{equation}
  L^{\mathrm{direct}}-L^{\mathrm{AI},*}
  =\ell_i^{\mathrm{direct}}-\ell_i^{\mathrm{AI},*}
  =A_\gamma h^2\operatorname{Var}(\theta)>0
  \label{eq:exchange-direct-strict-gain}
 \end{equation}
 The same worker gain holds conditional on each supported own private state $(T_i,X_i)$. At $\gamma=\sqrt2-1$, local-condition reporting specifies the selection among firm-indifferent equilibria.

 \item If $0<\gamma<\sqrt2-1$ and $\overline\theta\leq\mu_c$, direct communication uses full worker revelation. Let $p^*=\Pr_{\sigma^*}(\mu(s)\leq\mu_c)$ be the probability of full worker revelation under AI. Then
 \begin{align}
  L^{\mathrm{direct}}-L^{\mathrm{AI},*}
  &=A_\gamma h^2\operatorname{Var}(\mu(s))
       -4b_\gamma\beta^2\sum_iq_i(1-q_i)(1-p^*)\geq0
  \label{eq:exchange-direct-firm-tradeoff}\\
  \overline \ell^{\mathrm{direct}}-\overline \ell^{\mathrm{AI},*}
  &=A_\gamma h^2\operatorname{Var}(\mu(s))
       -2\omega_\gamma\beta^2\sum_iq_i(1-q_i)(1-p^*)
  \label{eq:exchange-direct-worker-tradeoff}
 \end{align}
 In particular, AI benefits the firm but harms workers on average when
 \begin{equation}
  4b_\gamma\beta^2\sum_iq_i(1-q_i)(1-p^*)
  <A_\gamma h^2\operatorname{Var}(\mu(s))
  <2\omega_\gamma\beta^2\sum_iq_i(1-q_i)(1-p^*)
  \label{eq:exchange-direct-worker-loss-region}
 \end{equation}
 With common beliefs $q_1=q_2$, both workers have the same ex-ante loss under the selected symmetric reporting rules, so both lose in this region.
\end{enumerate}
Here $A_\gamma$ and $b_\gamma$ are defined in~\eqref{eq:exchange-aggregate-coefficient} and~\eqref{eq:exchange-firm-coefficients}, and
\begin{equation}
 \omega_\gamma=\frac{\gamma^2(1-\gamma)(3-\gamma^2)}{(1+\gamma)^2}>0,
 \qquad \omega_\gamma-2b_\gamma=\frac{2\gamma^2(1-\gamma)}{1+\gamma}>0
 \label{eq:exchange-worker-preference-value}
\end{equation}
\end{proposition}

In the first case, AI adds project context without changing what workers communicate, benefiting both parties. In the second case, better aggregate information can come at the expense of preference disclosure. Workers value that disclosure more than the firm does, as $\omega_\gamma>2b_\gamma$ shows, so the firm can accept a tradeoff that leaves workers worse off.

The proof is in Appendix~\ref{subsec:proof-main-direct-comparison}.

\paragraph{Worker acceptance.}
The above results show that workers do not necessarily benefit from a firm-optimal AI policy. Suppose both workers must agree to use the assistant before learning their private states. Rejection restores direct exchange without project information, and the firm cannot compensate workers with transfers. Each worker's expected loss under the proposed policy must then be no greater than their direct-exchange loss. The firm can satisfy this constraint by changing the information policy, which alters the extent to which each worker's preference is accommodated. We consider intervals centered at the reporting threshold and ask how the acceptance constraint changes their width.

Proposition~\ref{prop:exchange-worker-acceptance} in Appendix~\ref{subsec:exchange-acceptance-proof} characterizes this adjustment for an aligned uniform prior with common type beliefs. Within the stated centered-interval family, acceptance leaves the firm's unconstrained half-width $w_{\mu_c}$ from~\eqref{eq:exchange-pooling-width} unchanged when workers already prefer adoption. Otherwise, the firm chooses the smallest wider half-width $w^A$ at which the worker constraint binds. The parameter restrictions ensure feasibility and a strictly positive firm gain relative to direct exchange.

Workers value preference accommodation more than the firm does. Once pooling reaches the firm's unconstrained optimum, widening it can therefore improve worker welfare while reducing the firm's gain. The firm chooses the first width at which workers accept. Feasibility follows because the largest admissible pool preserves preference reporting in every state while revealing some project information.

The comparative statics in Appendix~\ref{subsec:exchange-acceptance-numerical} show how this adjustment depends on uncertainty about preferences. In that example, additional pooling is needed near the balanced prior, where the coworker's type is most uncertain. Near either extreme, the unconstrained policy already satisfies acceptance. Acceptance does not make preference reporting strict. Shifting the pool slightly below the reporting threshold and adjusting its width can provide a positive incentive while preserving participation; Appendix~\ref{subsec:exchange-strict-reporting} gives the construction.

\section{Endogenous forwarding}\label{sec:exchange-relay-extension}

In the baseline, AI forwards worker reports unchanged. We now allow AI to choose which message to deliver after receiving a worker's report. The firm chooses the initial public signal policy $\sigma$ before information is realized, while AI chooses the delivered message after observing the aggregate state and report. Workers anticipate AI's forwarding strategy and interpret the delivered messages using Bayes' rule. Reports and delivered messages use the same set $\mathcal R=\{1,2,3,4\}$.

The delivery architecture consists of two separate processors with the same firm-loss objective. At the delivery stage, processor $i$ chooses only $y_i$; no decision maker can revise both deliveries together. We study Nash behavior between these processors. This is an institutional restriction on permissible deviations, in addition to the restriction on their information. When deciding what to convey about worker $i$, it observes $(\theta,q,s,r_i)$ but not the other reply $r_j$ or the other delivery decision. Let $y_i$ denote the message about worker $i$ privately delivered to worker $j$, so
\begin{equation}
 y_i=g_i(\theta,q,s,r_i)\in\mathcal R
 \label{eq:exchange-general-relay}
\end{equation}
Each forwarding decision minimizes conditional expected firm loss, holding the other forwarding rule fixed. Local information alone does not justify this restriction: when both deliveries change, the loss difference also contains the cross term $-2\gamma\,\mathbb E[\Delta a_1\Delta a_2]$, where $\Delta a_i$ is the induced action change with worker strategies held fixed. Separate delivery checks therefore do not establish stability against joint revisions.

Worker $i$ remembers their own reply $r_i$ and receives $y_j$, but does not observe $y_i$. The worker's operating action satisfies
\begin{equation}
 a_i=(1-\gamma)\mathbb E_i[D_i\mid T_i,X_i,q,s,r_i,y_j]
          +\gamma\mathbb E_i[a_j\mid T_i,X_i,q,s,r_i,y_j]
 \label{eq:exchange-private-relay-action}
\end{equation}
The expectations use the forwarding rules and reporting strategies of the equilibrium. Delivered messages can convey information about both the worker's private state and the aggregate state.

\subsection{Why unchanged forwarding remains an equilibrium}\label{subsec:exchange-general-relay}

Consider a reporting equilibrium of the baseline after signal $s$, and retain its operating actions and Bayesian interpretation of every label. A worker can change their own report and then re-optimize their action. AI can instead change the delivered label without the sender observing that change. These are different deviations and use different information sets. The next result connects their losses directly.

\begin{proposition}[No incentive to replace an incentive-compatible reply]\label{prop:exchange-no-replacement}
Take any baseline worker reporting equilibrium and let AI forward each reply unchanged. For an original report $r_i$ and alternative label $y_i$, let $\Delta_i^\ell(r_i,y_i;q,s)$ be the increase in worker $i$'s expected loss from reporting $y_i$ and re-optimizing their action, averaged over their private states assigned report $r_i$. Let $\Delta_i^{\mathrm{AI}}(\theta,r_i,y_i;q,s)$ be the increase in firm loss from replacing that report in the separate delivery decision. Then
\begin{equation}
 \Delta_i^{\mathrm{AI}}(\theta,r_i,y_i;q,s)
 =\frac12\Delta_i^\ell(r_i,y_i;q,s)
 +\frac{\gamma^2(1+\gamma^2)}{2(1+\gamma)^2}
       \{\widehat D_i(y_i)-\widehat D_i(r_i)\}^2\geq0
 \label{eq:exchange-independent-forwarding-identity}
\end{equation}
Thus unchanged forwarding is an equilibrium of the separate-processor game even when both processors know the actual $\theta$. The result allows arbitrary independent beliefs and does not require directional alignment. The inequality is strict whenever the alternative label changes the inferred target.
\end{proposition}

Appendix~\ref{subsec:exchange-replacement-proof} derives both loss changes from the original loss functions. The worker incentive averages the state under the worker's posterior; the AI incentive holds the actual state fixed. Independence and the separate delivery decisions make the difference in~\eqref{eq:exchange-independent-forwarding-identity} nonnegative.

Let $L^{\mathrm{rep},*}(q)$ be the lowest firm loss among equilibria under endogenous forwarding, also optimizing the initial signal. Let $L^{\mathrm{truth}}(q;\sigma)$ and $L^{\mathrm{rep}}(q;\sigma)$ denote the lowest firm losses under unchanged and endogenous forwarding, respectively, holding the initial signal policy $\sigma$ fixed. Write $\sigma_T^*$ for an optimal unchanged-forwarding policy, so $L^{\mathrm{truth}}(q;\sigma_T^*)=L^{\mathrm{AI},*}(q)$.

\begin{corollary}[Replacement after full initial disclosure]\label{cor:exchange-full-disclosure-replacement}
If the initial signal fully reveals $\theta$, replacement and unchanged forwarding have the same firm-preferred value under that signal. In particular, under alignment, if $\gamma\geq\sqrt2-1$, $\mu_c\notin(\theta_L,\theta_H)$, or $\sum_iq_i(1-q_i)=0$, take $\sigma_T^*$ to reveal $\theta$. Then
\begin{equation}
 L^{\mathrm{rep}}(q;\sigma_T^*)=L^{\mathrm{AI},*}(q)
 \label{eq:exchange-baseline-full-disclosure}
\end{equation}
No equilibrium under endogenous forwarding improves on unchanged forwarding while this initial policy is held fixed.
\end{corollary}

Once $\theta$ is public, a worker knows which delivered label each report will produce. Reporting followed by replacement can then be represented as a baseline reporting strategy with the same outcomes and incentives. Proposition~\ref{prop:exchange-no-replacement} gives the reverse inclusion: every baseline reporting equilibrium remains an equilibrium of the separate-processor game, with both processors forwarding unchanged. Appendix~\ref{subsec:proof-full-disclosure-replacement} proves the equivalence.

\begin{corollary}[Firm loss under endogenous forwarding]\label{cor:exchange-replacement-firm-value}
Under firm-preferred equilibrium selection,
\begin{equation}
 L^{\mathrm{rep},*}(q)\leq L^{\mathrm{AI},*}(q)\leq L^{\mathrm{direct}}(q)
 \label{eq:exchange-three-regime-ordering}
\end{equation}
\end{corollary}

The proof is in Appendix~\ref{subsec:proof-main-firm-ordering}.

\subsection{When selective forwarding improves firm performance}\label{subsec:exchange-baseline-optimal-comparison}

The firm chooses its initial signal anticipating endogenous forwarding. The next result identifies a region in which selective forwarding lowers firm loss relative to optimal unchanged forwarding.

\begin{proposition}[Sufficient conditions for gains from selective forwarding]\label{prop:exchange-baseline-optimal-reduction}
There exists a threshold $\bar\gamma\approx0.652$, the root in $(0,1)$ of $3\gamma^3+5\gamma^2-3\gamma-1=0$, such that, for every $0<\gamma<\bar\gamma$, there is a nonempty open region of independent type priors and aligned uniform aggregate supports for which
\begin{equation}
 L^{\mathrm{rep},*}(q)<L^{\mathrm{AI},*}(q)
 \label{eq:exchange-replacement-region-gain}
\end{equation}
A sufficient region has common high-type priors close to one and a narrow aggregate support centered at $\beta u/h$, with $u$ chosen as described below. Small perturbations, including asymmetric type priors, preserve the strict improvement.
\end{proposition}

To construct the region, let $q_1=q_2=1-\varepsilon$ for sufficiently small $\varepsilon>0$. The number $u$ is the center of the aggregate support measured in units of $\beta/h$. Choose
\begin{equation}
 u>\max\left\{1+\gamma,\frac{2\gamma}{1-\gamma}\right\},\qquad
 (1-2\gamma)u<1+4\gamma
 \label{eq:exchange-replacement-center-incentives}
\end{equation}
When $0<\gamma<1/2$, take $u$ sufficiently close to, but below, $(1+4\gamma)/(1-2\gamma)$. When $1/2\leq\gamma<\bar\gamma$, take $u$ sufficiently large. These choices make the firm gain positive as well as satisfying~\eqref{eq:exchange-replacement-center-incentives}. Set
\begin{equation}
 \theta\sim U\left[\frac{\beta}{h}(u-w_\varepsilon),
                   \frac{\beta}{h}(u+w_\varepsilon)\right],\qquad
 w_\varepsilon=
 \frac{4\gamma^2\{(1-\gamma)u-2\gamma\}}
      {\gamma(1-\gamma)u+1+2\gamma-\gamma^2}\varepsilon+O(\varepsilon^2)
 \label{eq:exchange-allgamma-support}
\end{equation}
Appendix~\ref{subsec:exchange-baseline-optimal-proof} defines the width exactly and proves the incentive and welfare comparisons. The expansion describes its size; the incentive conditions are checked at the exact width. The center is chosen first, and then $\varepsilon$ is taken sufficiently small.

AI initially sends an uninformative signal. Workers report both $T_i$ and $X_i$, and AI always conveys $X_i$. What changes with the aggregate state is whether the delivered label also distinguishes types. To write the rule using the same four labels, identify $\mathcal R$ with $\{-1,1\}^2$. After a report $r_i=(\widetilde T_i,\widetilde X_i)$, AI delivers
\begin{equation}
 g_i(\theta,q,s,r_i)=
 \begin{cases}
  (1,\widetilde X_i),&\widetilde T_i=1\text{ and }\theta\geq\beta u/h\\
  (-1,\widetilde X_i),&\text{otherwise}
 \end{cases}
 \label{eq:exchange-allgamma-delivery}
\end{equation}
The worker's equilibrium report is $r_i=(T_i,X_i)$. A delivered label with first component $1$ identifies a high type and an aggregate state in the upper half of the support. A label with first component $-1$ can come from a low type or a high type in the lower half. Workers interpret these labels using Bayes' rule. Both worker reporting and each processor's separate forwarding choice are incentive compatible.

For sufficiently small $\varepsilon$, the entire support in~\eqref{eq:exchange-allgamma-support} lies above $\mu_c$. Optimal unchanged forwarding therefore reveals $\theta$ fully and elicits factual replies without type disclosure. Corollary~\ref{cor:exchange-full-disclosure-replacement} rules out improvement if that initial signal is retained. Proposition~\ref{prop:exchange-baseline-optimal-reduction} instead withholds aggregate information initially and conveys some of it after workers reply. A worker reporting their type to AI thereby selects a state-dependent delivery rule, rather than directly choosing the coworker's type belief. This changes reporting incentives while preserving all factual information. The proof compares the resulting firm loss, including the loss from less precise aggregate information, with the globally optimized unchanged-forwarding value.

\subsection{When selective forwarding does not improve firm performance}\label{subsec:exchange-replacement-implications}

Coordination motives and preference biases are both necessary for selective forwarding to improve on optimal unchanged forwarding.

\begin{proposition}[No improvement without coordination or preference bias]\label{prop:exchange-no-gain-redesign}
Extend the model to the boundary cases $\gamma=0$ and $\beta=0$, and suppose that either $\gamma=0$ or $\beta=0$. Then, even after optimizing initial disclosure and all admissible equilibria under endogenous forwarding,
\begin{equation}
 L^{\mathrm{rep},*}(q)=L^{\mathrm{AI},*}(q)
 \label{eq:exchange-no-gain-redesign}
\end{equation}
The result allows arbitrary independent type priors. In the first case the common value is $\beta^2$. In the second case it is $2\gamma(1-\gamma)h^2\mathbb E[\theta^2]/(1+\gamma)^2$.
\end{proposition}

When $\gamma=0$, a coworker's information cannot change the worker's desired coordination response, and full aggregate disclosure removes all avoidable firm loss. When $\beta=0$, full aggregate disclosure and factual communication already minimize firm loss. With both coordination motives and preference biases present, selective forwarding can improve performance even when $\gamma>\sqrt2-1$, as the preceding construction shows. The proof is in Appendix~\ref{subsec:proof-no-gain-redesign}.

\section{Repeated collaboration}\label{sec:exchange-dynamics}

We now consider several decision stages within one task, indexed by $t$, and call them periods. The preference type $T_i$ is drawn once and remains fixed, while $X_{it}$ and $\theta_t$ are independently redrawn each period with the distributions in Section~\ref{sec:exchange-setup}. Workers and the firm discount losses by $\delta\in[0,1]$ in a finite horizon. Initial type beliefs $q^0=(q_1^0,q_2^0)$ are arbitrary and independent. We first compare the optimal two-period policy with the static policy. Under low coordination incentives, future losses can sustain full worker revelation without the aggregate-state pooling needed in the static model. Under high coordination incentives and alignment, the static preference for factual disclosure without type information extends to both periods.

This extension returns to committed unchanged forwarding in every period. The firm chooses the sequence of public disclosure policies before the task begins; the separate delivery decisions of Section~\ref{sec:exchange-relay-extension} are not part of this dynamic game.

Neither actual operating actions nor realized payoffs are observed by the coworker or AI before the task ends. Workers know their own actions, but there is no intermediate payoff feedback. An operating action therefore does not affect future information. Reports can reveal persistent types, and fresh local conditions must still be communicated in later periods.

Let $\mathcal H_{t-1}$ be the public history of initial signals and forwarded report pairs, and let $q_{it}=\Pr(T_i=1\mid\mathcal H_{t-1})$ and $q_t=(q_{1t},q_{2t})$, with $q_1=q^0$; in this section, a single subscript on $q$ indexes the period. The firm commits to the complete disclosure policy, allowing later signals to depend on public history. These histories can prescribe different continuations even when they imply the same type beliefs. Worker reports update the two type posteriors independently. Appendix~\ref{subsec:exchange-updating} specifies the histories, updates, and continuation values. We write $\ell_{it}(T,\mathcal H_{t-1})$ for worker $i$'s minimum continuation loss conditional on their actual type, and $L_t(\mathcal H_{t-1})$ for firm loss. A worker who has misreported their type continues to optimize using their actual type.

\subsection{Two periods: Reporting incentives}\label{subsec:exchange-dynamic-learning-ic}

Set $\ell_{i3}=L_3=0$. Period 2 is the static game at the inherited $q_2$, so we first solve the second-period problem and then check first-period reporting incentives. In period 1, an operating action still solves Lemma~\ref{lem:exchange-actions}, but the worker's preceding reply affects continuation losses. After signal $s_1$, the worker chooses their reply to minimize
\begin{equation}
 \begin{aligned}
 &\gamma(1-\gamma)\left\{(h\mu_1+\beta T_i)X_{i1}
                     -\frac{\gamma}{1+\gamma}\widehat D_i(r_i)\right\}^2\\
 &\qquad+\delta\sum_{r_j}P_j(r_j\mid q_1,s_1)
                 \ell_{i2}(T_i,(s_1,r_1,r_2))
 \end{aligned}
 \label{eq:exchange-dynamic-replycost}
\end{equation}
Here $P_j(r_j\mid q_1,s_1)$ is the probability of coworker report $r_j$, and report pairs are ordered as $(r_1,r_2)$. The first term is current operating loss after re-optimization, omitting terms independent of the worker's report. The second averages future loss over the coworker's report and the future draws already integrated in $\ell_{i2}$. A current report changes the public history and can therefore change both future type beliefs and the firm's subsequent disclosure.

We first ask what the firm should disclose in period 2 if first-period replies reveal both workers' types. We then use that optimal continuation to determine when full first-period reporting of $(T_i,X_{i1})$ is an equilibrium, rather than assuming workers report truthfully. Write $\overline\theta=\mathbb E[\theta]$.

\begin{lemma}[Second-period disclosure and first-period reporting]\label{lem:exchange-dynamic-full-reporting}
Suppose $q^0\in(0,1)^2$ and $h\theta_L>\beta$.
\begin{enumerate}
 \item If first-period replies fully reveal both persistent types, then $q_2\in\{0,1\}^2$. An optimal second-period policy fully reveals $\theta_2$, with workers reporting their fresh local conditions $X_{i2}$.

 \item Under this second-period policy, full first-period reporting of $(T_i,X_{i1})$ is an equilibrium after signal $s_1$ if and only if
\begin{equation}
 \mu_1+\delta\overline\theta\leq(1+\delta)\mu_c
 \label{eq:exchange-dynamic-full-reporting-condition}
\end{equation}
\end{enumerate}
\end{lemma}

The first part follows directly from Proposition~\ref{prop:exchange-aligned-signal}: once types are known, the firm no longer needs to withhold aggregate information to encourage preference disclosure. Reporting fresh $X_{i2}$ remains useful because local conditions are redrawn. At these known-type histories, we write the worker's continuation loss as $\ell_{i2}(T,q_2)$ because the specified optimal continuation depends on history only through $q_2$.

The second part checks whether workers want to reveal their types in the first place.

The condition accounts for how a false high-type claim changes coworker responses to both the current local condition and fresh future conditions. If $\overline\theta<\mu_c$, that future response is too strong for an actual low type, so patience deters exaggeration and expands the current full-reporting region. If $\overline\theta>\mu_c$, patience instead encourages exaggeration. Aggregate-state variance cancels from this particular incentive comparison. Appendix~\ref{subsec:exchange-two-period-proofs} derives the deviation gain and checks all four reply deviations, retaining the worker's actual type when public beliefs exclude it.

\subsection{Low coordination: Full disclosure instead of static pooling}\label{subsec:exchange-dynamic-optimum}

The static firm sometimes withholds aggregate information to make full worker revelation incentive compatible. The next result shows that future reporting incentives can remove that need, even with the same aggregate-state distribution in both periods.

\begin{proposition}[Full disclosure with early preference learning]\label{prop:exchange-dynamic-optimum}
Suppose $q^0\in(0,1)^2$, $h\theta_L>\beta$, $0<\gamma<\sqrt2-1$, and
\begin{equation}
 \theta_H+\delta\overline\theta\leq(1+\delta)\mu_c
 \label{eq:exchange-dynamic-optimum-region}
\end{equation}
A jointly optimal committed policy fully reveals $\theta_t$ in both periods. Workers fully reveal $(T_i,X_{i1})$ in period 1 and report the fresh $X_{i2}$ in period 2. At the initial history we write the values as functions of $q^0$. The optimal firm loss and each worker's loss conditional on their actual type are
\begin{align}
 L_1^*(q^0)
 &=\frac{(1+\delta)(1-\gamma)}{(1+\gamma)^2}
   \left\{2\gamma h^2\mathbb E[\theta^2]
      +\gamma(1-\gamma)h\beta\overline\theta\sum_k(2q_k^0-1)
      +(1+2\gamma-\gamma^2)\beta^2\right\}
 \label{eq:exchange-dynamic-optimal-firm}\\
 \ell_{i1}^*(T,q^0)
 &=\frac{(1+\delta)\gamma(1-\gamma)}{(1+\gamma)^2}
   \left\{2h^2\mathbb E[\theta^2]
      +2h\beta\overline\theta(T+2q_j^0-1)+2\beta^2\right\}
 \label{eq:exchange-dynamic-optimal-worker}
\end{align}
Compared with firm-preferred direct communication without AI, the firm and every supported worker type gain
\begin{equation}
 L_1^{\mathrm{direct}}-L_1^*
 =\ell_{i1}^{\mathrm{direct}}(T,q^0)-\ell_{i1}^*(T,q^0)
 =(1+\delta)A_\gamma h^2\operatorname{Var}(\theta)>0
 \label{eq:exchange-dynamic-direct-gain}
\end{equation}

If, in addition, $\theta_L<\mu_c<\theta_H$, full aggregate-state revelation is suboptimal in the static model: pooling a sufficiently small interval around $\mu_c$ strictly reduces static firm loss. In the two-period model, the policy above instead fully reveals the aggregate state in both periods and elicits full worker revelation in period 1.
\end{proposition}

The policy attains a full-information lower bound in both periods, as shown in Appendix~\ref{subsec:exchange-two-period-proofs}. Future accommodation deters a false preference claim when $\overline\theta<\mu_c$, sustaining reporting where the static incentive fails. Learning then removes the need to elicit preferences again, although fresh local conditions still require communication.

When $\theta_H>\mu_c>\overline\theta$, condition~\eqref{eq:exchange-dynamic-optimum-region} is equivalent to $\delta\geq(\theta_H-\mu_c)/(\mu_c-\overline\theta)$, provided this bound is at most one.

Outside this optimality region, learning reduces the expected squared pooling width under a statically optimal continuation. Full type learning eliminates the need to pool project states to elicit preferences. Appendix~\ref{subsec:exchange-learning-pooling} derives the posterior-variance identity and states the interiority restrictions. A jointly committed policy may instead choose a different continuation to influence first-period reports.

\subsection{High coordination: Factual disclosure without type learning}\label{subsec:exchange-dynamic-high-coordination}

Under alignment, the static firm prefers workers to reveal their local conditions without also revealing their types when $\gamma>\sqrt2-1$. The question is whether future learning can reverse that preference. The next result shows that it does not.

\begin{proposition}[Optimal disclosure without preference learning]\label{prop:exchange-dynamic-high-coordination}
Suppose $q^0\in(0,1)^2$, $h\theta_L>\beta$, and $\sqrt2-1\leq\gamma<1$. For every $\delta\in[0,1]$, a jointly optimal committed policy fully reveals $\theta_t$ in each period and selects worker reporting of $X_{it}$ without revealing $T_i$. The type beliefs remain $q_t=q^0$, and
\begin{equation}
 L_1^*(q^0)=(1+\delta)L^{\mathrm{AI},*}(q^0)
 \label{eq:exchange-dynamic-high-coordination-value}
\end{equation}
Here $L^{\mathrm{AI},*}(q^0)$ is the optimal static firm loss defined in~\eqref{eq:exchange-direct-ai-ordering}. For $\gamma>\sqrt2-1$, any policy producing positive-probability type learning in period 1 has strictly greater total expected firm loss when $\delta>0$. At $\gamma=\sqrt2-1$, the stated policy is one optimum; concealing type need not be strictly preferred.
\end{proposition}

The firm fully reveals its aggregate information, but prefers workers to reveal only factual information. These are different disclosure decisions. Knowing $X_{it}$ allows a coworker to respond to the productive condition; knowing $T_i$ additionally changes their response to the worker's private objective. At high coordination incentives, this additional accommodation hurts the firm in the current period, as in the static model. Learning the persistent type also carries that cost into the next period. In fact,
\begin{equation}
 \mathbb E[L^{\mathrm{AI},*}(q_2)]-L^{\mathrm{AI},*}(q^0)
 =-4b_\gamma\beta^2\sum_i\operatorname{Var}(q_{i2})\geq0
 \label{eq:exchange-dynamic-high-coordination-learning}
\end{equation}
The proof in Appendix~\ref{subsec:exchange-two-period-proofs} uses this future-loss bound together with a current-loss bound that does not impose static reporting incentives on period-1 replies.

Thus there is no static-to-dynamic reversal under the alignment assumptions of this proposition. Full worker revelation can still be an equilibrium when Lemma~\ref{lem:exchange-dynamic-full-reporting} permits it, but it is not the firm's preferred equilibrium. The static cases in which type disclosure can help at high $\gamma$ allow the low type to favor the opposite direction from the firm, as characterized in Appendix~\ref{app:exchange-general}; they are outside $h\theta_L>\beta$.

\section{The value and timing of project information}\label{sec:exchange-simulations}

An organization can provide project context before employees report, or retain information for subsequent delivery. We compare the gains from disclosure design and distinguish the timing of project information from the concealment of employee reports.

\subsection{Gains across information regimes}\label{subsec:exchange-simulation-gains}

We set $h=\beta=1$ and compare $\theta\sim U[1,2]$, where workers agree with the firm's productive direction, with $U[0,2]$, which includes opposing directions. Each support uses $\gamma\in\{0.1,0.2,\ldots,0.9\}$ and $q_1=q_2\in\{0.05,0.2,0.5,0.8,0.95\}$. Workers can still disagree with the firm on intensity under alignment.

\begin{figure}[htbp]
\centering
\includegraphics[width=\textwidth]{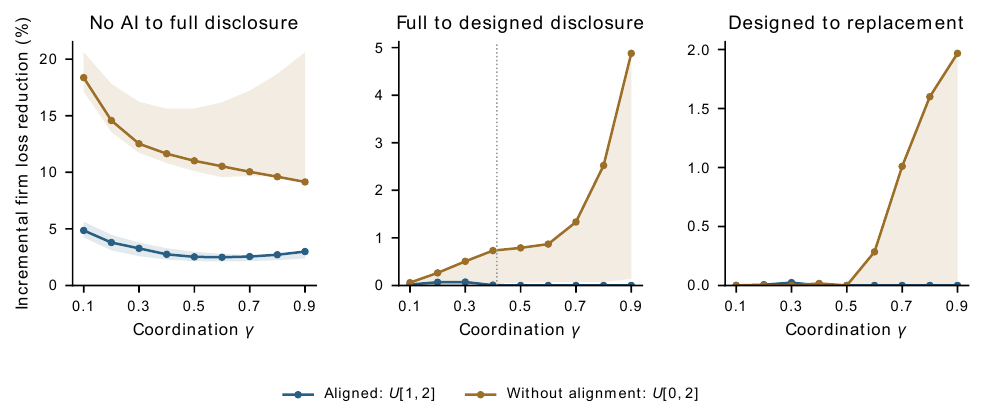}
\caption{Incremental reductions in firm loss across information regimes}\label{fig:exchange-simulation-incremental-gains}
\par\vspace{8pt}\noindent
\begin{minipage}{\textwidth}\footnotesize
Solid curves use $q_1=q_2=0.5$; shaded ranges span the five tested priors and are not confidence intervals. The vertical guide marks $\gamma=\sqrt2-1$. Each panel uses the preceding regime's loss as denominator. Forwarding gains are evaluated at the equilibria found by the numerical search. The two supports differ in their means and variances as well as preference disagreement.
\end{minipage}
\end{figure}

Figure~\ref{fig:exchange-simulation-incremental-gains} compares direct communication without project information, full initial disclosure, designed initial disclosure, and endogenous forwarding. The middle two regimes forward reports unchanged; all four retain strategic reporting and decentralized actions. Separate processors choose deliveries under the restriction in Section~\ref{sec:exchange-relay-extension}. Each percentage uses the preceding regime's loss as denominator and is not additive across panels. The aligned unchanged-forwarding policy is analytically optimal; other comparisons concern restricted searches described in Appendix~\ref{app:exchange-simulation-details}.

Under alignment, full initial disclosure accounts for most of the improvement. At $\gamma=0.3$ and $q_1=q_2=0.5$, it reduces firm loss by about $3.3\%$ relative to direct communication. Optimizing initial disclosure adds about $0.07\%$, and selective forwarding adds about $0.02\%$ relative to optimal unchanged forwarding. In this aligned comparison, workers can already communicate productive facts, so further design mainly changes preference revelation and accommodation.

The gains are larger when the support includes opposing targets and coordination is strong. Under alignment, designed initial disclosure adds nothing at $\gamma\geq\sqrt2-1$, where full disclosure is optimal (Proposition~\ref{prop:exchange-direct-worker-loss}). On $U[0,2]$, at $\gamma=0.9$ and $q_1=q_2=0.5$, the reported initial-disclosure policy improves on full disclosure by about $4.9\%$, and the forwarding candidate improves on that policy by a further $2.0\%$. At $\gamma=0.8$, the corresponding gains are about $2.5\%$ and $1.6\%$. The policies can now affect whether workers communicate productive facts. Workers need not share the firm's ranking: in the $\gamma=0.9$ example, designed initial disclosure raises average worker loss by about $3.7\%$ relative to full disclosure. Endogenous forwarding partly offsets that increase, but worker loss remains about $2.6\%$ above the full-disclosure level.

\subsection{Project information after elicitation}\label{subsec:exchange-timing-implications}

The selective-forwarding example at $\gamma=0.9$, $q_1=q_2=0.5$, and $\theta\sim U[0,2]$ preserves workers' factual reports. Its initial signal pools approximately $[0.35,1.48]$. Workers report $X_i$ using two messages; subsequent delivery identifies both that report and whether $\theta$ lies below or above approximately $0.91$. The recipient therefore learns the local condition and an additional project-state indicator. Forwarding the factual report unchanged and releasing that indicator afterward provides the same information. Appendix~\ref{subsec:exchange-timing-proof} formalizes this informational equivalence. In this example, the gain of about $2\%$ comes from providing additional project context after reporting, not from concealing worker information.

To isolate the effect of timing, consider a separate disclosure protocol. The initial signal is uninformative, type priors are balanced, and the firm chooses a public project signal $C$ to release after workers report their local signs. The policy generating $C$ is known before reporting. Write $M=\mathbb E[\theta\mid C]$ and $\mu=\mathbb E[\theta]>0$. After optimally adjusting its operating action, a low type prefers its factual report to the opposite sign if and only if
\begin{equation}
 h\mathbb E[M^2]\geq\beta\mu.
 \label{eq:exchange-delayed-context-ic}
\end{equation}
Releasing the same signal before elicitation instead requires $hM\geq\beta$ at each positive posterior mean. Delayed disclosure permits favorable and unfavorable project contexts to enter the reporting incentive together, while operating decisions still use the eventual context. 

\begin{figure}[htbp]
\centering
\subfloat[Initial and eventual project information]{\includegraphics[width=0.485\textwidth]{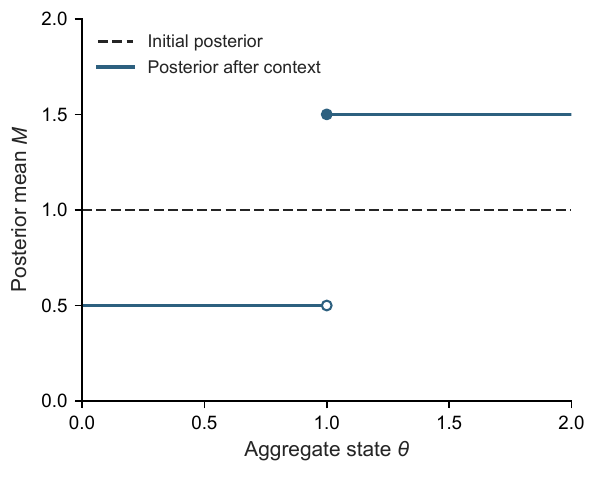}}\hfill
\subfloat[Regions supporting factual reporting]{\includegraphics[width=0.485\textwidth]{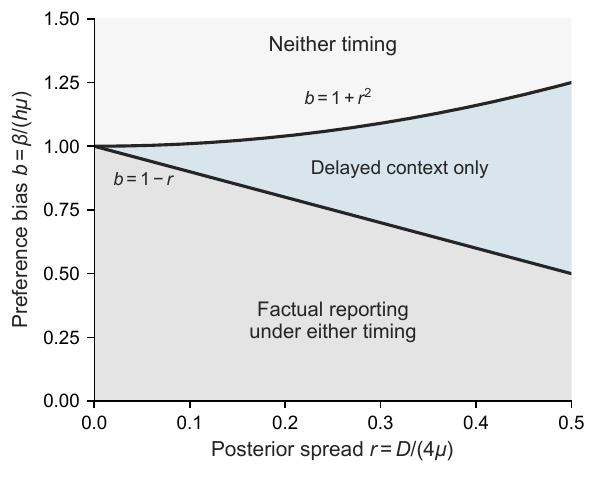}}
\caption{Project disclosure and the timing of factual reports}\label{fig:exchange-timing-mechanism}
\par\vspace{8pt}\noindent\begin{minipage}{\textwidth}\footnotesize
Panel (a) uses $U[0,2]$: a midpoint signal changes the posterior mean from 1 to 0.5 or 1.5. Panel (b) uses a nonnegative uniform prior with mean $\mu>0$ and width $D$, balanced type priors, and $0<\gamma<1$. Define $r=D/(4\mu)\leq0.5$ and $b=\beta/(h\mu)$. Midpoint context supports factual reporting before its release when $b\leq1+r^2$; releasing it first instead requires $b\leq1-r$.
\end{minipage}
\end{figure}

For $\theta\sim U[0,2]$, midpoint disclosure gives $\mathbb E[M^2]=1.25$. With $h=\beta=1$, factual reporting is strict when context arrives after reports but fails for low types following the lower signal if context arrives first. Figure~\ref{fig:exchange-timing-mechanism}(b) extends this comparison across prior spread and preference bias. At $r=0.5$, early disclosure supports factual reporting up to normalized bias $b=0.5$, whereas delayed disclosure supports it up to $b=1.25$. Appendix~\ref{subsec:exchange-delayed-binary} proves the midpoint's optimality within the deterministic binary-signal family and illustrates the reporting margins. If the later project message is unrestricted, revealing $\theta$ fully maximizes $\mathbb E[M^2]$ and minimizes both losses within the same factual-reporting family whenever that equilibrium is feasible. The binary comparison fixes the capacity of the later message; it is not a global optimum over delayed disclosure.

Selective forwarding can also conceal local conditions reported by workers, as the example in Appendix~\ref{subsec:exchange-concealing-simulation} shows. There, AI uses its private project information to decide which reports to combine. This differs from the preceding example, where every reported local condition remains recoverable. The two examples distinguish concealing worker information from conveying additional project context after reporting.

\FloatBarrier
\section{Conclusion}\label{sec:exchange-conclusion}

An assistant that supports teamwork participates in a strategic communication process. Project information changes both what employees know and what they are willing to tell one another. Our model distinguishes productive local facts from private objectives and shows how disclosure affects their communication. The firm must therefore anticipate the reports and actions induced by its information policy, as well as the direct value of the information it releases.

The appropriate policy depends on coordination needs and on the content of employees' private information. When workers share the firm's productive direction, pooling selected project states under weak coordination can preserve useful preference reporting at a cost in adaptation. Under strong coordination, additional preference information encourages accommodation that the firm finds costly; full project disclosure with factual reporting then serves it better. These conclusions concern objectives that shift workers' targets away from the productive objective, not productive expertise.

Worker acceptance creates a separate design constraint. A firm-optimal disclosure policy can improve organizational performance while leaving workers worse off than under direct exchange. Within the centered-interval family, widening the pool restores ex-ante acceptance while retaining a positive firm gain. A small shift below the reporting threshold can also provide a strict preference-reporting incentive. Participation and credible reporting are distinct requirements, with distinct costs. The numerical examples illustrate those costs under common primitives.

Timing provides another instrument. In the balanced-type benchmark, releasing project context after factual reports can sustain communication that releasing the same context beforehand would prevent. Every reported fact can still reach the coworker. This distinction matters when evaluating an intermediary: a delivery gain may arise from delayed context rather than concealment. Under the separate-processor restriction, selective forwarding can improve on unchanged forwarding. Repeated collaboration introduces a further incentive effect because a preference claim changes accommodation in future periods.

The analysis uses firm-preferred reporting equilibria, and worker acceptance is imposed ex ante within a restricted disclosure family. The forwarding results require separate delivery decisions; the delayed-context result fixes factual reporting and balanced type priors. These restrictions identify settings in which the mechanisms operate while leaving broader communication design open. Intermediate performance feedback could change learning and reporting incentives, and alternative equilibrium selection could change the disclosure policy workers implement.

\putbib[main]
\clearpage

\ifdefined\APPENDICES
\begin{APPENDICES}
\else
\appendix
\setcounter{section}{0}
\renewcommand{\thesection}{\Alph{section}}
\fi

\renewcommand{\theHsection}{appendix.\Alph{section}}
\renewcommand{\theHsubsection}{\theHsection.\arabic{subsection}}
\renewcommand{\theHsubsubsection}{\theHsubsection.\arabic{subsubsection}}

\section{Proofs of main-text results}\phantomsection\label{app:exchange-main-proofs}

This appendix collects the proofs of the main-text results.

\subsection{Operating actions}\phantomsection\label{subsec:proof-main-actions}

\proofheading{Proof of Lemma~\ref{lem:exchange-actions}.}
The best response is $a_i=(1-\gamma)d_i+\gamma\mathbb E_i[a_j]$. Conditional independence in~\eqref{eq:exchange-factorization} implies that the coworker's expected action equals its public conditional mean. Those means satisfy $\overline a_i=(1-\gamma)\widehat D_i+\gamma\overline a_j$. Solving gives $\overline a_j=(\widehat D_j+\gamma\widehat D_i)/(1+\gamma)$ and hence~\eqref{eq:exchange-actions}. The best-response operator has contraction modulus $\gamma<1$, establishing uniqueness.
\proofend

\subsection{Reporting incentives}\phantomsection\label{app:exchange-reply-derivation}

Fix a proposed reporting profile $(\lambda_1,\lambda_2)$ and the posteriors it induces through~\eqref{eq:exchange-posterior}. We first calculate the worker's full expected operating loss under optimal action. Conditional on their actual private state and the observed replies, the coworker's action has mean $[\widehat D_j(r_j)+\gamma\widehat D_i(r_i)]/(1+\gamma)$ and variance $(1-\gamma)^2\operatorname{Var}(d_j\mid q,s,r_j)$. Substituting their own best response into~\eqref{eq:exchange-workerloss} gives
\[
 \begin{aligned}
 \min_{a_i}\mathbb E_i[\ell_i\mid T_i,X_i,q,s,r_i,r_j]
 &=(1-\gamma)h^2\operatorname{Var}(\theta\mid q,s)
   +\gamma(1-\gamma)^2\operatorname{Var}(d_j\mid q,s,r_j)\\
 &\quad+\gamma(1-\gamma)\left\{d_i-\frac{\widehat D_j(r_j)+\gamma\widehat D_i(r_i)}{1+\gamma}\right\}^2
 \end{aligned}
\]
The first term is the loss from the remaining uncertainty about the aggregate state. The second is the loss from uncertainty about the coworker's action. The last term measures the difference between the worker's expected preferred target and their prediction of the coworker's action, after adjusting their own action optimally.

The coworker's reply is independent of worker $i$'s private state and chosen reply, and $\sum_{r_j}P_j(r_j\mid q,s)\widehat D_j(r_j)=\mathbb E[d_j\mid q,s]=0$. Averaging the full loss over the coworker's reply therefore gives
\[
 \begin{aligned}
 &\sum_{r_j\in\mathcal R}P_j(r_j\mid q,s)
       \min_{a_i}\mathbb E_i[\ell_i\mid T_i,X_i,q,s,r_i,r_j]\\
 &\quad=(1-\gamma)h^2\operatorname{Var}(\theta\mid q,s)
   +\gamma(1-\gamma)^2\sum_{r_j\in\mathcal R}P_j(r_j\mid q,s)\operatorname{Var}(d_j\mid q,s,r_j)\\
 &\qquad+\frac{\gamma(1-\gamma)}{(1+\gamma)^2}
       \sum_{r_j\in\mathcal R}P_j(r_j\mid q,s)\widehat D_j(r_j)^2
   +\gamma(1-\gamma)\left\{d_i-\frac{\gamma}{1+\gamma}\widehat D_i(r_i)\right\}^2
 \end{aligned}
\]
Only the last term depends on the chosen reply, establishing~\eqref{eq:exchange-message-cost} and~\eqref{eq:exchange-terminal-ic}.

\subsection{Firm-loss decomposition}\phantomsection\label{app:exchange-firm}

This subsection derives the expected firm loss in~\eqref{eq:exchange-expected-firm-loss}. Fix the current type beliefs $q$, initial signal $s$, and worker reporting strategies. In this subsection, $\mathbb E$ denotes the firm's expectation conditional on $(q,s)$. It averages worker types, local conditions, and the resulting replies; expressions involving $H_i$ also average the remaining uncertainty about $\theta$. Worker actions use $d_i=(h\mu+\beta T_i)X_i$, while the firm's realized target remains $H_i=h\theta X_i$. All replies are forwarded unchanged.

\proofheading{Proof.}
First, the coworker's prediction is $\widehat D_i=\mathbb E[d_i\mid q,s,r_i]$. The independent local condition has mean zero, so $\mathbb E[d_i]=0$. Iterated expectations then give $\mathbb E[\widehat D_i]=0$, and conditioning on the reply gives
\begin{equation}
 \begin{aligned}
 \mathbb E[d_i]&=\mathbb E[\widehat D_i]=0\\
 \mathbb E[d_i\widehat D_i]
 &=\mathbb E[\widehat D_i\mathbb E[d_i\mid q,s,r_i]]
   =\mathbb E[\widehat D_i^2]
 \end{aligned}
 \label{eq:exchange-app-posterior-moments}
\end{equation}
The two workers' private states and replies are independent conditional on $(q,s)$. The expectation of a product involving one variable from each worker therefore equals the product of their expectations. Since these expectations are zero,
\begin{equation}
 \mathbb E[d_i d_j]=\mathbb E[d_i\widehat D_j]
                  =\mathbb E[\widehat D_i\widehat D_j]=0,\qquad i\ne j
 \label{eq:exchange-app-cross-moments}
\end{equation}
No balanced-type assumption is needed: the zero means follow from the independent symmetric local conditions, even if $q_1\ne q_2$ or either type is much more likely.

Next substitute the operating action $a_i=(1-\gamma)d_i+\gamma^2\widehat D_i/(1+\gamma)+\gamma\widehat D_j/(1+\gamma)$. In its squared expectation, all products involving different workers vanish by~\eqref{eq:exchange-app-cross-moments}. The product of $d_i$ and $\widehat D_i$ remains, and~\eqref{eq:exchange-app-posterior-moments} replaces its expectation by $\mathbb E[\widehat D_i^2]$. Its coefficient, together with the square of the prediction term, is
\[
 \frac{2(1-\gamma)\gamma^2}{1+\gamma}
       +\frac{\gamma^4}{(1+\gamma)^2}
 =\frac{\gamma^2(2-\gamma^2)}{(1+\gamma)^2}
\]
Consequently,
\begin{equation}
 \mathbb E[a_i^2]=(1-\gamma)^2\mathbb E[d_i^2]
       +\frac{\gamma^2(2-\gamma^2)}{(1+\gamma)^2}\mathbb E[\widehat D_i^2]
       +\frac{\gamma^2}{(1+\gamma)^2}\mathbb E[\widehat D_j^2]
 \label{eq:exchange-app-action-square}
\end{equation}
For the product of the two actions, cross-worker products again vanish. The remaining terms pair $d_i$ with $\widehat D_i$, or $\widehat D_i$ with itself, giving
\begin{equation}
 \mathbb E[a_1a_2]
   =\left\{\frac{\gamma(1-\gamma)}{1+\gamma}+\frac{\gamma^3}{(1+\gamma)^2}\right\}\sum_i\mathbb E[\widehat D_i^2]
   =\frac{\gamma}{(1+\gamma)^2}\sum_i\mathbb E[\widehat D_i^2]
 \label{eq:exchange-app-action-product}
\end{equation}
The last equality uses $(1-\gamma)(1+\gamma)+\gamma^2=1$.

We also need the expected product of an action and the firm's target $H_i=h\theta X_i$. Conditional on $(q,s)$, the remaining uncertainty about $\theta$ is independent of worker private information and replies. The action depends on the signal's posterior mean, not on the unobserved actual state. Therefore $\mathbb E[H_i^2]=h^2\eta$ and $\mathbb E[a_iH_i]=h\mu\mathbb E[a_iX_i]$. Substituting the action yields
\begin{equation}
 \mathbb E[a_iH_i]
   =(1-\gamma)h\mu\mathbb E[d_iX_i]
      +\frac{\gamma^2}{1+\gamma}h\mu\mathbb E[X_i\widehat D_i]
 \label{eq:exchange-app-action-target}
\end{equation}
The product $X_i\widehat D_j$ has expectation zero by independence across workers and $\mathbb E[X_i]=0$.

We can now substitute these expectations into the firm's loss. Expanding the two squared terms in~\eqref{eq:exchange-firmloss} and averaging gives
\begin{equation}
 \begin{aligned}
 \mathbb E[L_O]&=\frac{1+\gamma}{2}\sum_i\mathbb E[a_i^2]
                  -2\gamma\mathbb E[a_1a_2]\\
 &\quad-(1-\gamma)\sum_i\mathbb E[a_iH_i]+\frac{1-\gamma}{2}\sum_i\mathbb E[H_i^2]
 \end{aligned}
 \label{eq:exchange-app-expanded-loss}
\end{equation}
Substituting~\eqref{eq:exchange-app-action-square}--\eqref{eq:exchange-app-action-target}, the coefficient on each $\mathbb E[X_i\widehat D_i]$ is $-(1-\gamma)\gamma^2h\mu/(1+\gamma)$. The coefficient on each $\mathbb E[\widehat D_i^2]$ is
\begin{equation}
 \frac{\gamma^2(3-\gamma^2)}{2(1+\gamma)}
       -\frac{2\gamma^2}{(1+\gamma)^2}
 =-\frac{(1-\gamma)\gamma^2(1-2\gamma-\gamma^2)}{2(1+\gamma)^2}=-b_\gamma
 \label{eq:exchange-app-loss-coefficients}
\end{equation}
Collecting the remaining terms and using the definition of $\Delta_i$ in~\eqref{eq:exchange-firm-decomposition} gives
\begin{equation}
 \begin{aligned}
 \mathbb E[L_O]
 &=\frac{(1+\gamma)(1-\gamma)^2}{2}\sum_i\mathbb E[d_i^2]
   -(1-\gamma)^2h\mu\sum_i\mathbb E[d_iX_i]\\
 &\quad+(1-\gamma)h^2\eta-\sum_i\Delta_i(q,s)
 \end{aligned}
 \label{eq:exchange-app-final-decomposition}
\end{equation}
Since $\mathbb E[T_i]=2q_i-1$, $T_i^2=X_i^2=1$, and $d_i=(h\mu+\beta T_i)X_i$, the two remaining expectations are
\begin{equation}
 \mathbb E[d_i^2]=h^2\mu^2+2h\mu\beta(2q_i-1)+\beta^2,\qquad
 \mathbb E[d_iX_i]=h\mu+\beta(2q_i-1)
 \label{eq:exchange-app-private-moments}
\end{equation}
Substituting these moments into~\eqref{eq:exchange-app-final-decomposition} and collecting the squared, linear, and constant terms in $\mu$ gives~\eqref{eq:exchange-expected-firm-loss}. At balanced type beliefs, the term proportional to $\sum_i(2q_i-1)$ is zero. This completes the derivation.
\proofend

\subsection{Factual and preference disclosure}\phantomsection\label{subsec:proof-main-aligned-reporting}

\proofheading{Proof of Proposition~\ref{prop:exchange-aligned-reporting}.}
Under full revelation, the coworker's prediction is $d_i=(h\mu+\beta T_i)X_i$. Under reporting $X_i$, it is $[h\mu+\beta(2q_i-1)]X_i$. Both rules therefore give
$\mathbb E[X_i\widehat D_i]=h\mu+\beta(2q_i-1)$, whereas the difference in $\mathbb E[\widehat D_i^2]$ is $4\beta^2q_i(1-q_i)$. Substitution into~\eqref{eq:exchange-firm-decomposition} proves~\eqref{eq:exchange-preference-disclosure-value}.

Reporting $X_i$ satisfies the reply incentive~\eqref{eq:exchange-terminal-ic}: each type prefers the prediction with its own local-condition sign. At $h\mu=\beta$, the low type is indifferent. Under full revelation, the only potentially profitable deviation is for an inner target $v_l=h\mu-\beta$ to imitate the larger prediction $v_u=h\mu+\beta$. Its incentive condition is
\begin{equation}
 \left(v_l-\frac{\gamma v_l}{1+\gamma}\right)^2
 \leq\left(v_l-\frac{\gamma v_u}{1+\gamma}\right)^2
 \quad\Longleftrightarrow\quad
 \frac{v_l}{v_u}\leq\frac{\gamma}{2+\gamma}
 \label{eq:exchange-full-revelation-comparison}
\end{equation}
Substituting $v_l=h\mu-\beta$ and $v_u=h\mu+\beta$ gives $\mu\leq\mu_c$. The remaining comparisons follow by ordering and symmetry.

To establish optimality among all reporting equilibria, not just these two, the aligned-target argument in the proof of Proposition~\ref{prop:exchange-frontier} bounds the two moments in~\eqref{eq:exchange-firm-decomposition} using Bayesian consistency and reply incentives. Full revelation attains the bound when $b_\gamma>0$ and it is feasible. Reporting $X_i$ attains it when $b_\gamma\leq0$ or full revelation is infeasible. That argument applies to mixed as well as pure replies and does not rely on the equilibrium classification. The sign of $b_\gamma$ gives the stated comparison.
\proofend

\subsection{Optimal AI disclosure on an aligned support}\phantomsection\label{subsec:proof-main-aligned-signal}

\proofheading{Proof of Proposition~\ref{prop:exchange-aligned-signal}.}
When the additional preference-disclosure benefit in~\eqref{eq:exchange-aggregate-coefficient} is zero or constant on the interior of the support, $G$ is convex there. For any signal, convexity gives
$\mathbb E[G(\mathbb E[\theta\mid s],q)]\leq\mathbb E[G(\theta,q)]$.
Full revelation attains this bound. Values at a support endpoint do not change the conclusion because the continuous prior gives that endpoint zero probability, and no positive-probability posterior can have its mean there.

In the remaining case, the benefit of preference disclosure falls by
$4b_\gamma\beta^2\sum_iq_i(1-q_i)>0$ immediately above $\mu_c$. Pooling $[\mu_c-w,\mu_c+w]$ induces mean $\mu_c$, where full revelation satisfies the weak reply incentives. Relative to revealing those states, its increase in expected benefit is
\[
 \frac{4b_\gamma\beta^2\sum_iq_i(1-q_i)w
                -2A_\gamma h^2w^3/3}{\theta_H-\theta_L}
\]
This is positive for small $w$. Its unique positive maximizer is~\eqref{eq:exchange-pooling-width}; substitution gives~\eqref{eq:exchange-strict-gain}.

The width calculation above compares centered pooling intervals. To establish optimality among all public signal policies, fix $q$ and construct an upper bound $\overline G(x,q)$. Outside the proposed interval, set $\overline G(x,q)=G(x,q)$. Inside, replace $G$ by the line joining its endpoint values. At the chosen width, this line also passes through $(\mu_c,G(\mu_c,q))$. It lies above both convex quadratic branches, and its slope is between the derivatives at the interval endpoints. Therefore $\overline G$ is convex on the entire support, satisfies $\overline G(x,q)\geq G(x,q)$ everywhere, and agrees with $G$ at $\mu_c$.

Now consider any public signal policy, including a randomized policy. After signal $s$, let $\mu(s)=\mathbb E[\theta\mid s]$. The upper bound and its convexity give
\[
 G(\mu(s),q)\leq\overline G(\mu(s),q)
 \leq\mathbb E[\overline G(\theta,q)\mid s]
\]
The first inequality uses $\overline G\geq G$; the second is Jensen's inequality applied to the posterior after $s$. Averaging over signals gives
\[
 \mathbb E_s[G(\mu(s),q)]\leq
 \mathbb E_\theta[\overline G(\theta,q)]
\]
The left side averages the benefits generated by the candidate signal policy. The right side averages the fixed upper bound under the prior of $\theta$, so it is the same for every policy. This argument places no restriction on the posterior support: a signal can pool an interval or a disconnected set of states.

Policy~\eqref{eq:exchange-aligned-policy} attains this bound. Outside the pool, it reveals $\theta$, so the posterior is a point mass and $\overline G=G$. Inside the pool, the uniform prior gives posterior mean $\mu_c$, where $\overline G=G$; moreover, $\overline G$ is affine throughout that interval, so Jensen's inequality holds with equality. Thus both inequalities are equalities at every signal generated by the proposed policy. No disconnected-pooling or randomized policy can achieve a larger expected benefit, although another policy could tie.
\proofend

\subsection{Comparison with direct worker communication}\phantomsection\label{subsec:proof-main-direct-comparison}

\proofheading{Proof of Proposition~\ref{prop:exchange-direct-worker-loss}.}
We first calculate worker $i$'s loss conditional on a fixed initial signal $s$ and an incentive-compatible reporting profile. Write $\mu=\mathbb E[\theta\mid s]$ and $\eta=\mathbb E[\theta^2\mid s]$. All expectations in this calculation condition on $(q,s)$ and average worker types, local conditions, replies, and the remaining aggregate uncertainty. Averaging the minimized operating loss in Section~\ref{subsec:exchange-reply-ic} gives
\[
 \begin{aligned}
 \mathbb E[\ell_i\mid q,s]
 &=(1-\gamma)h^2(\eta-\mu^2)
  +\gamma(1-\gamma)^2\{\mathbb E[d_j^2]-\mathbb E[\widehat D_j^2]\}
  +\frac{\gamma(1-\gamma)}{(1+\gamma)^2}\mathbb E[\widehat D_j^2]\\
 &\quad+\gamma(1-\gamma)
       \mathbb E\left[\left\{d_i-\frac{\gamma}{1+\gamma}\widehat D_i\right\}^2\right]
 \end{aligned}
\]
Bayesian consistency implies $\mathbb E[d_i\widehat D_i]=\mathbb E[\widehat D_i^2]$. Expanding the square and collecting terms therefore yields
\begin{equation}
 \begin{aligned}
 \mathbb E[\ell_i\mid q,s]
 &=(1-\gamma)h^2(\eta-\mu^2)
  +\gamma(1-\gamma)\{\mathbb E[d_i^2]+(1-\gamma)\mathbb E[d_j^2]\}\\
 &\quad-\frac{\gamma^2(1-\gamma)}{(1+\gamma)^2}
  \{(2+\gamma)\mathbb E[\widehat D_i^2]
                    +(1-\gamma-\gamma^2)\mathbb E[\widehat D_j^2]\}
 \end{aligned}
 \label{eq:exchange-general-worker-loss}
\end{equation}
Here $\mathbb E[d_i^2]=h^2\mu^2+2h\mu\beta(2q_i-1)+\beta^2$. Averaging the formula over the two workers gives coefficient $-\omega_\gamma/2$ on $\sum_i\mathbb E[\widehat D_i^2]$, with $\omega_\gamma$ defined in~\eqref{eq:exchange-worker-preference-value}.

For $\sqrt2-1\leq\gamma<1$, Propositions~\ref{prop:exchange-aligned-reporting} and~\ref{prop:exchange-aligned-signal} give local-condition reporting and optimal full aggregate disclosure. At $\gamma=\sqrt2-1$, use the stated factual-reporting selection among firm-indifferent equilibria. The firm gain is the increase in the aggregate-information benefit $A_\gamma h^2\mathbb E[\mu^2]$, proving~\eqref{eq:exchange-direct-strict-gain} for the firm.

For the worker, hold their own $(T_i,X_i)$ fixed and substitute the local-condition operating actions into their loss. Remaining aggregate uncertainty contributes $(1-\gamma)h^2(\eta-\mu^2)$; the remaining terms have coefficient $2\gamma(1-\gamma)h^2/(1+\gamma)^2$ on $\mu^2$ and are otherwise constant or linear in $\mu$. Full revelation raises $\mathbb E[\mu^2]$ by $\operatorname{Var}(\theta)$, leaves the constant and linear terms unchanged, and eliminates the uncertainty term. Hence
\begin{equation}
 \begin{aligned}
 &\mathbb E_i[\ell_i\mid q,T_i,X_i,\mathrm{direct}]
       -\mathbb E_i[\ell_i\mid q,T_i,X_i,\sigma^*]\\
 &\qquad=\left[(1-\gamma)-\frac{2\gamma(1-\gamma)}{(1+\gamma)^2}\right]
                          h^2\operatorname{Var}(\theta)
       =A_\gamma h^2\operatorname{Var}(\theta)
 \end{aligned}
 \label{eq:exchange-direct-worker-gain}
\end{equation}
This also proves the first case after averaging own private information.

Now suppose $0<\gamma<\sqrt2-1$ and $\overline\theta\leq\mu_c$.
Under local-condition reporting, $\widehat D_i=[h\mu+\beta(2q_i-1)]X_i$. Full revelation raises its second moment by $4\beta^2q_i(1-q_i)$ without changing how accurately $X_i$ is predicted. Thus full revelation reduces firm loss by $4b_\gamma\beta^2\sum_iq_i(1-q_i)$ and average worker loss by $2\omega_\gamma\beta^2\sum_iq_i(1-q_i)$ at any fixed signal.

For local-condition reporting, both the firm and average worker benefit have coefficient $A_\gamma h^2$ on $\mu^2$. Their constant and linear terms have unchanged expectations under any signal policy because $\mathbb E[\mu]=\overline\theta$ and $\mathbb E[\eta]=\mathbb E[\theta^2]$. Direct communication uses full reporting at $\overline\theta\leq\mu_c$. Under the optimal AI policy, full reporting occurs with probability $p^*$. Subtracting proves~\eqref{eq:exchange-direct-firm-tradeoff}--\eqref{eq:exchange-direct-worker-tradeoff}. The firm's nonnegative gain follows also from~\eqref{eq:exchange-direct-ai-ordering}. The two strict inequalities in~\eqref{eq:exchange-direct-worker-loss-region} give the claimed signs. With common beliefs and the stated symmetric reporting selection, the worker losses coincide.

For unequal beliefs, the individual preference-disclosure benefit at a fixed signal is
\[
 \frac{4\gamma^2(1-\gamma)\beta^2}{(1+\gamma)^2}
   \{(2+\gamma)q_i(1-q_i)+(1-\gamma-\gamma^2)q_j(1-q_j)\}
\]
Replacing the average coefficient in~\eqref{eq:exchange-direct-worker-tradeoff} by this coefficient gives the individual ex-ante comparison. For $h=\beta=1$, $\gamma=1/5$, $q_1=q_2=1/2$, and $\theta\sim U[1.01,1.35]$, the interior pooling formula gives half-width approximately $0.104$. Substitution into the original losses reduces firm loss by about $0.24\%$ while increasing each worker's ex-ante loss by about $0.86\%$, so the strict worker-loss region is nonempty.
\proofend

\subsection{Why the separate replacement decision forwards unchanged}\phantomsection\label{subsec:exchange-replacement-proof}

\proofheading{Proof of Proposition~\ref{prop:exchange-no-replacement}.}
Fix $(q,s)$, original report $r_i$, and alternative delivered label $y_i$. Let $\Delta\widehat D_i=\widehat D_i(y_i)-\widehat D_i(r_i)$. Under the candidate unchanged-forwarding strategies, changing only this delivery changes the recipient's action by $\gamma\Delta\widehat D_i/(1+\gamma)$ and leaves the sender's action unchanged. The sender does not observe the replaced message.

Conditional on the actual state and the observed original report, independence gives
\[
 \begin{aligned}
 \mathbb E[d_i\mid\theta,q,s,r_i]&=\widehat D_i(r_i),\qquad
 \mathbb E[d_j\mid\theta,q,s,r_i]=\mathbb E[\widehat D_j(r_j)\mid\theta,q,s,r_i]=0\\
 \mathbb E[a_i\mid\theta,q,s,r_i]&=\frac{\widehat D_i(r_i)}{1+\gamma},\qquad
 \mathbb E[a_j\mid\theta,q,s,r_i]=\frac{\gamma\widehat D_i(r_i)}{1+\gamma},\qquad
 \mathbb E[H_j\mid\theta,q,s,r_i]=0
 \end{aligned}
\]
The first equality does not identify the actual state with the posterior mean. The numerical target $d_i=(h\mu+\beta T_i)X_i$ depends on the public posterior mean. Conditional on $s$, the worker's private information and report are independent of the actual $\theta$. The coworker's target has mean zero because their local condition is independent and symmetric.

Write $\Delta a_j=\gamma\Delta\widehat D_i/(1+\gamma)$. Direct expansion of firm loss gives
\[
 L_O'-L_O
 =\{(1-\gamma)(a_j-H_j)+2\gamma(a_j-a_i)\}\Delta a_j
                  +\frac{1+\gamma}{2}(\Delta a_j)^2
\]
Taking AI's expectation at its actual $\theta$ and substituting the conditional means yields
\begin{equation}
 \Delta_i^{\mathrm{AI}}
 =-\frac{\gamma^2(1-\gamma)}{(1+\gamma)^2}
            \widehat D_i(r_i)\Delta\widehat D_i
       +\frac{\gamma^2}{2(1+\gamma)}(\Delta\widehat D_i)^2
 \label{eq:exchange-replacement-loss-expanded}
\end{equation}
The actual aggregate state cancels from this difference; it has not been averaged under the worker's beliefs.

Now evaluate the different deviation in which worker $i$ sends $y_i$ and then optimally adjusts their own action. The worker's report-dependent minimized loss is~\eqref{eq:exchange-message-cost}. Therefore, averaging over the actual private states that were assigned $r_i$,
\[
 \begin{aligned}
 \Delta_i^\ell
 &=\gamma(1-\gamma)\mathbb E_i\left[
 \left\{d_i-\frac{\gamma}{1+\gamma}\widehat D_i(y_i)\right\}^2
 -\left\{d_i-\frac{\gamma}{1+\gamma}\widehat D_i(r_i)\right\}^2
                       \,\middle|\,q,s,r_i\right]\\
 &=-\frac{2\gamma^2(1-\gamma)}{(1+\gamma)^2}
          \widehat D_i(r_i)\Delta\widehat D_i
       +\frac{\gamma^3(1-\gamma)}{(1+\gamma)^2}(\Delta\widehat D_i)^2
 \end{aligned}
\]
The second equality uses $\mathbb E_i[d_i\mid q,s,r_i]=\widehat D_i(r_i)$. Each supported private state prefers its assigned report, so this conditional average satisfies $\Delta_i^\ell\geq0$. Subtracting half this expression from~\eqref{eq:exchange-replacement-loss-expanded} proves~\eqref{eq:exchange-independent-forwarding-identity}.

This checks every output in the common space $\mathcal R$. Unused labels may duplicate used-label beliefs at both reporting and delivery. Holding the other delivery rule fixed, no separate replacement decision can lower firm loss. Workers face exactly their original communication process, so their reporting and action incentives remain valid. Hence the entire candidate profile is an equilibrium. Any deterministic state-dependent deviation is also ruled out, because the inequality holds at each actual state. A randomized output deviation cannot improve on all the pure outputs it mixes.
\proofend

\subsection{Replacement after full initial disclosure}\phantomsection\label{subsec:proof-full-disclosure-replacement}

\proofheading{Proof of Corollary~\ref{cor:exchange-full-disclosure-replacement}.}
Fix a public signal $s=\theta$ and any deterministic equilibrium under endogenous forwarding. The worker knows the delivery rule conditional on this signal, so their report and delivery can be combined into the baseline reporting rule
\[
 \widetilde\lambda_i(T_i,X_i,q,s)
 =g_i(\theta,q,s,\lambda_i(T_i,X_i,q,s))
\]
Each used delivered label has an input report that produces it. A deviation to that label can therefore be reproduced by a reporting deviation in the replacement game, followed by the same best operating response. The worker also knows their own outgoing delivery, since $\theta$ is public. Thus the combined rule preserves information, Bayesian beliefs, and operating outcomes; unused labels can duplicate used-label interpretations. Every replacement outcome is consequently a baseline reporting-equilibrium outcome. Proposition~\ref{prop:exchange-no-replacement} gives the reverse inclusion through unchanged forwarding. The two firm-preferred values at full initial disclosure are therefore equal. Under the additional aligned-support conditions in the corollary, Proposition~\ref{prop:exchange-aligned-signal} makes this initial policy optimal under unchanged forwarding, giving~\eqref{eq:exchange-baseline-full-disclosure}. This is a comparison at the stated policy, not at every other initial signal. The same representation argument applies whenever delivery uses no aggregate information beyond $s$.
\proofend

\subsection{Firm loss under endogenous forwarding}\phantomsection\label{subsec:proof-main-firm-ordering}

\proofheading{Proof of Corollary~\ref{cor:exchange-replacement-firm-value}.}
Proposition~\ref{prop:exchange-no-replacement} makes every baseline worker equilibrium available in the endogenous-forwarding game. Minimizing over its equilibria and the initial signal gives the first inequality. A constant initial signal reproduces direct communication, giving the second inequality.
\proofend

\subsection{Sufficient conditions for gains from selective forwarding}\phantomsection\label{subsec:exchange-baseline-optimal-proof}

\proofheading{Proof of Proposition~\ref{prop:exchange-baseline-optimal-reduction}.}
We take $\bar\gamma$ to be the unique root in $(0,1)$ of
\begin{equation}
 3\bar\gamma^3+5\bar\gamma^2-3\bar\gamma-1=0
 \label{eq:exchange-replacement-gamma-bound}
\end{equation}
This root is approximately $0.652$. The argument below establishes a strict improvement region for every $0<\gamma<\bar\gamma$.

\proofstep{Candidate policies and operating actions.}
Normalize the aggregate state as $z=h\theta/\beta$ and actions as $a_i/\beta$, and divide losses by $\beta^2$. Fix $0<\gamma<1$, choose a center $u$ satisfying~\eqref{eq:exchange-replacement-center-incentives}, and put $q_1=q_2=\bar q=1-\varepsilon$. Consider
\[
 z\sim U[u-w,u+w]
\]
The candidate has an uninformative initial signal, truthful worker reports $(T_i,X_i)$, and the delivery rule~\eqref{eq:exchange-allgamma-delivery}. We first verify incentives at a general center and then choose centers giving a positive firm gain. Let $k=0$ denote a delivered label whose first component is $-1$, and let $k=1$ denote first component $1$. The second component reveals $X_i$ in either case.

The category probabilities are $(2-\bar q)/2$ and $\bar q/2$. Write $\mu(k)=\mathbb E[z\mid k]$ for the aggregate posterior mean after category $k$. Bayes' rule gives
\begin{equation}
 \begin{aligned}
  \mu(0)&=u-\frac{\bar q w}{2(2-\bar q)},&
  \mu(1)&=u+\frac w2\\
  \Pr(T_i=1\mid k_i=0)&=\frac{\bar q}{2-\bar q},&
  \Pr(T_i=1\mid k_i=1)&=1
 \end{aligned}
 \label{eq:exchange-allgamma-posteriors}
\end{equation}
The delivered category can make the sender's type and aggregate state correlated. The action calculation retains their joint posterior.

Because both local-condition signs are known after delivery, write the operating action as
\begin{equation}
 a_i(T_i,X_i;k_j,X_j)=A_{T_i k_j}X_i+B_{T_i k_j}X_j
 \label{eq:exchange-allgamma-action}
\end{equation}
Order the coefficient vectors by $(-1,0),(-1,1),(1,0),(1,1)$. The Bayesian best-response equations are
\begin{equation}
 A=(1-\gamma)d+\gamma PB,\qquad B=\gamma PA,
 \qquad A=(I-\gamma^2P^2)^{-1}(1-\gamma)d
 \label{eq:exchange-allgamma-action-system}
\end{equation}
where
\begin{equation}
 d=\begin{pmatrix}\mu(0)-1\\\mu(1)-1\\\mu(0)+1\\\mu(1)+1\end{pmatrix},
 \qquad
 P=\begin{pmatrix}
  \dfrac{2(1-\bar q)}{2-\bar q}&0&\dfrac{\bar q}{2-\bar q}&0\\
  0&0&1&0\\
  \dfrac{1-\bar q}{2-\bar q}&\dfrac{1-\bar q}{2-\bar q}&\dfrac{\bar q}{2-\bar q}&0\\
  0&0&0&1
 \end{pmatrix}
 \label{eq:exchange-allgamma-transition}
\end{equation}
Each row of $P$ gives the coworker's type and received category conditional on the worker's own type and received category. For example, a high worker receiving category 1 knows that the coworker is high and the aggregate state is in the upper half. Their own report therefore also generates category 1, giving the last row. Since $P$ is stochastic and $\gamma<1$, the inverse in~\eqref{eq:exchange-allgamma-action-system} exists. The best-response operator is a contraction, so these are the unique Bayesian Nash operating actions, not an approximation to worker behavior.

\proofstep{Choice of width and AI delivery incentives.}
After a high-type report, let $\Delta_-(z)$ be AI's expected loss from delivering category 1 minus category 0, with the reported local-condition sign unchanged, when the actual aggregate state is in the lower half. Define $\Delta_+(z)$ for the upper half. AI conditions on its source report and actual $z$, averaging over the other worker's private state. The two formulas may differ at the cutoff because the other delivery rule changes there as well.

Let $p_{-1}=1-\bar q$, $p_1=\bar q$, $k_-(v)=0$, and $k_+(v)=\mathbf 1\{v=1\}$. Taking the source's condition to be $+1$ by symmetry, subtraction of the original firm losses gives
\begin{equation}
 \begin{aligned}
 \Delta_s(z)={}&\sum_{v\in\{-1,1\}}p_v\left[
  \frac{1+\gamma}{2}(A_{v1}^2+B_{v1}^2-A_{v0}^2-B_{v0}^2)
  -(1-\gamma)z(A_{v1}-A_{v0})\right]\\
 &-2\gamma\sum_vp_v\left[
  A_{1,k_s(v)}(B_{v1}-B_{v0})+B_{1,k_s(v)}(A_{v1}-A_{v0})\right],
  \qquad s\in\{-,+\}
 \end{aligned}
 \label{eq:exchange-allgamma-ai-difference}
\end{equation}
The source does not observe their outgoing delivery, so their action stays fixed in this comparison. Only the recipient's action changes. The other delivery rule is held fixed, as required by the separate-decision restriction.

Define $w_\varepsilon$ to be the root near zero of
\begin{equation}
 \frac{\Delta_-(u)+\Delta_+(u)}2=0
 \label{eq:exchange-allgamma-exact-width}
\end{equation}
The left side is an analytic function of $(\varepsilon,w)$ near $(0,0)$, equals zero there, and has width derivative
\[
 \left.\frac{\partial}{\partial w}
  \frac{\Delta_-(u)+\Delta_+(u)}2\right|_{(0,0)}
 =\frac{(1-\gamma)D_\gamma(u)}{(1+\gamma)^2}>0
\]
where $D_\gamma(u)=\gamma(1-\gamma)u+1+2\gamma-\gamma^2>0$ abbreviates the common denominator in the following calculations.
The implicit function theorem gives a smooth root, and expansion yields
\begin{equation}
 w_\varepsilon=
 \frac{4\gamma^2\{(1-\gamma)u-2\gamma\}}{D_\gamma(u)}\varepsilon
 +O(\varepsilon^2)>0
 \label{eq:exchange-allgamma-width-expansion}
\end{equation}
All expansions in this proof take $\varepsilon\downarrow0$ with $\gamma$ and $u$ fixed. The first inequality in~\eqref{eq:exchange-replacement-center-incentives} makes this width positive.

Substituting the action system and this exact root into the two one-sided delivery comparisons gives
\begin{equation}
 \begin{aligned}
  \Delta_-(u)&=
   \frac{32\gamma^4(1-\gamma)^2(u+1)(\gamma u+2\gamma+1)}
        {D_\gamma(u)^2}\varepsilon^2+O(\varepsilon^3)>0\\
  \Delta_+(u)&=-\Delta_-(u)<0
 \end{aligned}
 \label{eq:exchange-allgamma-cutoff-incentives}
\end{equation}
Within either half, the difference is affine in the actual state, with slope
\begin{equation}
 \frac{d\Delta_s(z)}{dz}
 =-\frac{4\gamma^2(1-\gamma)^2(u+1)}{D_\gamma(u)}
   \varepsilon+O(\varepsilon^2)<0
 \label{eq:exchange-allgamma-ai-slope}
\end{equation}
Thus after a high report AI strictly prefers category 0 throughout the lower half and category 1 throughout the upper half. Equation~\eqref{eq:exchange-allgamma-exact-width} chooses a width with strict one-sided incentives; it does not impose indifference at the cutoff.

After a low report, the loss from changing to the same-sign category 1 is, uniformly over the support,
\begin{equation}
 \frac{16\gamma^2(1-\gamma)^2(\gamma u+2\gamma+1)}{D_\gamma(u)}
 \varepsilon+O(\varepsilon^2)>0
 \label{eq:exchange-allgamma-low-delivery}
\end{equation}
Wrong-sign deliveries have limiting loss differences
\[
 \frac{4\gamma^2(u+1)(u-1+2\gamma^2)}{(1+\gamma)^2}
 \quad\text{after a low report},\qquad
 \frac{4\gamma^2(u+1)^2}{(1+\gamma)^2}
 \quad\text{after a high report}
\]
These are positive since $u>1+\gamma$, and remain so for small $\varepsilon$. This checks all four delivered labels after every report and at every actual aggregate state.

\proofstep{Worker reporting incentives.}
A worker can change their report and then re-optimize their action after receiving $y_j$. Let $a_j(r)$ denote the coworker's action when AI applies its delivery rule to the deviating report $r$. Completing the square in the worker's operating loss gives their optimized expected loss:
\begin{equation}
 \begin{aligned}
 \mathbb E\Big[&(1-\gamma)D_i^2+\gamma a_j(r)^2\\
 &-\{(1-\gamma)\mathbb E[D_i\mid T_i,X_i,y_j]
      +\gamma\mathbb E[a_j(r)\mid T_i,X_i,y_j]\}^2
      \,\Big|\,T_i,X_i\Big]
 \end{aligned}
 \label{eq:exchange-allgamma-worker-loss}
\end{equation}
The outer expectation averages the aggregate state, coworker private state, and resulting incoming label. The coworker retains their equilibrium interpretation of delivered labels. The deviator uses their actual type and local condition when re-optimizing, not the type named by their false report.

Substitution of~\eqref{eq:exchange-allgamma-action-system} gives the loss from a false type report with the same local-condition sign, relative to truthful reporting:
\begin{equation}
 \begin{aligned}
  \Delta_{-1}^{\mathrm{worker}}
  &=\frac{4\gamma^2(1-\gamma)^2\{1+4\gamma-(1-2\gamma)u\}}{D_\gamma(u)}
      \varepsilon+O(\varepsilon^2)>0\\
  \Delta_1^{\mathrm{worker}}
  &=\frac{4\gamma^2(1-\gamma)^2(u+1)}{D_\gamma(u)}
      \varepsilon+O(\varepsilon^2)>0
 \end{aligned}
 \label{eq:exchange-allgamma-worker-incentives}
\end{equation}
The second inequality in~\eqref{eq:exchange-replacement-center-incentives} makes the low-type comparison positive. For actual type $T$, a wrong-sign report has limiting deviation loss $4\gamma^2(1-\gamma)(u+T)(u+1)/(1+\gamma)>0$. Changing both type and sign has the same positive limit. Every reporting deviation is therefore unprofitable for small positive $\varepsilon$. Truthful worker reporting and the candidate replacement rule, together with the Bayesian Nash actions, constitute an equilibrium.

\proofstep{Comparison with optimal unchanged forwarding.}
For sufficiently small $\varepsilon$, $w_\varepsilon<u-(1+\gamma)$, so the entire normalized support lies above $1+\gamma$. Every posterior mean induced by an initial public signal also lies above that threshold. Proposition~\ref{prop:exchange-aligned-reporting} makes factual reporting firm-preferred under unchanged forwarding, and Proposition~\ref{prop:exchange-aligned-signal} makes full aggregate disclosure optimal over all initial signals. The comparison below is therefore with the globally optimized unchanged-forwarding benchmark.

Integrating the original firm loss under the candidate and under that benchmark, then restoring units, gives
\begin{equation}
 L^{\mathrm{AI},*}-L^{\mathrm{candidate}}
 =\beta^2 C_\gamma(u)\varepsilon^2+O(\varepsilon^3),\qquad
 C_\gamma(u)=\frac{4\gamma^2(1-\gamma)R_\gamma(u)}
                  {3(1+\gamma)^2D_\gamma(u)^2}
 \label{eq:exchange-allgamma-firm-gain}
\end{equation}
where
\begin{equation}
 \begin{aligned}
 R_\gamma(u)={}&-2\gamma^2(1-\gamma)^2
                 (3\gamma^3+5\gamma^2-3\gamma-1)u^2\\
 &+(-18\gamma^7+2\gamma^6+70\gamma^5-40\gamma^4
                 -26\gamma^3+6\gamma^2+6\gamma)u\\
 &-12\gamma^7+5\gamma^6+66\gamma^5-19\gamma^4
                 -60\gamma^3-21\gamma^2+6\gamma+3
 \end{aligned}
 \label{eq:exchange-allgamma-gain-polynomial}
\end{equation}
The loss difference has zero first derivative in both $\varepsilon$ and $w$ at $(0,0)$. Consequently, the $O(\varepsilon^2)$ correction to the width does not change the displayed second-order gain coefficient.

\proofstep{Centers giving a positive gain.}
First suppose $0<\gamma<1/2$. The low-type reporting inequality requires $u<(1+4\gamma)/(1-2\gamma)$. This upper bound strictly exceeds both $1+\gamma$ and $2\gamma/(1-\gamma)$, so there are centers satisfying all the strict incentive conditions. Substitution into the gain coefficient at the upper bound gives
\begin{equation}
 C_\gamma\left(\frac{1+4\gamma}{1-2\gamma}\right)
 =-\frac{4\gamma^2(1-\gamma)
       (12\gamma^5+7\gamma^4-18\gamma^3+4\gamma^2+6\gamma-3)}
       {3(1+\gamma)^2(1-2\gamma^2)^2}>0
 \label{eq:exchange-replacement-upper-center-gain}
\end{equation}
For an exact sign check, the polynomial in parentheses has the representation
\begin{equation}
 \sum_{j=0}^5 b_j\binom5j(2\gamma)^j(1-2\gamma)^{5-j},\qquad
 (b_0,\ldots,b_5)=
 \left(-3,-\frac{12}{5},-\frac{17}{10},-\frac98,-\frac{13}{16},-\frac7{16}\right)
 \label{eq:exchange-allgamma-polynomial-certificate}
\end{equation}
All coefficients are negative. Hence the polynomial is negative on $[0,1/2]$. Continuity makes $C_\gamma(u)>0$ for centers sufficiently close to, but strictly below, the upper bound. We choose such a center before taking $\varepsilon$ small; the low type is not left indifferent at the boundary.

Next suppose $1/2\leq\gamma<\bar\gamma$. The low-type inequality in~\eqref{eq:exchange-replacement-center-incentives} holds at every positive $u$. Also,
\begin{equation}
 \lim_{u\to\infty}C_\gamma(u)
 =-\frac{8\gamma^2(1-\gamma)}{3(1+\gamma)^2}
          (3\gamma^3+5\gamma^2-3\gamma-1)>0
 \label{eq:exchange-replacement-large-center-gain}
\end{equation}
The cubic is strictly increasing on $[1/2,1]$, is negative at $1/2$, and vanishes at $\bar\gamma$. A sufficiently large finite center therefore satisfies the width and reporting inequalities and has a positive gain coefficient.

For each $0<\gamma<\bar\gamma$, fix a center just constructed. Sufficiently small positive $\varepsilon$ then gives
\[
 L^{\mathrm{rep},*}\leq L^{\mathrm{candidate}}<L^{\mathrm{AI},*}
\]
The first inequality uses firm-preferred equilibrium selection.

Finally, all nontrivial reporting and delivery incentives are strict at each constructed positive-$\varepsilon$ environment. The exact action system, worker loss comparisons, and endpoint AI loss comparisons are continuous in the priors and support endpoints. Small perturbations preserve these inequalities, the support condition above $\mu_c$, and the strict firm gain. Since delivery loss differences are affine within each aggregate interval, the endpoint checks continue to cover the whole support. This proves a nonempty open improvement region, including asymmetric priors near the symmetric construction.
\proofend

\subsection{When selective forwarding does not improve firm performance}\phantomsection\label{subsec:proof-no-gain-redesign}

\proofheading{Proof of Proposition~\ref{prop:exchange-no-gain-redesign}.}
\proofstep{No coordination motive.}
If $\gamma=0$, let $\widehat\theta_i$ be worker $i$'s posterior mean at their operating decision, using all their information under an arbitrary initial signal and equilibrium under endogenous forwarding. Worker $i$'s optimal action is $a_i=(h\widehat\theta_i+\beta T_i)X_i$. Because $T_i$ belongs to that information, posterior forecast errors are orthogonal to $T_i$. Consequently, firm loss is
\[
 L=\frac12\sum_i\mathbb E[
       \{h(\widehat\theta_i-\theta)+\beta T_i\}^2]
   =\beta^2+\frac{h^2}{2}\sum_i
             \mathbb E[(\widehat\theta_i-\theta)^2]\geq\beta^2
\]
Full initial revelation attains this bound with unchanged forwarding.

\proofstep{No preference bias.}
Suppose $\beta=0$ and $0<\gamma<1$. We prove a lower bound for arbitrary deterministic local delivery rules and reporting rules, without using either reporting or delivery incentives. Thus the bound also applies to every equilibrium under endogenous forwarding.

For an interior type prior, define an orthonormal basis of worker $i$'s four private states by
\[
 \psi_i^0=1,\qquad \psi_i^1=X_i,\qquad
 \psi_i^2=\frac{T_i-(2q_i-1)}{2\sqrt{q_i(1-q_i)}},\qquad
 \psi_i^3=X_i\psi_i^2
\]
When a type is known, omit the last two basis functions. The incoming delivery depends on $(\theta,s,T_j,X_j)$ but not worker $i$'s private state, so the operating action has a unique expansion
\[
 a_i=\sum_k f_i^k(s,y_j)\psi_i^k
     =\sum_{k,l} A_i^{kl}(\theta,s)\psi_i^k\psi_j^l
\]
The first equality expands over own private information; the second expands each coefficient over the coworker's private information conditional on $(\theta,s)$. These expansions do not restrict the delivery rules.

Multiply the best response $a_i=(1-\gamma)h\mathbb E_i[\theta]X_i+\gamma\mathbb E_i[a_j]$ by $f_i^k(s,y_j)\psi_i^k$ and take expectations. Orthonormality and independence of the workers' private variables give, for every $i,k$,
\begin{equation}
 \mathbb E\left[\sum_l(A_i^{kl})^2
       -\gamma\sum_l A_i^{kl}A_j^{lk}
       -(1-\gamma)h\theta A_i^{10}\mathbf1\{k=1\}\right]=0
 \label{eq:exchange-zero-bias-row-moments}
\end{equation}
Summing these identities and expanding the original firm loss gives
\begin{equation}
 L=(1-\gamma)h^2\mathbb E[\theta^2]
           -\frac{1-\gamma}{2}\mathbb E[a_1^2+a_2^2]
 \label{eq:exchange-zero-bias-firm-loss}
\end{equation}

Under full aggregate disclosure and factual communication, the only nonzero action coefficients are $A_i^{10,*}=h\theta/(1+\gamma)$ and $A_i^{01,*}=\gamma h\theta/(1+\gamma)$. Define $E_i^{kl}=A_i^{kl}-A_i^{kl,*}$ and weights
\[
 \lambda_1=\frac{2(1+\gamma^2)}{1-\gamma^2},\qquad
 \lambda_k=\frac4{1-\gamma^2}\quad(k\ne1)
\]
Multiply~\eqref{eq:exchange-zero-bias-row-moments} by $\lambda_k$, sum over workers and basis terms, and complete the square around $A^*$. This gives
\begin{equation}
 \begin{aligned}
 &\frac{2h^2(1+\gamma^2)}{(1+\gamma)^2}\mathbb E[\theta^2]
                -\mathbb E[a_1^2+a_2^2]\\
 &=\mathbb E\sum_{k,l}\left[
   (\lambda_k-1)(E_1^{kl})^2+(\lambda_l-1)(E_2^{lk})^2
   -\gamma(\lambda_k+\lambda_l)E_1^{kl}E_2^{lk}\right]\geq0
 \end{aligned}
 \label{eq:exchange-zero-bias-certificate}
\end{equation}
To check the inequality, each summand is a quadratic form in $(E_1^{kl},E_2^{lk})$. If the two weights coincide, its smallest eigenvalue is either $(3-\gamma)/(1+\gamma)$ or $(1-\gamma+2\gamma^2)/(1+\gamma)$, both positive. If the weights differ, the diagonal entries are positive and the determinant is $(3+\gamma^2)(1+\gamma^2)/(1-\gamma^2)>0$. Every summand is therefore nonnegative.

Combining~\eqref{eq:exchange-zero-bias-certificate} with~\eqref{eq:exchange-zero-bias-firm-loss} proves
\[
 L\geq\frac{2\gamma(1-\gamma)}{(1+\gamma)^2}h^2\mathbb E[\theta^2]
\]
Full aggregate disclosure and unchanged factual forwarding attain this value, with operating actions $a_i=h\theta(X_i+\gamma X_j)/(1+\gamma)$. Factual reporting is incentive compatible when $\beta=0$, and Proposition~\ref{prop:exchange-no-replacement} makes unchanged forwarding credible. The lower bound and this feasible policy prove equality.

\proofend

\subsection{Type beliefs and private histories}\phantomsection\label{subsec:exchange-dynamic-beliefs-proof}\label{app:exchange-dynamic-proofs}

The argument conditions throughout on the public history and the publicly selected reporting tables. Initially the types and local conditions are independent across workers. At a later history, suppose the private-history posterior factors across workers and is independent of past actual aggregate states after conditioning on their public signals. The next aggregate state and local conditions are fresh independent draws. The public signal depends on that aggregate state and the public history, but not directly on either worker's private variables. Conditional on the public history and current signal, the two report likelihoods consequently factor by worker. Bayes' rule multiplies each worker's prior private-history likelihood by only their own report likelihood. Summing the resulting product over local histories gives independent type posteriors. This proves the factorization inductively.

In particular, at a positive-probability current reply,
\begin{equation}
 q^+_{i\mid r_i}
 =\frac{q_i\sum_x\mathbf1\{\lambda_i(1,x)=r_i\}}
 {q_i\sum_x\mathbf1\{\lambda_i(1,x)=r_i\}
            +(1-q_i)\sum_x\mathbf1\{\lambda_i(-1,x)=r_i\}}
 \label{eq:exchange-dynamic-update-explicit}
\end{equation}
The common probability $1/2$ of each local condition cancels. The reporting tables in this equation are the ones selected at the current public history and signal. Each numerator count and corresponding low-type count is zero, one, or two, giving the finite updated-belief set in Appendix~\ref{subsec:exchange-updating}.

The same factorization shows that conditioning additionally on a coworker's private type, local-condition history, and previous private reporting deviations does not change the posterior over the other worker. That worker's own likelihood factor cancels. AI's extra knowledge of past and current actual aggregate states likewise supplies no additional type information, because workers react to those states only through public signals. These conclusions concern unchanged forwarding and the stated selected reporting class. They need not survive private replacement that hides from the coworker a report known to AI.

The current $q_t$ is therefore sufficient for the current distribution of persistent types. It need not be sufficient for continuation policies chosen under initial commitment. The latter can depend on the observed history even when two histories produce the same type posterior.

\subsection{Continuation losses, reporting incentives, and optimality}\phantomsection\label{subsec:exchange-two-period-proofs}

\proofheading{Proof of Lemma~\ref{lem:exchange-dynamic-full-reporting}.}
\proofstep{Second-period disclosure.}
If first-period replies fully reveal both types, Bayes' rule gives $q_{i2}=\mathbf1\{T_i=1\}$ for each worker. Period 2 is therefore the static problem with known types and fresh independent aggregate and local conditions. Proposition~\ref{prop:exchange-aligned-signal} implies that full revelation of $\theta_2$ is optimal for the firm. With types known, local-condition reporting reveals all remaining worker private information and gives the firm-preferred static outcome. This proves the first part.

\proofstep{First-period reporting incentives.}
Now consider candidate first-period replies that fully reveal $(T_i,X_{i1})$, followed by the optimal second-period policy just derived. At a known-type belief vector $q$, write $S_k=2q_k-1$ for the publicly inferred type. If worker $i$ earlier lied, their actual type $T$ may differ from $S_i$. The second-period reporting strategy uses the two condition labels interpreted according to $S_i$. Unused labels duplicate a used label, so they cannot change the publicly inferred type.

All target coefficients are positive under $h\theta_L>\beta$. The actual worker's best report therefore gives their actual condition sign even after a false type claim. Against the coworker's specified strategy, the worker then best responds using their actual target. Averaging both fresh condition signs gives the actual-type continuation loss
\begin{equation}
 \begin{aligned}
 \ell_{i2}(T,q)
 &=\gamma(1-\gamma)\mathbb E_\theta\left[
 \frac{(h\theta+\beta S_j)^2}{(1+\gamma)^2}
 +\left\{h\theta+\beta T-\frac{\gamma(h\theta+\beta S_i)}{1+\gamma}\right\}^2
 \right]
 \end{aligned}
 \label{eq:exchange-corner-continuation}
\end{equation}
Only the fresh aggregate state remains to be averaged. In particular, a deviating worker is not forced to act like the type they claimed.

For an actual low type, compare its true and false type claims with the same first-period condition sign. Subtracting the two current squares in~\eqref{eq:exchange-message-cost} gives
\[
 \frac{4\gamma^2(1-\gamma)\beta h}{(1+\gamma)^2}(\mu_1-\mu_c)
\]
Comparing~\eqref{eq:exchange-corner-continuation} at $S_i=-1$ and $S_i=1$, holding actual $T=-1$, gives the continuation gain
\[
 \frac{4\gamma^2(1-\gamma)\beta h}{(1+\gamma)^2}(\overline\theta-\mu_c)
\]
Multiplying the latter by $\delta$ and adding gives the low type's total gain from a false high-type claim with the same local-condition sign:
\begin{equation}
 \Delta_i^{-\to+}(\mu_1)
 =\frac{4\gamma^2(1-\gamma)\beta h}{(1+\gamma)^2}
       \{\mu_1-\mu_c+\delta(\overline\theta-\mu_c)\},\qquad
       \overline\theta=\mathbb E[\theta]
 \label{eq:exchange-dynamic-exaggeration-gain}
\end{equation}
Positive values favor deviation. The terms involving the coworker's type and $\theta^2$ cancel.

A high type's gain from falsely claiming low type is
\[
 -\frac{4\gamma^2(1-\gamma)\beta h}{(1+\gamma)^2}
    \{\mu_1+\mu_c+\delta(\overline\theta+\mu_c)\}<0
\]
For each claimed type, reversing the condition sign raises current loss and leaves the type-learning continuation unchanged. Hence a deviation changing both type and sign is worse than changing only type. These comparisons exhaust all four labels and establish necessity and sufficiency of~\eqref{eq:exchange-dynamic-full-reporting-condition}, proving the second part.
\proofend

\proofheading{Proof of Proposition~\ref{prop:exchange-dynamic-optimum}.}
The full-disclosure policy has $\mu_1=\theta_1$. Its first-period reporting incentives hold at every state by~\eqref{eq:exchange-dynamic-optimum-region} and Lemma~\ref{lem:exchange-dynamic-full-reporting}. In period 2, known types and fresh condition reporting give a static equilibrium.

For optimality, relax reporting incentives and allow full private revelation in each period. Under alignment, $X_i$ is the sign of $d_i=(h\mu+\beta T_i)X_i$. In any conditional report group, $\operatorname{Cov}(X_i,d_i\mid r_i)\geq0$. Bayesian consistency consequently implies
\[
 \mathbb E[X_i\widehat D_i]\leq\mathbb E[X_id_i],\qquad
 \mathbb E[\widehat D_i^2]\leq\mathbb E[d_i^2]
\]
Full private revelation attains both bounds. Both coefficients in~\eqref{eq:exchange-firm-decomposition} are positive for $0<\gamma<\sqrt2-1$, so it maximizes the reporting benefit without using a static reporting incentive in period 1. Under full private revelation, the remaining aggregate-information benefit has positive coefficient $A_\gamma h^2$ on $\mathbb E[\mu^2]$ and is maximized by full aggregate revelation.

This is a lower bound for each period at every inherited type posterior. Type posteriors can change, but the unconditional distribution of actual persistent types remains the initial distribution. Integrating the stage bounds over histories therefore gives a lower bound on the total commitment value, including history-dependent continuation policies. The proposed policy attains every stage bound. Under it, actions are $a_{it}=(D_{it}+\gamma D_{jt})/(1+\gamma)$. Substituting into the original losses and averaging the actual types and fresh conditions gives~\eqref{eq:exchange-dynamic-optimal-firm} and~\eqref{eq:exchange-dynamic-optimal-worker}.

Condition~\eqref{eq:exchange-dynamic-optimum-region} implies $\overline\theta\leq\mu_c$. Without AI, use mean $\overline\theta$ in both periods and full worker revelation in period 1. The low-type deviation gain is proportional to $(1+\delta)(\overline\theta-\mu_c)\leq0$, so this direct-communication equilibrium is feasible and attains the corresponding no-aggregate-information bound. Full AI disclosure raises expected squared posterior means by $\operatorname{Var}(\theta)$ in each period; constant and linear terms have unchanged expectations. The coefficient for both firm loss and actual-type worker loss is $A_\gamma h^2$, yielding~\eqref{eq:exchange-dynamic-direct-gain}.

Finally, if $\theta_L<\mu_c<\theta_H$, Proposition~\ref{prop:exchange-aligned-signal} shows that a sufficiently small static pool around $\mu_c$ strictly improves on full aggregate-state revelation. Its reporting gain is positive to first order in the pool width, whereas its aggregate-information loss is third order. This comparison does not require the optimal interior-width formula to fit inside the support. The two-period policy just proved optimal fully reveals the same aggregate-state draws, establishing the stated contrast with the static model.
\proofend

\proofheading{Proof of~\eqref{eq:exchange-uncertainty-reduction}--\eqref{eq:exchange-expected-pooling-width}.}
The posterior is a martingale, so $\mathbb E[q_{i2}\mid q_1,s_1]=q_{i1}$. Hence
\[
 \mathbb E[q_{i2}(1-q_{i2})\mid q_1,s_1]
 =q_{i1}(1-q_{i1})-\operatorname{Var}(q_{i2}\mid q_1,s_1)
\]
Summing gives~\eqref{eq:exchange-uncertainty-reduction}. Squaring~\eqref{eq:exchange-pooling-width} and substituting gives~\eqref{eq:exchange-expected-pooling-width}. At a known-type posterior the additional preference-disclosure value is zero, so a statically optimal continuation fully reveals the aggregate state. Local conditions remain fresh and their reports remain useful.
\proofend

\proofheading{Proof of Proposition~\ref{prop:exchange-dynamic-high-coordination}.}
\proofstep{A current-loss bound without static reporting constraints.}
The aligned moment bound in the proof of Proposition~\ref{prop:exchange-frontier} applies even after relaxing reporting incentives when $b_\gamma\leq0$. Revealing the local-condition sign weakly raises the firm's reporting benefit; conditional on that sign, concealing all remaining type information minimizes the prediction second moment with its nonpositive coefficient $b_\gamma$. Thus factual reporting minimizes current expected firm loss even among reporting rules that might be sustained only by future incentives. Under factual reporting, the coefficient of $\mu^2$ in the information benefit is $A_\gamma h^2>0$, so full aggregate disclosure attains a lower bound for each period at every inherited type posterior. Consequently,
\[
 \mathbb E[L_{O1}]\geq L^{\mathrm{AI},*}(q^0),\qquad
 \mathbb E[L_{O2}\mid\mathcal H_1]\geq L^{\mathrm{AI},*}(q_2)
\]
These bounds allow arbitrary history-dependent committed continuation policies.

\proofstep{The effect of learning on the future bound.}
Under full aggregate disclosure and factual reporting, the static comparison~\eqref{eq:exchange-preference-disclosure-value} gives
\begin{equation}
 \begin{aligned}
 L^{\mathrm{AI},*}(q)
 &=\frac{1-\gamma}{(1+\gamma)^2}
   \left\{2\gamma h^2\mathbb E[\theta^2]
     +\gamma(1-\gamma)h\beta\overline\theta\sum_i(2q_i-1)
     +(1+2\gamma-\gamma^2)\beta^2\right\}\\
 &\quad+4b_\gamma\beta^2\sum_iq_i(1-q_i)
 \end{aligned}
 \label{eq:exchange-dynamic-high-coordination-static-value}
\end{equation}
The expression in braces is affine in $q$. Bayesian updating gives $\mathbb E[q_{i2}]=q_i^0$ and
\[
 \mathbb E[q_{i2}(1-q_{i2})]
 =q_i^0(1-q_i^0)-\operatorname{Var}(q_{i2})
\]
Substitution proves~\eqref{eq:exchange-dynamic-high-coordination-learning}. Combining the two period bounds yields
\begin{equation}
 \mathbb E[L_{O1}+\delta L_{O2}]
 \geq(1+\delta)L^{\mathrm{AI},*}(q^0)
       -4\delta b_\gamma\beta^2\sum_i\operatorname{Var}(q_{i2})
 \geq(1+\delta)L^{\mathrm{AI},*}(q^0)
 \label{eq:exchange-dynamic-high-coordination-bound}
\end{equation}
For $\gamma>\sqrt2-1$, $\beta>0$, and $\delta>0$, the first lower bound is strictly above the proposed value whenever some $q_{i2}$ has positive variance.

\proofstep{Attainment and reporting incentives.}
Fully reveal the aggregate state in each period and select factual reporting. Since $X_{it}$ is independent of type, either factual reply leaves the type posterior at $q_i^0$. Let the second-period policy and strategies ignore past local conditions. A first-period report then has no effect on continuation losses, so the static factual-reporting incentives from Proposition~\ref{prop:exchange-aligned-reporting} also establish first-period optimality. Unused labels can copy a used label and its continuation. The policy attains the lower bound in both periods, proving~\eqref{eq:exchange-dynamic-high-coordination-value}. At $b_\gamma=0$, the same policy attains the bound, but the argument gives no strict preference against type learning.
\proofend

\subsection{Disclosure with ex-ante worker acceptance}\phantomsection\label{subsec:exchange-acceptance-proof}

Let the workers have common priors $q_1=q_2=q\in(0,1)$, an aligned uniform aggregate support, and low coordination. Put $D=\theta_H-\theta_L$, $S=2q(1-q)$, and $c=\mu_c$. Suppose
\[
 h\theta_L>\beta,\quad 0<\gamma<\sqrt2-1,\quad
 \frac{\theta_L+\theta_H}{2}<c<\theta_H,\quad
 w^*=\sqrt{\frac{2b_\gamma\beta^2S}{A_\gamma h^2}}
       <\overline w=\theta_H-c.
\]
Restrict disclosure to a single interval $[c-w,c+w]$ pooled at posterior mean $c$, with all other states revealed and $0\leq w\leq\overline w$. Every interval fits inside the support. The full-reporting probability and variance of the posterior mean are
\begin{equation}
 p(w)=\frac{c+w-\theta_L}{D},\qquad
 v(w)=\frac{D^2}{12}-\frac{2w^3}{3D}.
 \label{eq:exchange-acceptance-statistics}
\end{equation}
Relative to direct exchange, the firm and each worker obtain expected loss reductions
\begin{align}
 G_L(w)&=A_\gamma h^2v(w)-4b_\gamma\beta^2S\{1-p(w)\},\\
 G_\ell(w)&=A_\gamma h^2v(w)-2\omega_\gamma\beta^2S\{1-p(w)\}.
 \label{eq:exchange-acceptance-gains}
\end{align}

\begin{proposition}[Disclosure subject to ex-ante worker acceptance]\label{prop:exchange-worker-acceptance}
Within this interval family, the firm's unconstrained optimum is $w^*$. If $G_\ell(w^*)\geq0$, worker acceptance does not change that choice. Otherwise, the constrained optimum $w^A$ is the unique root of $G_\ell(w)=0$ in
\begin{equation}
 w^*<w^A<\min\left\{\overline w,
          \sqrt{\frac{\omega_\gamma\beta^2S}{A_\gamma h^2}}\right\}.
 \label{eq:exchange-acceptance-width}
\end{equation}
The firm strictly benefits relative to direct exchange under $w^A$, and both workers are indifferent ex ante. The optimization is over the stated interval family.
\end{proposition}

\proofheading{Proof of Proposition~\ref{prop:exchange-worker-acceptance}.}
The assumptions imply $0<\overline w<c-\theta_L$, so every proposed interval lies in the support. Its posterior mean is $c$. Full reporting occurs at revealed states below the pool and throughout the pool, giving $p(w)$. The law of total variance gives $v(w)$ because the pool has probability $2w/D$ and conditional variance $w^2/3$. The no-information posterior is below $c$, so direct exchange selects full preference reporting. Proposition~\ref{prop:exchange-direct-worker-loss} then gives the two gain functions in~\eqref{eq:exchange-acceptance-gains}.

Differentiate:
\[
 G_L'(w)=\frac{2A_\gamma h^2}{D}\{(w^*)^2-w^2\},
 \qquad
 G_\ell'(w)=\frac{2}{D}\{\omega_\gamma\beta^2S-A_\gamma h^2w^2\}.
\]
Thus $G_L$ increases up to $w^*$ and decreases strictly afterward. Since $\omega_\gamma>2b_\gamma$, the worker's unconstrained preferred width $w^\ell=\sqrt{\omega_\gamma\beta^2S/(A_\gamma h^2)}$ exceeds $w^*$. If $G_\ell(w^*)\geq0$, the unconstrained optimum is feasible and hence remains optimal.

Suppose $G_\ell(w^*)<0$. It is strictly increasing on $[0,\min\{w^\ell,\overline w\})$. Moreover, $p(\overline w)=1$ and
\[
 G_\ell(\overline w)=G_L(\overline w)=A_\gamma h^2v(\overline w)>0,
\]
where strictness follows because a positive interval below $c-\overline w$ is revealed. If $w^\ell<\overline w$, $G_\ell$ decreases after $w^\ell$, so its value at $w^\ell$ is at least its positive endpoint value. A unique zero therefore lies strictly between $w^*$ and $\min\{w^\ell,\overline w\}$. No smaller width is feasible, and $G_L$ decreases after $w^*$, proving optimality of this first feasible width. Finally, $G_L(w^A)>G_L(\overline w)>0$. Symmetric priors and policies make the two workers' ex-ante losses identical, so both participation constraints bind. The proof optimizes only the specified interval family and maintains the stated equilibrium selection.

\proofstep{Strict reporting with acceptance.}
For a pool centered at $c-e$, the uniform posterior mean equals that center and the conditional variance remains $w^2/3$. Its upper endpoint falls by $e$ at a fixed width, reducing the probability of full reporting by $e/D$. These observations give~\eqref{eq:exchange-strict-pool-gains}. Substituting $h(c-e)= (1+\gamma)\beta-he$ into the difference of the squared terms in~\eqref{eq:exchange-full-revelation-comparison}, and multiplying by the report-loss coefficient $\gamma(1-\gamma)$ from~\eqref{eq:exchange-message-cost}, gives~\eqref{eq:exchange-strict-reporting-margin}.

In the binding case, $G_\ell'(w^A)>0$. The implicit function theorem applied to $G_\ell(w;e)=0$ therefore supplies a nearby width with
\[
 \left.\frac{d w^A(e)}{de}\right|_{e=0}
 =\frac{\omega_\gamma\beta^2S}
        {\omega_\gamma\beta^2S-A_\gamma h^2(w^A)^2}>1.
\]
The original pool is strictly interior and straddles $c$, so these properties persist for sufficiently small $e>0$. Its firm gain remains positive by continuity. Since $G_L'(w^A)<0$ and the direct $e$ term in~\eqref{eq:exchange-strict-pool-gains} is negative, the firm sacrifices some gain to obtain the margin; that sacrifice tends to zero as $e\downarrow0$.
\proofend

\section{Reporting equilibria and AI disclosure}\phantomsection\label{app:exchange-general}\label{app:exchange-reporting}

This appendix characterizes reporting equilibria and optimal AI disclosure for all nonnegative aggregate-state posteriors and independent type beliefs. It includes both the aligned case analyzed in the main text and the case in which the low type favors the opposite direction from the firm.

\subsection{Existence of reporting equilibria}\phantomsection\label{subsec:exchange-reporting-existence}

We first characterize the pure reporting equilibria and then identify the firm-preferred outcome in Proposition~\ref{prop:exchange-frontier}. The classification distinguishes reports that induce the same target prediction but reveal different private information. It applies to each worker separately, so a joint equilibrium can use different valid reporting rules for the two workers.

\paragraph{High-type revelation.}
The high type uses separate replies for $X_i=1$ and $X_i=-1$, while the low type uses a third reply for both values of $X_i$. This reveals the worker's type in every case but reveals the local condition only for the high type.

\begin{proposition}[Existence of pure reporting equilibria]\label{prop:exchange-communication}
Fix worker $i$ with $0<q_i<1$, and suppose $\mu>0$ and $h\mu\ne\beta$. Write $v_l=|h\mu-\beta|$ and $v_u=h\mu+\beta$, so their four distinct targets are $-v_u,-v_l,v_l,v_u$. Each extreme target has probability $q_i/2$, and each smaller target has probability $(1-q_i)/2$.

After combining replies that induce the same target prediction, every pure reporting equilibrium has one of the following forms. Each set contains targets sending the same reply; vertical bars separate replies. Reflecting all signs gives an additional version of either of the last two rows.
\begin{center}
\begingroup\small\normalfont
\renewcommand{\arraystretch}{1.35}
\begin{tabular}{@{}p{0.25\linewidth}p{0.40\linewidth}p{0.28\linewidth}@{}}
\toprule
Reporting outcome & Groups of private targets & Existence condition \\
\midrule
Babbling & $\{-v_u,-v_l,v_l,v_u\}$ & Always \\
Direction only & $\{-v_u,-v_l\}\mid\{v_l,v_u\}$ & Always \\
Full revelation & $\{-v_u\}\mid\{-v_l\}\mid\{v_l\}\mid\{v_u\}$ & $\displaystyle\frac{v_l}{v_u}\leq\frac{\gamma}{2+\gamma}$ \\[8pt]
High-type revelation & $\{-v_u\}\mid\{-v_l,v_l\}\mid\{v_u\}$ & $\displaystyle\frac{v_l}{v_u}\leq\frac{\gamma}{2(1+\gamma)}$ \\[8pt]
One extreme is revealed & $\{-v_u\}\mid\{-v_l,v_l,v_u\}$ & $\displaystyle\frac{v_l}{v_u}\leq\frac{\gamma(1-q_i)}{(1+\gamma)(2-q_i)}$ \\[8pt]
One sign reveals type & $\{-v_u\}\mid\{-v_l\}\mid\{v_l,v_u\}$ & $\displaystyle\frac{\gamma q_i}{2(1+\gamma)+\gamma q_i}\leq\frac{v_l}{v_u}\leq\frac{\gamma}{2+\gamma}$ \\[8pt]
\bottomrule
\end{tabular}
\endgroup
\end{center}
\smallskip
Conversely, every listed rule satisfying its condition can be supported as an equilibrium. A group can also be divided into several replies without changing incentives if every new reply has the same conditional mean target as the original group. Together, the table and these divisions give all pure reporting equilibria, up to reply labels. The additional equal-prediction cases and coincident targets are detailed below.

In particular, type-only reporting, with one reply for $T_i=1$ and another for $T_i=-1$, is always an equilibrium: both predictions are zero. It reveals $T_i$ but not $X_i$, and gives the same operating actions as babbling. The full-revelation condition in the table is equivalent to~\eqref{eq:exchange-revelation-region}.
This interval condition concerns four distinct targets. If $q_i\in\{0,1\}$, the type is already known and either pooling or revealing the two local conditions is an equilibrium. If $\mu=0$ or $h\mu=\beta$, targets coincide and the additional rules are characterized below; full private-state revelation is feasible at both values. Any pair of worker rules satisfying the respective conditions forms a reporting equilibrium with the Bayesian Nash operating actions.
\end{proposition}

\proofheading{Proof.}
For two predictions $z<z'$, the loss difference relevant to~\eqref{eq:exchange-terminal-ic} is
\[
 \left(d-\frac{\gamma z}{1+\gamma}\right)^2
 -\left(d-\frac{\gamma z'}{1+\gamma}\right)^2
 =\frac{\gamma(z'-z)}{1+\gamma}
   \left\{2d-\frac{\gamma(z+z')}{1+\gamma}\right\}
\]
It increases with $d$. Once replies with the same prediction are combined, each remaining reply must therefore contain adjacent targets. Four ordered targets have only the six forms in the table, including sign reflections. If adjacent groups have predictions $z<z'$, their boundary must satisfy
\begin{equation}
 \max\{d:d\text{ is in the lower group}\}
 \leq\frac{\gamma(z+z')}{2(1+\gamma)}
 \leq\min\{d:d\text{ is in the upper group}\}
 \label{eq:exchange-partition-boundary}
\end{equation}
This condition is necessary and sufficient for every assigned target to prefer its own prediction to the neighboring predictions, and hence to all other predictions.

The full-revelation condition is~\eqref{eq:exchange-full-revelation-comparison}, applied to the ordered positive targets $v_l$ and $v_u$. The other conditions follow from~\eqref{eq:exchange-partition-boundary}, using the Bayesian conditional means displayed below. Dividing a group into replies with the same conditional mean leaves all these comparisons unchanged. Conversely, combining such replies in any equilibrium gives one of the ordered rules, establishing completeness. For unused labels, assigning the posterior of an existing reply prevents an additional deviation. Finally, the own-reply incentive does not depend on the coworker's reporting rule after their reply is averaged out, as shown in Section~\ref{subsec:exchange-reply-ic}. The two workers' valid rules can therefore be combined independently.
\proofend

For any partition of supported private states into reply groups, compute each group's conditional mean target using Bayes' rule and apply~\eqref{eq:exchange-terminal-ic} to every assigned state and alternative used reply. Unused replies can copy a used reply's posterior; other interpretations are also permissible if they create no profitable deviation. The same test applies when targets coincide or a type is known.

\subsubsection{Four distinct targets}

Suppose $0<q_i<1$, $\mu>0$, and $h\mu\ne\beta$. Order targets as $-v_u,-v_l,v_l,v_u$. There are fifteen partitions of these four private states, but their incentive conditions do not require fifteen separate policy comparisons. Eight partitions group adjacent targets and are represented by the six rows of Proposition~\ref{prop:exchange-communication}, counting the two sign-reflected versions. Of the remaining seven partitions, the following four, including a sign reflection in the last row, can be equilibria:

\begin{center}
\begingroup\small\normalfont
\renewcommand{\arraystretch}{1.4}
\begin{tabular}{@{}p{0.28\linewidth}p{0.34\linewidth}p{0.31\linewidth}@{}}
\toprule
Reporting outcome & Groups of private targets & Existence condition \\
\midrule
Type only & $\{-v_u,v_u\}\mid\{-v_l,v_l\}$ & Always \\
Two replies, both predicting zero & $\{-v_u,v_l\}\mid\{-v_l,v_u\}$ & $(1-q_i)v_l=q_i v_u$ \\
Three replies, two predictions & $\{-v_u,v_l\}\mid\{-v_l\}\mid\{v_u\}$ & $(2-q_i)v_l=q_i v_u$ and $\displaystyle q_i\leq\frac{\gamma}{1+2\gamma}$ \\[8pt]
\bottomrule
\end{tabular}
\endgroup
\end{center}
\smallskip
The first two rows induce prediction zero after either reply. The last induces predictions $-v_l,-v_l,v_u$; its sign reflection induces $-v_u,v_l,v_l$. Consequently, each row is obtained by dividing a group from the main table into replies with the same conditional mean. The conditions above describe exactly when those divisions are Bayesian-consistent. Reply labels can be changed without changing the reporting outcome.

\proofheading{Proof.}
For direction-only reporting, the two predictions are $\pm[q_i v_u+(1-q_i)v_l]$, and their midpoint is zero, so~\eqref{eq:exchange-partition-boundary} always holds. Full revelation uses~\eqref{eq:exchange-full-revelation-comparison}.

When high types reveal $X_i$ while low types pool, the predictions are $-v_u,0,v_u$. The positive boundary condition is
\[
 v_l\leq\frac{\gamma v_u}{2(1+\gamma)}\leq v_u
\]
The negative boundary is its reflection, giving the stated $v_l/v_u$ condition. If only target $-v_u$ is separated, the other reply predicts $q_i v_u/(2-q_i)$. The boundary between these replies is $-\gamma v_u(1-q_i)/[(1+\gamma)(2-q_i)]$. Requiring it to lie between $-v_u$ and $-v_l$ gives the fifth row of the main table.

When the negative targets are separated while positive targets pool, predictions are $-v_u,-v_l,q_i v_u+(1-q_i)v_l$. The first boundary gives the full-revelation upper bound on $v_l/v_u$. The second must lie between $-v_l$ and $v_l$:
\[
 -v_l\leq\frac{\gamma q_i(v_u-v_l)}{2(1+\gamma)}\leq v_l
 \quad\Longleftrightarrow\quad
 \frac{v_l}{v_u}\geq\frac{\gamma q_i}{2(1+\gamma)+\gamma q_i}
\]
This proves the asymmetric three-reply condition.

For the additional cases, the two type-only groups each have mean zero because the two local conditions have equal probability within either type. For the groups $\{-v_u,v_l\}$ and $\{-v_l,v_u\}$, the predictions are $-q_i v_u+(1-q_i)v_l$ and its negative. They can support these nonadjacent groups only if both predictions are zero, giving the second row of the appendix table.

For $\{-v_u,v_l\}\mid\{-v_l\}\mid\{v_u\}$, nonadjacency similarly requires the first prediction to equal $-v_l$, which gives $(2-q_i)v_l=q_i v_u$. Combining the first two replies then gives the ordered partition $\{-v_u,-v_l,v_l\}\mid\{v_u\}$. Its existence condition from the main table reduces to
\[
 \frac{q_i}{2-q_i}
 \leq\frac{\gamma(1-q_i)}{(1+\gamma)(2-q_i)}
 \quad\Longleftrightarrow\quad
 q_i\leq\frac{\gamma}{1+2\gamma}
\]
The sign-reflected rule has the same condition. The other three partitions isolate one inner target while pooling the other three, or reveal the two inner targets while pooling the extremes. They cannot be equilibria with four distinct targets: the first two violate the ordering of distinct predictions, while in the last the state with target $v_u$ prefers prediction $v_l$ to the pooled prediction zero. These cases exhaust the fifteen partitions.
\proofend

\subsubsection{Coincident targets and known types}

The reporting condition~\eqref{eq:exchange-terminal-ic} continues to apply when some targets coincide. The following descriptions make its implications explicit without treating a zero-probability state as a possible deviation.

\paragraph{Known type, $q_i\in\{0,1\}$.}
Only the two local conditions remain uncertain. They can send a common reply or separate replies, and either rule is an equilibrium. If their targets are nonzero, the separated predictions equal the two opposite targets. If both targets are zero, both predictions are zero and the two local conditions are indifferent between replies.

\paragraph{Zero aggregate posterior, $\mu=0$, with $0<q_i<1$.}
The two possible targets are $\pm\beta$, each generated by two private states. Every equilibrium has one of two forms. Either every used reply predicts zero, requiring each reply group to balance the probability-weighted positive and negative targets; or replies separate the two target signs, and states with the same target can be separated further in any way. In the second case, all replies used by a positive-target state predict $\beta$, and all replies used by a negative-target state predict $-\beta$. In particular, revealing all four private states is feasible. A positive-target state strictly prefers the largest available prediction, and a negative-target state the smallest, which rules out any additional form with distinct predictions.

\paragraph{Zero low-type target, $h\mu=\beta$, with $0<q_i<1$.}
The high type has targets $\pm2\beta$, while both low-type local conditions have target zero. There are three possibilities:
\begin{enumerate}
 \item Every reply predicts zero. The two high-type states must share a reply, while the zero-target states may join that reply, pool separately, or use separate replies
 \item The two high-type states use different replies, and both zero-target states use zero-prediction replies, either together or separately
 \item There are exactly two replies, one containing each high-type state. Both zero-target states can join the same reply, or one can join each reply
\end{enumerate}
These rules are equilibria for every $0<\gamma<1$ and $0<q_i<1$, and there are no others. In the last case, putting both zeros with one extreme gives prediction magnitude $2\beta q_i/(2-q_i)$ at that reply, below the other reply's magnitude $2\beta$, so both zero-target states prefer their assigned reply. Splitting the zeros gives predictions $\pm2\beta q_i$, making both zero-target states indifferent. If a zero-prediction reply exists, a zero-target state instead strictly prefers it to any nonzero prediction. The high-target states prefer the prediction in their own direction. These comparisons also show that revealing all four private states is feasible at $h\mu=\beta$.

\subsection{Worker reporting and conditional firm loss}\phantomsection\label{subsec:exchange-general-reporting}

We first derive an upper bound on the firm's gain under any incentive-compatible reporting strategy. We then show that a candidate rule attains it. This establishes the best-case outcome without classifying the other equilibria. Under direction-only reporting, the sign of the worker's preferred target is $X_i$ when $h\mu>\beta$ and $T_iX_i$ when $h\mu<\beta$.

The aggregate-state posterior determines which of these reporting rules satisfy worker incentives. Define the thresholds
\begin{equation}
 \mu_-=\frac{\beta}{h(1+\gamma)},\qquad
 \mu_3=\frac{(2+\gamma)\beta}{(2+3\gamma)h},\qquad
 \mu_0=\frac{\beta}{h},\qquad
 \mu_c=\frac{(1+\gamma)\beta}{h}
 \label{eq:exchange-reporting-thresholds}
\end{equation}
They satisfy $0<\mu_-<\mu_3<\mu_0<\mu_c$. The threshold $\mu_0$ separates opposite-direction and aligned targets. The proof below derives $\mu_-$ and $\mu_c$ from the full-revelation incentive condition, and $\mu_3$ from the condition for high-type revelation. Coincident targets and known types are treated directly in the same proof.

To compare feasible reporting outcomes, let $\Delta_i^{\mathrm{dir}}$, $\Delta_i^{\mathrm{full}}$, and $\Delta_i^{\mathrm{high}}$ denote the contribution in~\eqref{eq:exchange-firm-decomposition} under direction-only reporting, full revelation, and High-type revelation, respectively. A larger contribution means lower expected firm loss at the same aggregate-state posterior; these are not worker incentive constraints. Write $\widehat D_i^k$ for the prediction under reporting outcome $k$. The needed moments are
\begin{equation}
 \begin{array}{l|ll}
  \text{reporting outcome}&\mathbb E[X_i\widehat D_i^k]&\mathbb E[(\widehat D_i^k)^2]\\ \hline
  \text{direction only},\ \mu<\mu_0
    &(2q_i-1)[\beta+h\mu(2q_i-1)]&[\beta+h\mu(2q_i-1)]^2\\
  \text{direction only},\ \mu\geq\mu_0
    &h\mu+\beta(2q_i-1)&[h\mu+\beta(2q_i-1)]^2\\
  \text{full revelation}
    &h\mu+\beta(2q_i-1)&h^2\mu^2+2h\mu\beta(2q_i-1)+\beta^2\\
  \text{High-type revelation}
    &q_i v_u&q_i v_u^2
 \end{array}
 \label{eq:exchange-reporting-moments}
\end{equation}
At $\mu=\mu_0$, direction-only reporting in this table means that both types report $X_i$. A low type then has target zero and is indifferent between the two replies. The comparisons therefore allow weak incentive inequalities, as elsewhere in the model.

\begin{proposition}[Firm-preferred reporting equilibria]\label{prop:exchange-frontier}
Suppose $0<\gamma<1$, $\mu\geq0$, and the independent type beliefs are $q\in[0,1]^2$. For a worker whose type is uncertain, $0<q_i<1$, the largest attainable reduction in firm loss from their reply is
\begin{equation}
 G_i(\mu,q_i)=
 \begin{cases}
  \max\{0,\Delta_i^{\mathrm{dir}}(\mu,q_i)\},&0\leq\mu<\mu_-\\
  \max\{0,\Delta_i^{\mathrm{full}}(\mu,q_i)\},&\mu_-\leq\mu<\mu_3\\
  \max\{0,\Delta_i^{\mathrm{high}}(\mu,q_i)\},&\mu_3\leq\mu<\mu_0\\
  \Delta_i^{\mathrm{dir}}(\mu,q_i)
    +4\beta^2q_i(1-q_i)\max\{b_\gamma,0\}\mathbf1\{\mu\leq\mu_c\},
       &\mu\geq\mu_0
 \end{cases}
 \label{eq:exchange-best-reply-gain}
\end{equation}
Zero corresponds to babbling. In the last region, full revelation is preferred when $\gamma<\sqrt2-1$ and $\mu\leq\mu_c$; otherwise, reporting $X_i$ without revealing type is preferred. If $q_i\in\{0,1\}$, only $X_i$ remains private, and instead set $G_i(\mu,q_i)=\max\{0,\Delta_i^{\mathrm{full}}(\mu,q_i)\}$ at every posterior.

The lowest conditional expected firm loss is $L(\mu,\eta,q)$ in~\eqref{eq:exchange-frontier}, with $G$ defined by~\eqref{eq:exchange-reward} and $G_i$ given above. The workers may use different reporting rules if their beliefs differ. Allowing workers to randomize their replies does not lower the stated best-case loss: the bounds apply to mixed reporting strategies, while the stated values are attained by pure rules using at most four replies.
\end{proposition}

\proofheading{Proof.}
We bound the gain from any equilibrium reply strategy and then construct a rule attaining the bound. The argument uses Bayesian consistency and the reply incentive condition~\eqref{eq:exchange-terminal-ic}. All expectations below are conditional on the fixed $(q,s)$.

\emph{Implications of the reply incentive.}
First suppose there are four distinct targets $-v_u,-v_l,v_l,v_u$. Adding the incentive comparisons for two targets $d<d'$ shows that any predictions used by them satisfy $z\leq z'$. This monotonicity also holds for mixed replies: every prediction used with positive probability must be optimal for that target.

The extreme target $v_u$ strictly prefers the largest used prediction. Any lower used prediction excludes that state and is therefore at most $v_l$. If
\[
 2v_l>\frac{\gamma(v_l+v_u)}{1+\gamma}
\]
the target $v_l$ also strictly prefers the largest prediction to every lower one. The reflected argument applies to negative targets. Hence, when $v_l/v_u>\gamma/(2+\gamma)$, a nonconstant equilibrium prediction must equal
$\pm[q_i v_u+(1-q_i)v_l]$ according to the target's sign. If the prediction is constant, it is zero by iterated expectations. This bounds the induced actions without listing the reporting partitions.

A second implication will apply when $v_l/v_u>\gamma/[2(1+\gamma)]$. Suppose a used reply is sent by targets of both signs, and let $z$ be its prediction. Since the loss comparison between two predictions is affine in the true target, that prediction must also be optimal for both inner targets $-v_l$ and $v_l$. If $z\geq0$ and predictions are nonconstant, their unconditional mean zero implies that a smaller prediction $z_{\min}$ exists. The target $-v_l$ not preferring it requires
\[
 2v_l\leq-\frac{\gamma(z+z_{\min})}{1+\gamma}
       \leq\frac{\gamma v_u}{1+\gamma}
\]
The case $z<0$ follows by reflection. Therefore, if $v_l/v_u>\gamma/[2(1+\gamma)]$, any nonconstant equilibrium reply reveals the preferred-target sign $S_i=\operatorname{sgn}(d_i)$.

Babbling and direction-only reporting are feasible directly from the same incentive condition. Babbling makes every reply prediction zero. Under direction-only reporting, positive and negative targets strictly prefer the prediction with their own sign. Unused replies can be assigned the posterior of a used reply.

\emph{Opposite-direction low-type targets, $0<h\mu<\beta$.}
Here $v_l=\beta-h\mu$, $v_u=\beta+h\mu$, and
\[
 d_i=\beta S_i+h\mu X_i,\qquad S_i=\operatorname{sgn}(d_i)
\]
The firm's gain depends on the size of the worker's inferred target and its alignment with the worker's preferred direction. Write these moments as
\[
 Q_i=\mathbb E[\widehat D_i^2],\qquad
 M_i=\mathbb E[S_i\widehat D_i]
\]
Let $\Delta_i$ denote the firm's gain from worker $i$'s reply. Bayesian consistency gives $\mathbb E[d_i\widehat D_i]=Q_i$. Substituting the preceding expression for $d_i$ into~\eqref{eq:exchange-firm-decomposition} therefore gives
\begin{equation}
 \Delta_i=\frac{A_\gamma-(1-\gamma)^3}{2}Q_i
          -\frac{(1-\gamma)\gamma^2}{1+\gamma}\beta M_i
 \label{eq:exchange-direct-moment-objective}
\end{equation}
where $A_\gamma$ is defined in~\eqref{eq:exchange-aggregate-coefficient} and
\[
 \frac{A_\gamma-(1-\gamma)^3}{2}
 =\frac{(1-\gamma)\gamma^2}{1+\gamma}+b_\gamma>0
\]
The following two bounds hold for every incentive-compatible reporting strategy:
\begin{equation}
 Q_i\leq v_u M_i,\qquad Q_i-v_l M_i\leq q_i v_u(v_u-v_l)
 \label{eq:exchange-direct-moment-bounds}
\end{equation}
For the first, monotonicity of the induced prediction gives
\[
 v_u M_i-Q_i
 =\frac{(1-q_i)(v_u-v_l)}{2}
   \left\{\mathbb E[\widehat D_i\mid d_i=v_l]
          -\mathbb E[\widehat D_i\mid d_i=-v_l]\right\}\geq0
\]
For the second, $d_i-v_l S_i$ takes the ordered values $-(v_u-v_l),0,0,v_u-v_l$ as $d_i$ increases. Conditional covariance is therefore nonnegative, and Bayesian consistency gives
\[
 q_i v_u(v_u-v_l)-Q_i+v_l M_i
 =\mathbb E[\operatorname{Cov}(d_i,d_i-v_l S_i\mid r_i)]\geq0
\]
These are bounds on arbitrary reporting strategies, rather than comparisons between an enumerated set of equilibria.

If $\mu<\mu_-$, then $v_l/v_u>\gamma/(2+\gamma)$. The first incentive implication above makes every nonzero prediction outcome direction-only. Its gain is $\Delta_i^{\mathrm{dir}}$, while a zero prediction gives gain zero. This proves the first region.

If $\mu_-\leq\mu<\mu_3$, the second incentive implication makes every nonconstant reply reveal $S_i$. Conditional on such a reply,
$\widehat D_i=\beta S_i+h\mu\mathbb E[X_i\mid r_i]$. Thus the full-revelation gain bounds the gain from any such reporting strategy:
\[
 \Delta_i^{\mathrm{full}}-\Delta_i
 =\frac{A_\gamma-(1-\gamma)^3}{2}
    h^2\mu^2\,\mathbb E[\operatorname{Var}(X_i\mid r_i)]\geq0
\]
Full revelation attains the bound. Its only potentially profitable deviation is for target $v_l$ to imitate prediction $v_u$, so~\eqref{eq:exchange-full-revelation-comparison} requires $v_l/v_u\leq\gamma/(2+\gamma)$. Substituting $v_l=|h\mu-\beta|$ and $v_u=h\mu+\beta$ gives the full-revelation interval for four distinct targets:
\begin{equation}
 \frac{\beta}{1+\gamma}\leq h\mu\leq(1+\gamma)\beta
 \label{eq:exchange-revelation-region}
\end{equation}
Comparing this attainable bound with zero proves the second region.

If $\mu_3\leq\mu<\mu_0$, maximize~\eqref{eq:exchange-direct-moment-objective} over all nonnegative moment pairs satisfying~\eqref{eq:exchange-direct-moment-bounds}; the first bound already implies $M_i\geq0$ because $Q_i\geq0$. This enlarges the feasible set and therefore gives an upper bound without solving for the other equilibria. Since the coefficient on $Q_i$ is positive, for fixed $M_i$ the relaxed maximum uses
\[
 Q_i=\min\{v_u M_i,\ v_l M_i+q_i v_u(v_u-v_l)\}
\]
The two lines meet at $M_i=q_i v_u$, $Q_i=q_i v_u^2$. Above that point the objective's slope is
\[
 b_\gamma\beta-\frac{A_\gamma-(1-\gamma)^3}{2}h\mu<0
\]
Indeed, $2b_\gamma/[A_\gamma-(1-\gamma)^3]
=(1-2\gamma-\gamma^2)/(3-\gamma^2)<1/3$, whereas
$h\mu/\beta\geq h\mu_3/\beta>3/5$. Below the intersection the objective is linear and starts at zero. The maximum is therefore either zero or its value at the intersection, which is $\Delta_i^{\mathrm{high}}$.

High-type revelation attains the intersection moments. Its reply predictions are $-v_u,0,v_u$, and the inner targets prefer zero to imitating an extreme precisely when
\[
 v_l^2\leq\left(v_l-\frac{\gamma v_u}{1+\gamma}\right)^2
 \quad\Longleftrightarrow\quad
 \frac{v_l}{v_u}\leq\frac{\gamma}{2(1+\gamma)}
 \quad\Longleftrightarrow\quad
 \mu\geq\mu_3
\]
Thus high-type revelation or babbling attains the unrestricted moment bound, proving the third region.

\emph{Aligned targets, $h\mu\geq\beta$.}
If $b_\gamma>0$ and full revelation satisfies~\eqref{eq:exchange-revelation-region}, it maximizes both moments in~\eqref{eq:exchange-firm-decomposition}. Bayesian consistency gives
$\mathbb E[\widehat D_i^2]\leq\mathbb E[d_i^2]$, and the common ordering of $X_i$ and $d_i$ gives
\[
 \mathbb E[X_id_i]-\mathbb E[X_i\widehat D_i]
 =\mathbb E[\operatorname{Cov}(X_i,d_i\mid r_i)]\geq0
\]
If $h\mu>(1+\gamma)\beta$, the first incentive implication instead forces all nonzero prediction outcomes to be direction-only. Thus direction-only reporting attains the optimum there.

If $b_\gamma\leq0$, an upper bound can be obtained by temporarily relaxing reply incentives. Consider any reply pooling the two local-condition signs. Let $p$ be the conditional probability of $X_i=1$ and $v_+,v_-$ the conditional mean target magnitudes. Further revealing $X_i$ increases the firm-loss reduction by
\[
 p(1-p)(v_++v_-)
 \left\{\frac{2(1-\gamma)\gamma^2}{1+\gamma}h\mu
                +b_\gamma(v_++v_-)\right\}\geq0
\]
Here $v_++v_-\leq4h\mu$ and
\[
 \frac{(1-\gamma)\gamma^2}{1+\gamma}+2b_\gamma
 =\frac{(1-\gamma)^2\gamma^2(\gamma+2)}{(1+\gamma)^2}>0
\]
Once $X_i$ is revealed, the first information moment is fixed. Since $b_\gamma\leq0$, the second moment is best minimized by concealing all remaining type information. Reporting $X_i$ without type attains this bound and is incentive compatible; at $h\mu=\beta$, the low type is indifferent between the two sign replies. The full-versus-direction gain difference is
$4b_\gamma\beta^2q_i(1-q_i)$, giving the stated aligned-region formula.

\emph{Coincident targets and known types.}
At $\mu=0$, the gain is $b_\gamma\mathbb E[\widehat D_i^2]$ and
$0\leq\mathbb E[\widehat D_i^2]\leq\beta^2$. Revealing the preferred-target sign attains the upper endpoint and babbling attains zero, yielding the stated first-region value. The aligned argument already covers $h\mu=\beta$.
If $q_i\in\{0,1\}$, the coefficient $h\mu+\beta T_i$ in $d_i=(h\mu+\beta T_i)X_i$ is known. If it is nonzero, Bayesian consistency gives
$\mathbb E[X_i\widehat D_i]=\mathbb E[\widehat D_i^2]/(h\mu+\beta T_i)$, so the gain is linear in a second moment between zero and $(h\mu+\beta T_i)^2$. Babbling and revealing $X_i$ attain its two endpoints and are both incentive compatible. If the coefficient is zero, every prediction is zero.

Finally, the moments of the four attaining rules are given in~\eqref{eq:exchange-reporting-moments}. The contributions are additive across workers, so the individually attainable bounds can be combined. Substituting them into~\eqref{eq:exchange-expected-firm-loss} proves~\eqref{eq:exchange-frontier} and~\eqref{eq:exchange-reward}. No symmetry restriction on reporting was imposed.
\proofend

The proposition separates reporting incentives from the firm's preference over information. The thresholds determine which reporting rules workers are willing to use. The signs of the corresponding $\Delta_i^k$ determine whether the firm prefers that information to babbling. Equal beliefs $q_1=q_2=q$ need not equal $1/2$. For example, direction-only reporting below $\mu_0$ provides no information about $X_i$ at $q=1/2$, but becomes informative when either type is more likely. With unequal beliefs, one worker can reveal information while the other uses babbling at the same aggregate-state posterior.

\subsection{When should AI reveal the aggregate state?}\phantomsection\label{subsec:exchange-support-cases}

The conditional-loss formula turns the firm's signal problem into a comparison of posterior means. Since $q$ is fixed while AI chooses the current policy, iterated expectations give $\mathbb E[\eta]=\mathbb E[\theta^2]$ and $\mathbb E[\mu]=\mathbb E[\theta]$. Every term in~\eqref{eq:exchange-frontier} other than $-G(\mu,q)$ therefore has the same expectation under all signal policies. The firm minimizes expected loss by maximizing $\mathbb E[G(\mu,q)]$.

Between reporting thresholds, $G$ is convex in $\mu$. More precise AI information is then beneficial. Pooling can instead help at a threshold where the best attainable worker reporting outcome changes discretely. The support bounds matter because every posterior mean must lie in $[\theta_L,\theta_H]$.

\begin{proposition}[Full disclosure and pooling]\label{prop:exchange-support-cases}
Suppose $q\in[0,1]^2$ consists of independent type beliefs, $0<\gamma<1$, and $\theta\sim U[\theta_L,\theta_H]$, with $0\leq\theta_L<\theta_H<\infty$. Full aggregate-state revelation is optimal if and only if none of the thresholds satisfying the following conditions lies strictly inside $(\theta_L,\theta_H)$:
\begin{equation}
 \begin{array}{l|l}
  \text{threshold}&\text{condition for a strict improvement from pooling nearby states}\\ \hline
  \mu_-&\exists i:\ 0<q_i<1,\quad
           q_i>\dfrac{\gamma(1+\gamma^2)}{4(1-\gamma^2)}\\[5pt]
  \mu_3&2\gamma^2+3\gamma<4,\quad\sum_iq_i(1-q_i)>0\\
  \mu_0&\gamma>\sqrt2-1,\quad\sum_iq_i(1-q_i)>0\\
  \mu_c&\gamma<\sqrt2-1,\quad\sum_iq_i(1-q_i)>0
 \end{array}
 \label{eq:exchange-pooling-thresholds}
\end{equation}
If at least one qualifying threshold lies inside the support, pooling a sufficiently small interval centered at that threshold strictly improves expected firm loss relative to full revelation. These statements compare all initial public AI signal policies, including randomized ones.
\end{proposition}

\proofheading{Proof.}
For an uncertain type, the coefficients on $\mu^2$ in $\Delta_i^{\mathrm{dir}}$ below $\mu_0$, $\Delta_i^{\mathrm{full}}$, and $\Delta_i^{\mathrm{high}}$ are, respectively,
\[
 \frac{A_\gamma-(1-\gamma)^3}{2}h^2(2q_i-1)^2,\qquad
 \frac{A_\gamma-(1-\gamma)^3}{2}h^2,\qquad
 q_i\frac{A_\gamma-(1-\gamma)^3}{2}h^2
\]
They are nonnegative. Thus the first three regions of $G_i$ are maxima of convex quadratics, and the aligned region is a convex quadratic plus a constant reporting benefit below $\mu_c$. A known type contributes a convex maximum at every posterior.

At $\mu_-$, full revelation strictly improves on direction-only reporting for an uncertain type. It creates a positive jump precisely when $\Delta_i^{\mathrm{full}}(\mu_-,q_i)>0$. Direct substitution gives
\[
 \Delta_i^{\mathrm{full}}(\mu_-,q_i)
 =\frac{2(1-\gamma)^2\gamma^2(\gamma+2)\beta^2}{(1+\gamma)^3}
       \left\{q_i-\frac{\gamma(1+\gamma^2)}{4(1-\gamma^2)}\right\}
\]
This establishes the first row. At $\mu_3$, the three-reply rule strictly improves on full revelation for an uncertain type. Its gain is
\[
 \Delta_i^{\mathrm{high}}(\mu_3,q_i)
 =\frac{4q_i(1-\gamma)\gamma^2\beta^2}{(2+3\gamma)^2}
                       (4-3\gamma-2\gamma^2)
\]
Its positivity gives the second row. If either gain is nonpositive, the firm continues to prefer babbling and the corresponding threshold creates no discontinuity.

At $\mu_0$, the benefit is continuous for $b_\gamma\geq0$. Its right derivative minus its left derivative is $(1-\gamma)\gamma^2\beta h\sum_i(1-q_i)/(1+\gamma)\geq0$. If $b_\gamma<0$, switching from high-type-only local disclosure to reporting $X_i$ without type gives an upward jump of magnitude
\begin{equation}
 J_{\mu_0}=-4b_\gamma\beta^2\sum_iq_i(1-q_i)
 \label{eq:exchange-alignment-jump}
\end{equation}
At $\mu_c$, full worker revelation ceases to be feasible. There is a downward jump only if $b_\gamma>0$, of magnitude
\begin{equation}
 J_{\mu_c}=4b_\gamma\beta^2\sum_iq_i(1-q_i)
 \label{eq:exchange-upper-jump}
\end{equation}
These give the last two rows. At every threshold where there is no jump, the adjoining expressions either agree with the babbling benefit locally or form a continuous convex maximum. Hence $G$ is convex on the support if it has no interior discontinuity. For any AI signal,
\[
 \mathbb E[G(\mu,q)]
 =\mathbb E[G(\mathbb E[\theta\mid s],q)]
 \leq\mathbb E[G(\theta,q)]
\]
Full revelation attains this bound. An endpoint discontinuity does not affect the conclusion, because a continuous prior assigns that endpoint zero probability and a positive-probability posterior cannot have its mean there.

Conversely, let $\zeta$ be an interior threshold where $G$ jumps by magnitude $J_\zeta>0$. The higher one-sided value is attained at $\zeta$, because the corresponding reporting rule satisfies the weak incentive inequalities there. Pooling $[\zeta-w,\zeta+w]$ induces posterior mean $\zeta$. Its increase in expected benefit over full revelation is
\[
 \frac{2wG(\zeta,q)-\int_{\zeta-w}^{\zeta+w}G(x,q)\,dx}{\theta_H-\theta_L}
 =\frac{J_\zeta w+O(w^2)}{\theta_H-\theta_L}>0
\]
for sufficiently small $w>0$. This proves necessity as well as sufficiency.
\proofend

The support cases now follow from one comparison. If $\theta_L\geq\mu_0$, full revelation is optimal under high coordination incentives; under low coordination incentives, the only possible reason to pool is that the support crosses $\mu_c$. If $\theta_L<\mu_0<\theta_H$, high coordination incentives make pooling around $\mu_0$ beneficial whenever some type uncertainty remains. With lower coordination incentives, crossing $\mu_0$ alone does not justify pooling. If $\theta_H<\mu_0$, only the two lower thresholds can matter. In particular, when $2\gamma^2+3\gamma\geq4$, neither lower threshold qualifies, so full revelation is optimal for every support entirely below $\mu_0$.

The reasons for pooling differ across thresholds. Pooling around $\mu_c$ obtains additional worker type information by keeping the posterior within the full-revelation incentive region. Pooling around $\mu_0$ under high coordination incentives instead permits disclosure of local conditions without type: low types have target zero at that posterior and are indifferent between directional replies. Pooling around $\mu_-$ or $\mu_3$ can make a more useful reporting rule feasible when low types favor the opposite direction. These are all effects of AI's initial signal on voluntary worker communication.

\subsection{The optimal pooling policy}\phantomsection\label{subsec:exchange-optimal-signal}

Proposition~\ref{prop:exchange-support-cases} determines when full revelation is optimal. When pooling is beneficial, its optimal width also has a closed form if the resulting intervals stay within their adjacent reporting regions and do not overlap. Hold $q$ fixed and suppress it in the following derivatives of $G$.

\begin{proposition}[Optimal AI signal policy]\label{prop:exchange-optimal-signal}
Maintain the assumptions of Proposition~\ref{prop:exchange-support-cases}. At each qualifying threshold $\zeta$ inside the prior support, let $J_\zeta>0$ be the magnitude of the jump in $G$. Suppose the benefit expressions immediately to its left and right are quadratics, with coefficients $\alpha_-$ and $\alpha_+$ on $\mu^2$. Define
\begin{equation}
 \Delta_\zeta=G'_+(\zeta)-G'_-(\zeta),\qquad
 w_\zeta=\frac{\sqrt{\Delta_\zeta^2+4(\alpha_-+\alpha_+)J_\zeta}-\Delta_\zeta}
                    {2(\alpha_-+\alpha_+)}
 \label{eq:exchange-general-pooling-width}
\end{equation}
The derivatives are those of the right and left quadratic expressions. If all proposed intervals $[\zeta-w_\zeta,\zeta+w_\zeta]$ lie inside the prior support, remain within their respective two quadratic regions, and are pairwise disjoint, an optimal policy is
\begin{equation}
 s(\theta)=
 \begin{cases}
  \mathsf P_\zeta,&\theta\in[\zeta-w_\zeta,\zeta+w_\zeta]\\
  \theta,&\text{otherwise}
 \end{cases}
 \label{eq:exchange-optimal-policy}
\end{equation}
The labels $\mathsf P_\zeta$ are distinct across intervals. Each pooled signal induces mean $\zeta$, and workers use the firm-preferred reporting equilibrium from Proposition~\ref{prop:exchange-frontier}. The policy is optimal among all public AI signal policies, including randomized policies.

For an aligned support, Proposition~\ref{prop:exchange-aligned-signal} gives the corresponding single-pool policy and its gain. For high coordination incentives, a pool centered at $\mu_0$ instead uses
\begin{equation}
 \begin{aligned}
 J_{\mu_0}&=-4b_\gamma\beta^2\sum_iq_i(1-q_i),\qquad
 \Delta_{\mu_0}=\frac{(1-\gamma)\gamma^2}{1+\gamma}\beta h\sum_i(1-q_i)\\
 \alpha_-&=\left\{(1-\gamma)^3+
                  \frac{A_\gamma-(1-\gamma)^3}{2}\sum_iq_i\right\}h^2,
 \qquad \alpha_+=A_\gamma h^2
 \end{aligned}
 \label{eq:exchange-alignment-pooling-coefficients}
\end{equation}
If no threshold qualifies, including when both worker types are known, full aggregate-state revelation is optimal.
\end{proposition}

\proofheading{Proof.}
For one proposed interval, integrating the left and right quadratics gives the improvement over full revelation as a function of its half-width:
\begin{equation}
 \frac{J_\zeta w-\Delta_\zeta w^2/2-(\alpha_-+\alpha_+)w^3/3}
      {\theta_H-\theta_L}
 \label{eq:exchange-general-pooling-gain}
\end{equation}
Its derivative has the sign of $J_\zeta-\Delta_\zeta w-(\alpha_-+\alpha_+)w^2$. Because $J_\zeta>0$ and $\alpha_-+\alpha_+>0$, this expression has a unique positive root, given by~\eqref{eq:exchange-general-pooling-width}.

To compare all signal policies, let $\overline G(x,q)$ equal $G(x,q)$ outside the proposed intervals. On each interval, replace $G$ by the line joining its endpoint values. The first-order condition implies that this line also passes through $(\zeta,G(\zeta,q))$. Each quadratic piece is convex, and the value at the threshold is its higher one-sided value, so the line lies above both pieces. Its slope lies between the derivatives at the two interval endpoints. Between intervals, $G$ is continuous and convex. Thus $\overline G$ is convex on the entire prior support and satisfies $\overline G\geq G$.

Applying the conditional-expectation argument in Appendix~\ref{subsec:proof-main-aligned-signal} to this convex upper bound gives, for every public signal policy,
\[
 \mathbb E_s[G(\mu(s),q)]
 \leq\mathbb E_\theta[\overline G(\theta,q)]
\]
The bound depends only on the prior and $\overline G$, so it also applies to randomized policies and policies pooling disconnected sets.

The proposed policy attains the bound. It reveals the state outside the intervals, where $G=\overline G$. Within each interval, the posterior mean is its center $\zeta$, where $G=\overline G$, and $\overline G$ is affine throughout the posterior support. Its expected benefit therefore attains the common upper bound. No other public signal policy can improve on it under the stated interval conditions; ties are not excluded. Disjoint intervals contribute their gains additively.

The aligned-support specialization is proved in Proposition~\ref{prop:exchange-aligned-signal}. At $\mu_0$ under high coordination incentives, the left reporting rule has gains $\sum_i\Delta_i^{\mathrm{high}}$ and the right rule has gains $\sum_i\Delta_i^{\mathrm{dir}}$. Substitution of their moments gives~\eqref{eq:exchange-alignment-pooling-coefficients}. If both types are known, Proposition~\ref{prop:exchange-support-cases} has no qualifying threshold, so full revelation is optimal.
\proofend

The aligned-target pooling formula is one case of the same policy construction. If a proposed interval reaches a support boundary, crosses another quadratic region, or overlaps another pool, the width formula no longer determines the optimum. Proposition~\ref{prop:exchange-support-cases} still determines whether full revelation is optimal in those cases.

\FloatBarrier
\section{Numerical policy comparisons}\phantomsection\label{app:exchange-numerical}

\subsection{Disclosure benchmarks and worker welfare}

The illustrations evaluate the primitive loss functions at the equilibrium actions. Within each aggregate interval, actions use the posterior mean induced by its signal and losses use the realized state. Two-point Gaussian quadrature integrates the quadratic losses exactly. Averaging over the finite worker states reproduces the welfare formulas in Proposition~\ref{prop:exchange-direct-worker-loss}. Figure~\ref{fig:exchange-welfare} varies the common type prior over 197 equally spaced points from $0.01$ to $0.99$; the optimal pooling interval is interior at every displayed point.

\begin{figure}[htbp]
\centering\includegraphics[width=0.92\textwidth]{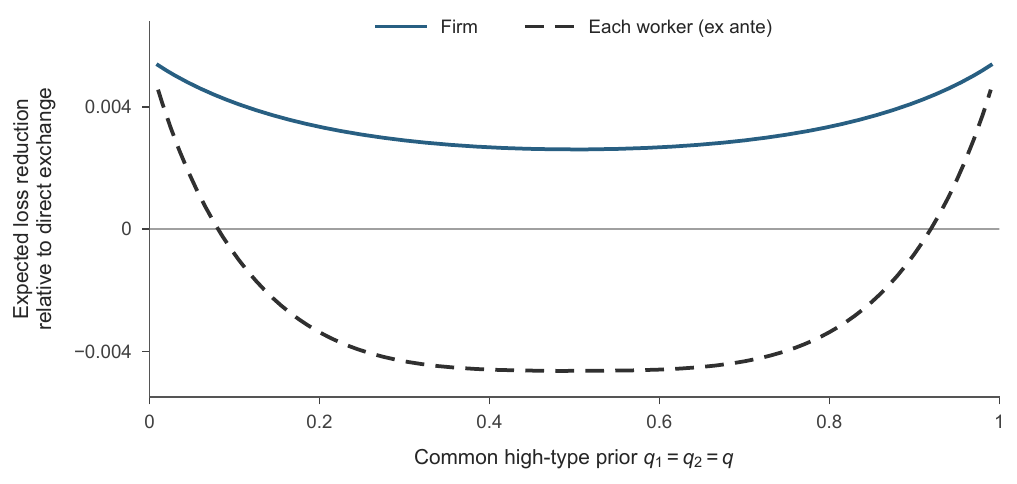}
\caption{Firm and worker gains from optimal project disclosure}\label{fig:exchange-welfare}
\par\vspace{8pt}\noindent
\begin{minipage}{\textwidth}\footnotesize
The reference is direct exchange without project context. Parameters are $h=\beta=1$, $\gamma=0.2$, and $\theta\sim U[1.01,1.35]$, with common type prior $q\in[0.01,0.99]$. Both policies use firm-preferred reporting equilibria. Positive values are reductions in expected loss; the worker curve applies to each worker ex ante by symmetry. The pooling interval is interior throughout the displayed range. Curves are model values without sampling uncertainty.
\end{minipage}
\end{figure}

\subsection{Worker acceptance and reporting incentives}\label{subsec:exchange-acceptance-numerical}

\begin{figure}[htbp]
\centering
\subfloat[Loss reductions as the pool widens]{\includegraphics[width=0.485\textwidth]{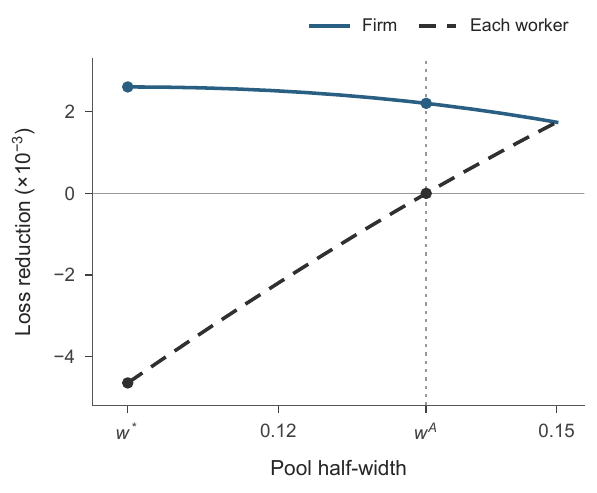}}\hfill
\subfloat[Disclosure with worker acceptance]{\includegraphics[width=0.485\textwidth]{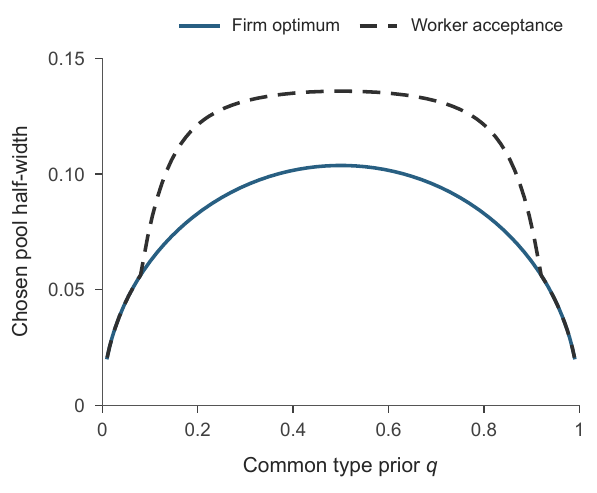}}
\caption{The disclosure cost of worker acceptance}\label{fig:exchange-acceptance}
\par\vspace{8pt}\noindent
\begin{minipage}{\textwidth}\footnotesize
Parameters are $h=\beta=1$, $\gamma=0.2$, and $\theta\sim U[1.01,1.35]$, with common type prior $q$. Panel (a) holds $q=0.5$ and varies the centered pool's half-width; positive values are reductions in expected loss relative to direct exchange. Panel (b) varies the common type prior while holding the other primitives fixed. Curves use firm-preferred reporting equilibria and optimize within the centered interval family in Proposition~\ref{prop:exchange-worker-acceptance}.
\end{minipage}
\end{figure}

\subsubsection{Strict preference-reporting incentives}\label{subsec:exchange-strict-reporting}

Acceptance does not by itself make preference reporting strict. A pool centered at $\mu_c$ leaves the moderate type indifferent between its truthful preference report and exaggeration. To provide a positive reporting incentive, move the center to $\mu_c-e$, where $e>0$, and adjust the width to preserve acceptance. Under the uniform prior, this shift changes the actual pooling interval and hence its posterior mean. The low type's loss increase from claiming the high type, after optimally adjusting its action, is
\begin{equation}
 M(e)=\frac{4\gamma^2(1-\gamma)h\beta}{(1+\gamma)^2}e>0.
 \label{eq:exchange-strict-reporting-margin}
\end{equation}
At a fixed width, the lower center reduces the range of project states supporting full preference reporting. Restoring acceptance requires a wider pool, which incurs an additional precision cost. The firm can obtain a small positive reporting margin at a cost that tends to zero with $e$.

Move the pool center to $c-e$ for $e>0$, and let its half-width be $w$. The interval is $[c-e-w,c-e+w]$, so its posterior mean is exactly $c-e$ under the uniform prior. For sufficiently small $e$, the interval still straddles $c$ and remains inside the support. Its reporting probability is $p(w)-e/D$, while its posterior-mean variance remains $v(w)$. The corresponding gains are
\begin{equation}
 G_L(w;e)=G_L(w)-\frac{4b_\gamma\beta^2S}{D}e,\qquad
 G_\ell(w;e)=G_\ell(w)-\frac{2\omega_\gamma\beta^2S}{D}e.
 \label{eq:exchange-strict-pool-gains}
\end{equation}
The preference-reporting margin is given by~\eqref{eq:exchange-strict-reporting-margin}.
When acceptance binds at $e=0$, a continuous nearby width $w^A(e)>w^A$ preserves $G_\ell(w^A(e);e)=0$. It retains a strictly positive firm gain for all sufficiently small $e$. The additional firm cost tends to zero with $e$. Appendix~\ref{subsec:exchange-acceptance-proof} establishes these local statements.

\begin{table}[htbp]
\caption{Worker acceptance with a strict reporting margin}\label{tab:exchange-strict-reporting}
\centering\small\setlength{\tabcolsep}{7pt}
\begin{tabular}{@{}rrrrr@{}}
\toprule
Posterior gap $e$ & Pool half-width & Full-reporting probability & \shortstack{Firm gain\\($10^{-3}$)} & \shortstack{Margin $M(e)$\\($10^{-3}$)} \\
\midrule
0 & 0.136 & 0.959 & 2.21 & 0.000 \\
0.001 & 0.137 & 0.960 & 2.13 & 0.089 \\
0.005 & 0.144 & 0.966 & 1.80 & 0.444 \\
0.010 & 0.152 & 0.975 & 1.31 & 0.889 \\
\bottomrule
\end{tabular}
\par\vspace{8pt}\noindent
\begin{minipage}{\textwidth}\footnotesize
Parameters are $h=\beta=1$, $\gamma=0.2$, $q_1=q_2=0.5$, and $\theta\sim U[1.01,1.35]$. Each worker's ex-ante gain is zero in every row. Firm gain is the reduction in expected loss relative to direct exchange. $M(e)$ is a worker-loss difference for preference exaggeration at the pooled signal. Nonzero-gap rows satisfy acceptance within the shifted interval family.
\end{minipage}
\end{table}

Table~\ref{tab:exchange-strict-reporting} shows how a lower pool center requires a wider interval to preserve acceptance. Both workers remain indifferent to direct exchange, while the firm retains a positive gain. Holding the width fixed would instead reduce full preference reporting and violate acceptance. Appendix~\ref{subsec:exchange-acceptance-proof} proves local feasibility and derives the reporting margin.

\subsection{Reporting in repeated collaboration}

\paragraph{Learning and later disclosure.}\phantomsection\label{subsec:exchange-learning-pooling}

The static results also describe how learning affects second-period disclosure outside the optimality region in Proposition~\ref{prop:exchange-dynamic-optimum}. For a statically optimal continuation, the pooling width around $\mu_c$ is given by~\eqref{eq:exchange-pooling-width} at the updated $q_2$, when its interval fits inside the support. Bayesian updating implies
\begin{equation}
 \mathbb E\left[\sum_iq_{i2}(1-q_{i2})\mid q_1,s_1\right]
 =\sum_iq_{i1}(1-q_{i1})-\sum_i\operatorname{Var}(q_{i2}\mid q_1,s_1)
 \label{eq:exchange-uncertainty-reduction}
\end{equation}
Therefore, when $\gamma<\sqrt2-1$, $\theta_L<\mu_c<\theta_H$, and the relevant intervals are interior,
\begin{equation}
 \mathbb E[w_{\mu_c}(q_2)^2\mid q_1,s_1]
 =w_{\mu_c}(q_1)^2
  -\frac{2b_\gamma\beta^2}{A_\gamma h^2}
                  \sum_i\operatorname{Var}(q_{i2}\mid q_1,s_1)
 \label{eq:exchange-expected-pooling-width}
\end{equation}
Learning reduces the expected squared pooling width, although a particular posterior can become more uncertain. With full type learning, the width is zero. These statements concern a statically optimal period-2 continuation. Initial commitment can instead choose a different continuation to influence period-1 reporting incentives, and that choice must be optimized jointly. The posterior-martingale proof is in Appendix~\ref{subsec:exchange-two-period-proofs}.

Under the continuation in Lemma~\ref{lem:exchange-dynamic-full-reporting}, the largest posterior supporting full first-period reporting is $\mu_c+\delta(\mu_c-\overline\theta)$. Figure~\ref{fig:exchange-dynamic-policy} illustrates the sufficient region in which the two-period policy reaches the full-information bound. 

\begin{figure}[htbp]
\centering\includegraphics[width=0.92\textwidth]{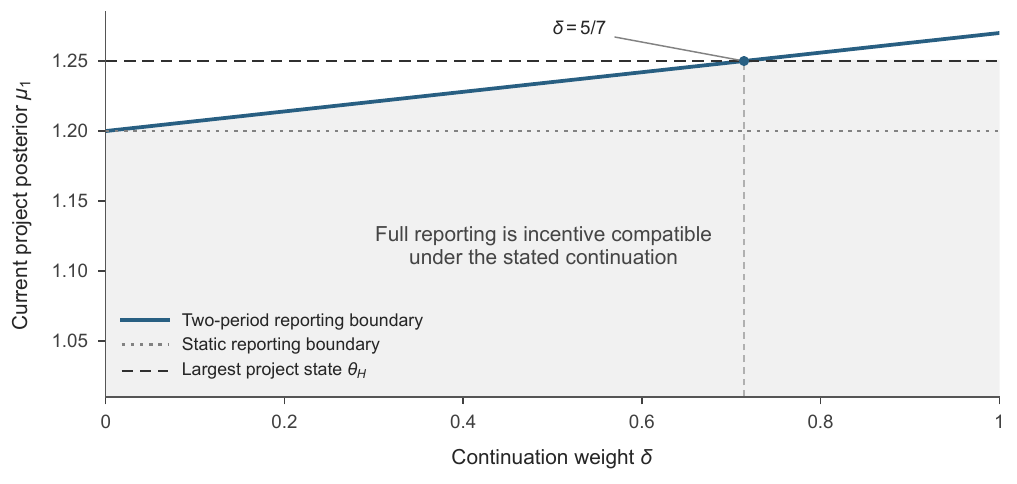}
\caption{The reporting boundary with repeated collaboration}\label{fig:exchange-dynamic-policy}
\par\vspace{8pt}\noindent
\begin{minipage}{\textwidth}\footnotesize
Parameters are $h=\beta=1$, $\gamma=0.2$, $\theta_t\sim U[1.01,1.25]$, and interior type priors. The boundary is $1.2+0.07\delta$ under the full-information second-period continuation in Lemma~\ref{lem:exchange-dynamic-full-reporting}. Shading marks current posteriors in the support satisfying the incentive condition. At $\delta\geq5/7$, the condition holds at every project state and Proposition~\ref{prop:exchange-dynamic-optimum} establishes joint optimality of full project disclosure and early preference learning.
\end{minipage}
\end{figure}

\subsection{Selective forwarding}

At $h=\beta=1$, $\gamma=0.1$, $q_1=q_2=0.65$, and $u=1.7$, the candidate reduces firm loss by about $0.05\%$ relative to optimized unchanged forwarding and also lowers each worker's loss. 

For the replacement construction, an operating information set consists of the worker's type, local condition, and incoming two-coordinate label. There are 16 such information sets. Conditional means and the coworker's action distribution determine the unique solution of the linear best-response system. A reporting deviation changes the outgoing delivery distribution, and the deviating worker re-optimizes their action at every incoming label. Separate delivery deviations instead hold the sending worker's action and the other delivery rule fixed. All four reports and all four delivered labels are considered. Within each half of the aggregate support, the difference in firm loss from a delivery deviation is affine in the realized aggregate state, so its endpoint values suffice. The half-width is the exact root defined in Appendix~\ref{subsec:exchange-baseline-optimal-proof}, rather than its first-order approximation.

Table~\ref{tab:exchange-relay-grid} reports all combinations of four coordination--center pairs and four common priors. Each row satisfies the reporting and separate-delivery inequalities, and the maximum action best-response residual is below $10^{-12}$. The whole support lies above $\mu_c$, so full initial disclosure with factual reporting is the globally optimized unchanged-forwarding benchmark. The reported gain is $100(L^{\mathrm{AI},*}-L^{\mathrm{candidate}})/L^{\mathrm{AI},*}$. A negative gain means that this particular candidate is inferior to the benchmark. The support width varies with the prior across rows.

\begin{table}[htbp]
\caption{Sensitivity of a selective-forwarding construction}\label{tab:exchange-relay-grid}
\centering\small\setlength{\tabcolsep}{5pt}
\begin{tabular}{@{}rrrrrr@{}}
\toprule
$\gamma$ & $u$ & $q$ & Half-width & Firm gain (\%) & \shortstack{Report margin\\($10^{-3}$)} \\
\midrule
0.1 & 1.7 & 0.50 & 0.275 & -0.899 & 4.89 \\
0.1 & 1.7 & 0.65 & 0.070 & 0.053 & 4.11 \\
0.1 & 1.7 & 0.80 & 0.020 & 0.043 & 0.98 \\
0.1 & 1.7 & 0.90 & 0.006 & 0.013 & 0.26 \\
0.2 & 2.8 & 0.50 & 0.501 & -1.219 & 16.56 \\
0.2 & 2.8 & 0.65 & 0.207 & -0.050 & 21.93 \\
0.2 & 2.8 & 0.80 & 0.071 & 0.045 & 6.24 \\
0.2 & 2.8 & 0.90 & 0.025 & 0.016 & 1.74 \\
0.3 & 5.0 & 0.50 & 0.691 & -0.701 & 31.92 \\
0.3 & 5.0 & 0.65 & 0.361 & -0.105 & 48.14 \\
0.3 & 5.0 & 0.80 & 0.148 & 0.012 & 15.35 \\
0.3 & 5.0 & 0.90 & 0.057 & 0.008 & 4.33 \\
0.4 & 10.0 & 0.50 & 0.862 & -0.256 & 49.74 \\
0.4 & 10.0 & 0.65 & 0.516 & -0.065 & 73.38 \\
0.4 & 10.0 & 0.80 & 0.243 & -0.004 & 26.54 \\
0.4 & 10.0 & 0.90 & 0.103 & 0.002 & 8.37 \\
\bottomrule
\end{tabular}
\par\vspace{8pt}\noindent
\begin{minipage}{\textwidth}\footnotesize
Parameters: $h=\beta=1$, $q_1=q_2=q$, and $\theta\sim U[u-w,u+w]$. Each row is a distinct environment; the support width changes with the prior to sustain the construction. Report margins are the smallest worker-loss increases among nontrivial reporting deviations; the final column is scaled as indicated. Positive firm gain denotes lower loss than optimal initial disclosure with unchanged forwarding.
\end{minipage}
\end{table}

\FloatBarrier
\section{Supplementary model details}\phantomsection\label{app:exchange-details}

\subsection{Posteriors induced by replies}\label{subsec:exchange-posteriors}

The probability of a reply averages the worker's actual type and local condition:
\begin{equation}
 P_i(r_i\mid q,s)=\frac12\sum_{T\in\{-1,1\}}\sum_{x\in\{-1,1\}}
             \Pr(T_i=T\mid q)\mathbf1\{\lambda_i(T,x,q,s)=r_i\}
 \label{eq:exchange-replyprob}
\end{equation}
For a positive-probability reply, the joint posterior over the worker's private variables is
\begin{equation}
 p_i(T,x\mid q,s,r_i)=
 \frac{\Pr(T_i=T\mid q)\mathbf1\{\lambda_i(T,x,q,s)=r_i\}}{2P_i(r_i\mid q,s)}
 \label{eq:exchange-posterior}
\end{equation}
For unused signals, off-equilibrium-path beliefs are specified as part of the equilibrium.

The aggregate state is independent of workers' private information, and their replies depend on it only through the public signal. Replies therefore convey no additional information about $\theta$. In particular, the posterior factorizes as
\begin{equation}
 \begin{aligned}
 &f(\theta,T_1,x_1,T_2,x_2\mid q,s,r_1,r_2)\\
 &\qquad=f(\theta\mid q,s)\,
                p_1(T_1,x_1\mid q,s,r_1)p_2(T_2,x_2\mid q,s,r_2)
 \end{aligned}
 \label{eq:exchange-factorization}
\end{equation}

\subsection{Recommendations of information-sharing strategies}\label{subsec:exchange-report-recommendations}

The signaling model can equivalently be written as recommendations of incentive-compatible reporting strategies. For a fixed signal policy, let $\lambda^*(q,s)$ denote the publicly specified pair of preferred reporting rules after $s$. AI sends the public message $m^{\mathrm{rec}}=(s,\lambda^*(q,s))$, recommending $r_i=\lambda_i^*(T_i,X_i,q,s)$. Each rule maps the worker's own $(T_i,X_i)$ to a reply, so AI need not observe the worker's private information. Workers continue to choose their operating actions themselves.

Under the aligned-support pooling policy in~\eqref{eq:exchange-aligned-policy}, the recommendation function is
\begin{equation}
 m^{\mathrm{rec}}(\theta,q)=
 \begin{cases}
  (\mathsf P_{\mu_c},\lambda^*(q,\mathsf P_{\mu_c})),
       &\theta\in[\mu_c-w_{\mu_c},\mu_c+w_{\mu_c}]\\
  (\theta,\lambda^*(q,\theta)),&\text{otherwise}
 \end{cases}
 \label{eq:exchange-report-recommendation}
\end{equation}
The pooling interval and its posterior mean are unchanged: every state in the pool induces mean $\mu_c$ and receives the same pair of recommended rules. In this case, AI recommends full worker revelation inside the pool around $\mu_c$ and at revealed states below $\mu_c$, and local-condition-only reporting at revealed states above $\mu_c$. Using the four reply labels, these rules can be written as
\[
 \begin{aligned}
 \lambda_i^{\mathrm{full}}(T_i,X_i)
   &=1+\mathbf1\{X_i=-1\}+2\mathbf1\{T_i=-1\}\\
 \lambda_i^{\mathrm{dir}}(T_i,X_i)
   &=1+\mathbf1\{X_i=-1\}
 \end{aligned}
\]
The second rule reveals $X_i$ but not $T_i$ because the support is aligned.

The equivalence follows because the appended rules are known functions of $(q,s)$: observing $m^{\mathrm{rec}}$ gives the same information as observing $s$. Posteriors, operating actions, and losses are unchanged. Following satisfies~\eqref{eq:exchange-terminal-ic} at every supported private state, including deviations followed by optimal action adjustment; no additional compliance assumption is imposed. Conversely, removing the appended recommendation maps any equilibrium back to the original signaling game.

Other reporting equilibria, including babbling, remain possible. The recommendation identifies a feasible reporting rule but does not establish unique implementation. Retaining $s$ is essential: a reporting rule alone can discard aggregate-state information and need not be equivalent.

\subsection{Histories, beliefs, and continuation losses}\label{subsec:exchange-updating}

Let $\mathcal H_{t-1}$ be the common history of public signals and forwarded reply pairs through period $t-1$. Define
\begin{equation}
 q_{it}=\Pr(T_i=1\mid\mathcal H_{t-1}),\qquad q_t=(q_{1t},q_{2t}),\qquad q_1=q^0
 \label{eq:exchange-dynamic-prior}
\end{equation}
The firm commits initially to the entire sequence of disclosure policies. We allow
\begin{equation}
 s_t\sim\sigma_t(\cdot\mid\theta_t,\mathcal H_{t-1}),\qquad
 r_{it}=\lambda_{it}(T_i,X_{it},\mathcal H_{t-1},s_t)
 \label{eq:exchange-dynamic-strategies}
\end{equation}
Current type beliefs summarize what is known about preferences. The disclosure policy can nevertheless promise different future signals after histories with the same $q_t$, changing the incentive to send a particular report today. The optimal policies characterized in Section~\ref{sec:exchange-dynamics} depend only on $q_t$ and current information.

After current replies, the updated belief is
\begin{equation}
 q^+_{i\mid r_i}=\sum_xp_i(1,x\mid q_t,s_t,r_i),\qquad
 q_r^+=(q^+_{1\mid r_1},q^+_{2\mid r_2})=q_{t+1}
 \label{eq:exchange-type-update}
\end{equation}
The posterior uses the reporting table selected at the current history and signal. Independence makes the two type likelihoods factor by worker. Knowing the actual aggregate state gives AI no additional type information, because the reporting rules depend on it only through the public signal. A worker's own private history likewise provides no additional information about their coworker's type conditional on the public history. Appendix~\ref{subsec:exchange-dynamic-beliefs-proof} proves these statements and the update formula.

Under pure reports and fair binary local conditions, each positive-probability reply gives an updated coordinate in
\[
 \left\{0,1,q_i,\frac{q_i}{2-q_i},\frac{2q_i}{1+q_i}\right\}
\]
Thus only finitely many type-belief vectors are reached at any fixed finite horizon. Histories leading to the same vector may nevertheless carry different continuation promises.

For a fixed policy and equilibrium, let $\ell_{it}(T,\mathcal H_{t-1})$ be worker $i$'s minimum expected discounted loss from period $t$, retaining their actual type $T$ and holding the coworker's strategy fixed. At supported types, the equilibrium strategy attains this value. Let $L_t(\mathcal H_{t-1})$ be expected firm loss from period $t$ onward. For supported types, these continuation losses satisfy
\begin{align}
 \ell_{it}(T,\mathcal H_{t-1})
 &=\mathbb E_i[\ell_{it}+\delta \ell_{i,t+1}(T,\mathcal H_t)
                         \mid T_i=T,\mathcal H_{t-1}]
 \label{eq:exchange-worker-value}\\
 L_t(\mathcal H_{t-1})
 &=\mathbb E_O[L_{Ot}+\delta L_{t+1}(\mathcal H_t)\mid\mathcal H_{t-1}]
 \label{eq:exchange-firm-value}
\end{align}
Here $\ell_{it}$ on the right-hand side is the period-$t$ operating loss in~\eqref{eq:exchange-workerloss}, while $\ell_{it}(T,\mathcal H_{t-1})$ on the left includes future losses. The worker expectation holds the worker's actual type fixed and averages the coworker's type under $q_{jt}$, the fresh aggregate state and local conditions, and the current signal. The firm also averages worker $i$'s type. These variables determine the reports and next history $\mathcal H_t=(\mathcal H_{t-1},s_t,r_t)$. If an earlier false report makes the public posterior assign zero probability to the worker's actual type, their continuation still uses that actual type and their best subsequent reports.

\ifdefined\APPENDICES
\clearpage
\setcounter{figure}{0}\renewcommand{\thefigure}{E.\arabic{figure}}\renewcommand{\theHfigure}{appendix.E.\arabic{figure}}
\section{Numerical disclosure and reporting policies}\label{app:exchange-simulation-details}

The calculations complement the performance comparisons in Section~\ref{sec:exchange-simulations}. They describe initial signals, the reports workers voluntarily supply, subsequent delivery, and learning over repeated periods. 

\subsection{Design and calculation}\label{subsec:exchange-simulation-setup}

We set $h=\beta=1$, so the aggregate state measures the weight on the productive condition relative to the magnitude of the worker's private bias. The aligned benchmark uses $\theta\sim U[1,2]$: both types favor the same direction as the firm's target, except that the low type's target is zero at the probability-zero endpoint $\theta=1$. We then use $\theta\sim U[0,2]$, which also includes states where the low type favors the opposite direction. Both environments use coordination weights $\gamma\in\{0.1,0.2,\ldots,0.9\}$ and common high-type probabilities $q_1=q_2\in\{0.05,0.2,0.5,0.8,0.95\}$, giving 45 static cases per support. Types remain independent, and each local condition has probability $1/2$. Numerical values are rounded for presentation; losses and incentive checks use the unrounded policies. Changing the support changes its mean and variance as well as the range of preference disagreement. The two-support comparison therefore concerns these environments as a whole. Without AI, the second support has posterior mean $\mu=1$, at which the low type's expected preferred target is zero.

We compare four information regimes. Under direct communication, workers exchange reports without aggregate project information. Under full initial disclosure, the firm reveals $\theta$ before workers report and forwards their reports unchanged. Optimal initial disclosure optimizes that public signal while retaining unchanged forwarding. Endogenous forwarding additionally lets AI choose a delivered message using the aggregate state and the source report. Workers anticipate these forwarding decisions when choosing their reports and actions.

Workers choose their reports and operating actions in every regime. When several reporting equilibria are feasible, the calculation selects the one with the lowest firm loss within the searched class. Expected firm and average worker losses integrate over the aggregate state and both workers' private information. Under unchanged forwarding, the aligned policy is the analytical global optimum, whereas the nonaligned policy is optimized over a finite set of signals. The delivery search identifies candidate improvements within a restricted policy family.

For unchanged forwarding without alignment, the calculation divides the aggregate support into 61 equal-width bins and uses a finite set of posterior means, including the reporting thresholds. Constant disclosure and exact full disclosure are available as comparison policies. A linear program allocates probability from each bin to each signal. The allocation is implemented deterministically by assigning each signal mirror-image subintervals within a bin. Symmetry about the bin midpoint preserves signal probabilities and posterior means; losses account for the continuous prior's second moment. Appendix~\ref{subsec:exchange-pure-implementation} establishes this deterministic implementation.

The unchanged-forwarding calculation considers every pure partition of the four worker states into reports. The forwarding search starts with rules defined on two aggregate-state intervals, then considers separate maps for the two workers on 3, 5, and 9 intervals, including cutoffs from the two-interval solutions. Each map uses only the aggregate state and source report. Maps can be asymmetric and nonmonotone. The search uses multiple starting points and local changes, retaining the two-interval policies as comparison candidates. Across the 90 static cases, it examines about 1.6 million profiles without finding a strict improvement over those candidates. Retained policies satisfy worker-reporting and AI-forwarding incentive conditions. Reported values are the best found within these numerical policy classes.

The two-period calculations use $q^0=(0.5,0.5)$, $\delta=0.8$, and five prior bins at all nine coordination weights. AI forwards reports unchanged. The firm chooses the first-period signal and history-dependent second-period signals jointly over a finite set of posterior means and reporting rules. Reporting constraints evaluate deviations using the worker's actual type, including after an earlier false report. Policies are deterministic at each public history, and losses incorporate second moments under the uniform prior. We measure learning by $\mathbb E[\sum_iq_{i,t+1}(1-q_{i,t+1})]$, the expected sum of posterior type variances after period-$t$ reports. This quantity starts at $0.5$ and reaches zero when both types are known.

The myopic-firm benchmark minimizes first-period firm loss and uses a statically optimal second-period signal at each updated belief. Workers remain forward-looking: their reporting incentives include both periods' losses under this policy. The jointly optimized policy instead minimizes total discounted firm loss and can choose future disclosure to influence first-period communication.

\subsection{Aligned benchmark}\label{subsec:exchange-simulation-performance}

\subsubsection{Initial disclosure and worker reporting}

Figure~\ref{fig:exchange-simulation-signals} checks the static analytical predictions under $U[1,2]$. At low coordination weights, the optimal unchanged-forwarding signal pools states around $\mu_c=1+\gamma$. The pooled signal induces mean $\mu_c$, at which full worker disclosure of $(T_i,X_i)$ is incentive compatible. At higher posterior means, workers report $X_i$ without revealing $T_i$. For $\gamma\geq\sqrt2-1$, the firm prefers disclosure of $X_i$ without type information, and full aggregate disclosure is optimal. The figure reproduces this distinction at $\gamma=0.2,0.3,0.6$.

Colors describe worker reporting rather than AI delivery. Full revelation distinguishes all four private states $(T_i,X_i)$; direction-only reporting reveals the sign of the worker's expected preferred target. Under alignment, that sign is $X_i$, so direction-only reporting communicates the productive condition but not the private bias. In a replacement panel, full revelation means that the worker reports their complete private state to AI. It need not mean that AI subsequently delivers all that information to the coworker.

\begin{figure}[!htbp]
 \centering
 \includegraphics[width=\textwidth]{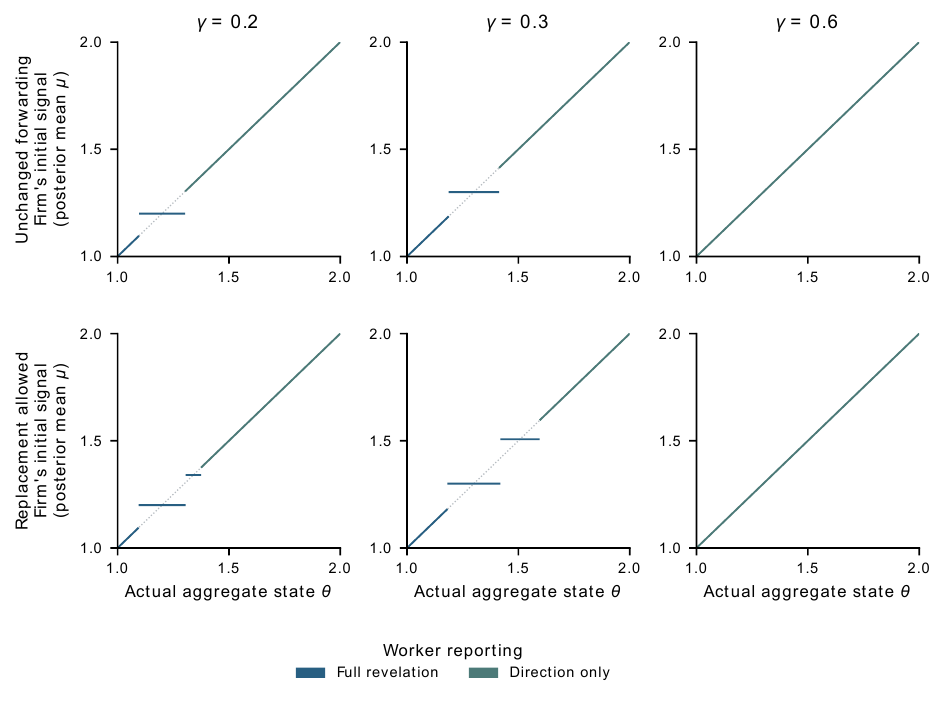}
 \caption{Initial disclosure and worker reporting under alignment}\label{fig:exchange-simulation-signals}
\par\vspace{8pt}\noindent
\begin{minipage}{\textwidth}\footnotesize
Parameters: $\theta\sim U[1,2]$, $h=\beta=1$, and $q_1=q_2=0.5$. Columns use $\gamma=0.2,0.3,0.6$; rows compare the analytical unchanged-forwarding optimum with the best selective-forwarding policies found. Signals are plotted at their posterior means. Diagonal segments reveal the state; horizontal segments pool it. Line styles and markers identify workers' reports before subsequent delivery; light gray dashed lines separate reporting regions.
\end{minipage}
\end{figure}

\subsubsection{Worker messages and AI forwarding}

Figure~\ref{fig:exchange-simulation-reporting-delivery} distinguishes workers' reports from the messages their coworkers receive. Its four rows correspond to the private states $(T_i,X_i)$. An entry $1\to3$ means that the worker reports message 1 and the assistant delivers message 3. The labels belong to the common space $\{1,2,3,4\}$; their informational content depends on the equilibrium reporting and delivery rules. The upper row shows the project information available before reporting.

At $\gamma=0.3$ and $q_1=q_2=0.5$, the best selective-forwarding policy found initially pools approximately $[1.42,1.60]$, giving posterior mean $1.51$. Workers report all four private states within that interval. Below approximately $\theta=1.49$, AI forwards unchanged; above it, AI merges each low-type report with the high-type report having the same $X_i$. Every delivered message therefore identifies $X_i$, but some no longer identify type. Messages 1 and 2 identify a low type, while messages 3 and 4 imply a high-type probability of about $62\%$. Recipients use the joint posterior over type and aggregate state when choosing actions. AI implements the rule using the source report and $\theta$; it does not independently observe the worker's private state.

\begin{figure}[!htbp]
 \centering
 \includegraphics[width=\textwidth]{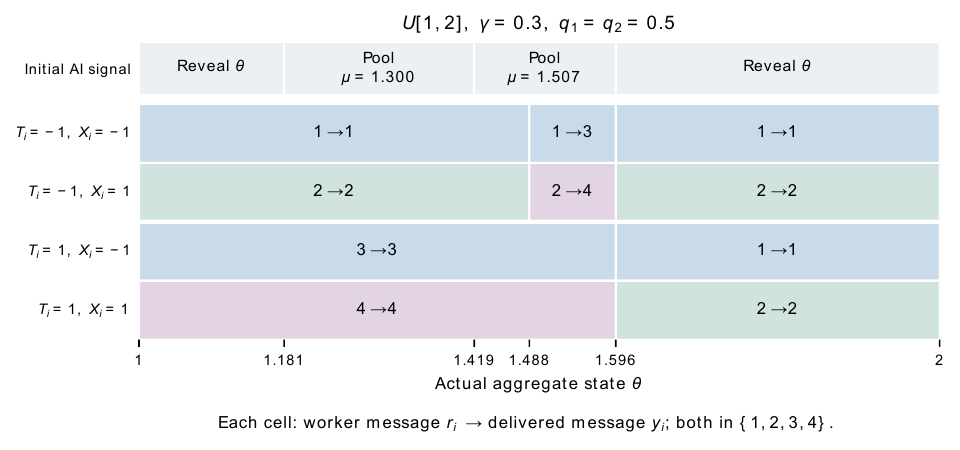}
 \caption{Worker reports and selective forwarding under alignment}\label{fig:exchange-simulation-reporting-delivery}
\par\vspace{8pt}\noindent
\begin{minipage}{\textwidth}\footnotesize
Parameters: $\theta\sim U[1,2]$, $\gamma=0.3$, and $q_1=q_2=0.5$. Each interval shows private states, sent messages, and received messages. The upper strip shows initial disclosure; blue delivery arrows and shaded received-message boxes mark replacements. Intervals are not drawn to scale. The assistant observes only the source report and aggregate state.
\end{minipage}
\end{figure}

\subsubsection{Two-period learning}

Figure~\ref{fig:exchange-aligned-dynamic-simulation} shows learning under the two-period commitment policy and compares firm losses with the myopic benchmark. The left panel measures type uncertainty after each period's reports, before the action and payoff observations available at the end of the task. At $\gamma=0.2$, uncertainty falls from $0.5$ initially to $0.4$ after period 1 and $0.24$ after period 2.

The right panel places the period-1 aggregate state $\theta_1$ on the horizontal axis and the period-2 state $\theta_2$ on the vertical axis. Each cell shows expected discounted firm loss under the myopic benchmark minus that under commitment, conditional on the displayed state intervals and averaging over private information and report histories. Blue indicates lower loss under commitment; gray indicates higher loss. At $\gamma=0.2$, commitment reduces ex-ante firm loss by about $0.12\%$. The differences are concentrated at $\theta_1<1.2$, where commitment induces earlier type learning. Myopic first-period reporting leaves type uncertainty at $0.5$, whereas commitment reduces it to $0.4$. Negative cells show that a policy minimizing expected loss need not improve performance at every state pair.

Two mechanisms can explain the difference. Learning about persistent types changes the information available for later decisions. In addition, committed future disclosure changes workers' incentives to report in period 1. For example, at $\gamma=0.1$, a first-period message with $\mu_1=1.3$ supports full worker reporting even though the static full-revelation threshold is only $\mu_c=1.1$. The anticipated continuation changes the payoff from a reporting deviation, making full reporting sustainable beyond the static threshold. This incentive need not produce type learning at every coordination weight: at the tested values $\gamma\geq0.5$, neither period reveals types.

The mean of $U[1,2]$ is $1.5$, above $\mu_c$ in the low-coordination region. This grid therefore lies outside the analytical parameter region in which full aggregate disclosure and complete first-period type learning are optimal. The static panels check the analytical optimum, while the dynamic panels examine how disclosure and reporting behave beyond that region.

\begin{figure}[!htbp]
 \centering
 \includegraphics[width=\textwidth]{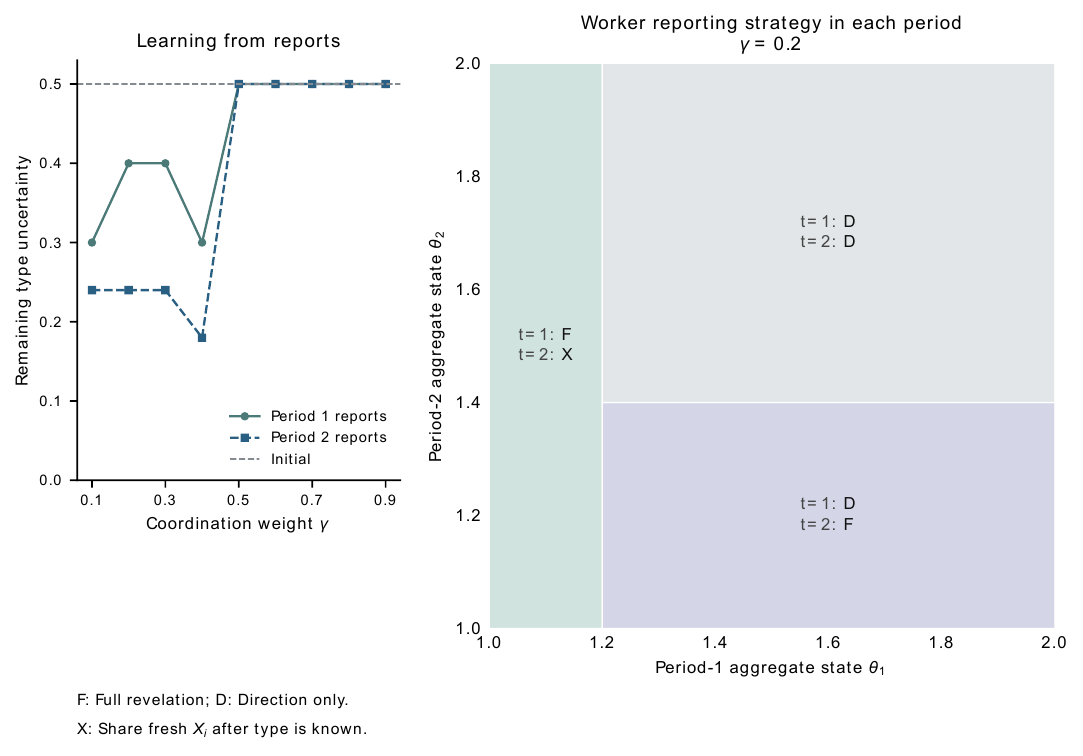}
 \caption{Learning and firm gains from dynamic commitment under alignment}\label{fig:exchange-aligned-dynamic-simulation}
\par\vspace{8pt}\noindent
\begin{minipage}{\textwidth}\footnotesize
Parameters: $\theta_t\sim U[1,2]$, $q^0=(0.5,0.5)$, and $\delta=0.8$. Left: type uncertainty after each period's reports under commitment. Right: expected myopic firm loss minus commitment loss, conditional on $(\theta_1,\theta_2)$ at $\gamma=0.2$. Loss is $L_{O1}+\delta L_{O2}$. Blue denotes a gain, gray a conditional loss, and white zero. Cells average private information, report histories, and states within the displayed intervals. Workers are forward-looking in both policies.
\end{minipage}
\end{figure}

\FloatBarrier
\subsection{Without alignment}\label{subsec:exchange-simulation-misalignment}\label{subsec:exchange-nonaligned-simulations}

\subsubsection{Initial disclosure and worker reporting}

We now repeat the analysis under $U[0,2]$. Figure~\ref{fig:exchange-nonaligned-signals} shows that initial pooling still changes reporting incentives, but can operate at additional thresholds. At $\gamma=0.3$, pooled signals induce means $\mu_3=(2+\gamma)/(2+3\gamma)$ and $\mu_c=1+\gamma$. The first supports high-type revelation: high types disclose their local condition, while low types send a common report. The second supports full revelation of both private variables. At $\gamma=0.6$ and $0.9$, pooling at $\mu_0=1$ supports disclosure of $X_i$ without $T_i$; at that mean the low type's expected preferred target is zero. The locations and associated reporting rules agree with Appendix~\ref{subsec:exchange-support-cases}. The numerical boundaries of the pooling intervals remain finite-grid approximations.

The change in reporting is economically important. Below $\mu_0$, the low type's preferred direction is opposite to the productive condition $X_i$. A report of preferred direction therefore need not identify the underlying fact. High-type revelation uses three messages and leaves low types' local conditions undisclosed; babbling reveals neither private variable. AI's initial information can now change whether workers communicate productive facts, rather than only whether they disclose preference intensity.

Replacement can also sustain disclosure of $X_i$ at an initial posterior below $\mu_0$. We label this rule ``Report $X_i$ only'': both types use one message for $X_i=-1$ and another for $X_i=1$. It differs from direction-only reporting because the low type's expected preferred direction is opposite to $X_i$. Its incentives must be checked under the replacement rule, rather than inferred from the unchanged-forwarding classification in Appendix~\ref{subsec:exchange-reporting-existence}. Within any given initial signal, the worker uses the same reporting rule at all pooled states because the worker does not observe $\theta$. The color in Figure~\ref{fig:exchange-nonaligned-signals} describes this worker rule; the subsequent AI delivery is shown separately below.

\begin{figure}[!htbp]
 \centering
 \includegraphics[width=\textwidth]{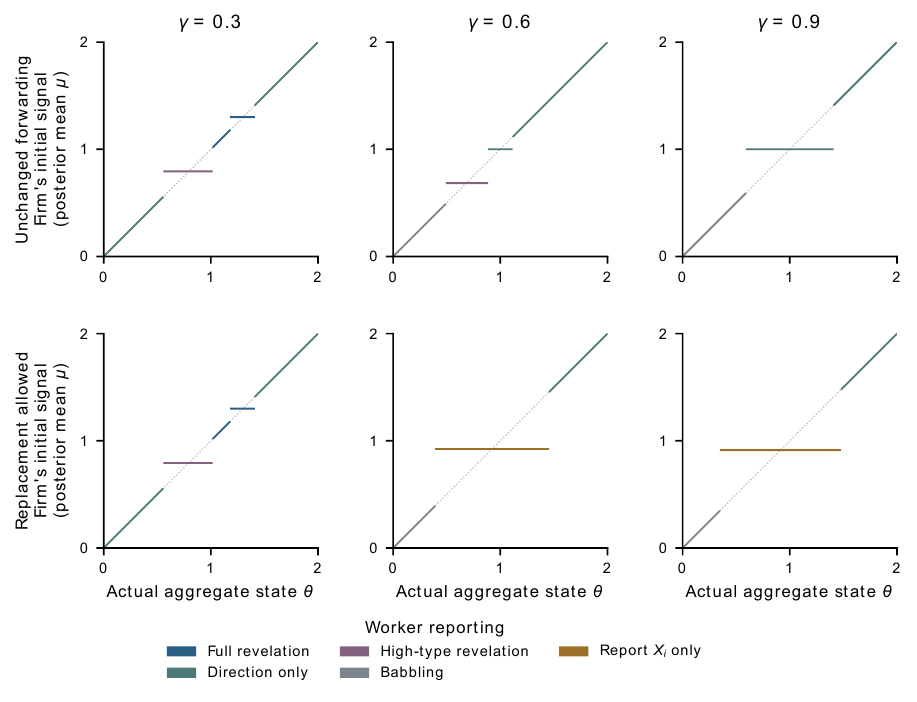}
 \caption{Initial disclosure when preferred directions can conflict}\label{fig:exchange-nonaligned-signals}
\par\vspace{8pt}\noindent
\begin{minipage}{\textwidth}\footnotesize
Parameters: $\theta\sim U[0,2]$ and $q_1=q_2=0.5$. Columns use $\gamma=0.3,0.6,0.9$; rows compare computed unchanged-forwarding policies with selective-forwarding candidates. Signals are plotted at their posterior means. Each finite bin displays the signal assigned to its largest portion; near-diagonal values are connected. Within-bin subdivisions are omitted only from the display. Line styles and markers identify worker reporting before delivery; light gray dashed lines separate reporting regions. Below $\mu_0=1$, reporting $X_i$ differs from reporting preferred direction.
\end{minipage}
\end{figure}

\subsubsection{Worker messages and AI forwarding}\label{subsec:exchange-concealing-simulation}

Replacement can conceal information that workers disclose to AI. Figure~\ref{fig:exchange-nonaligned-concealment} uses $\theta\sim U[0,2]$, $\gamma=0.4$, and $q_1=q_2=0.5$. The initial signal pools approximately $[0,0.48]$, inducing posterior mean $0.24$. Within that interval, workers use four messages to reveal $(T_i,X_i)$ to AI.

\begin{figure}[!htbp]
 \centering
 \includegraphics[width=\textwidth]{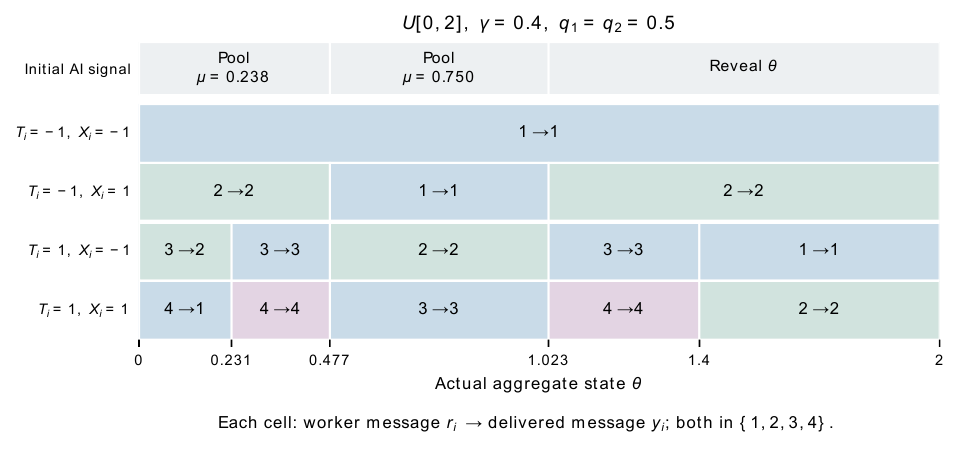}
 \caption{Selective forwarding that conceals worker information}\label{fig:exchange-nonaligned-concealment}
\par\vspace{8pt}\noindent
\begin{minipage}{\textwidth}\footnotesize
Parameters: $\theta\sim U[0,2]$, $\gamma=0.4$, and $q_1=q_2=0.5$. Each interval shows private states, sent messages, and received messages, with only used messages displayed. Blue delivery arrows and shaded boxes mark replacements; intervals are not drawn to scale. Within the first initial pool, delivery merges reports with opposite local conditions below approximately $\theta=0.23$ and leaves reports unchanged above that cutoff. Each delivery uses only the source report and aggregate state.
\end{minipage}
\end{figure}

Below approximately $\theta=0.23$, AI delivers message 2 instead of the high-type report 3, and message 1 instead of report 4. Delivered message 1 then pools $(-1,-1)$ with $(1,1)$, while message 2 pools $(-1,1)$ with $(1,-1)$. Each pair has the same preferred response direction but opposite local conditions. Even if the recipient additionally knew the actual aggregate state in this lower interval, either delivered message would leave probability $1/2$ on each sign of $X_i$. Delivery therefore conceals a reported local condition; it is not an invertible relabeling. Above the cutoff within the first pool, AI leaves the four reports unchanged.

The complete policy also pools approximately $[0.48,1.02]$ at mean $0.75$ and reveals higher states. Relative to the computed unchanged-forwarding benchmark, firm loss is nearly unchanged, falling by about $0.01\%$, while average worker loss rises by about $0.5\%$.

Revealing $\theta$ before workers report does not reproduce the first pool's complete worker disclosure. Under full aggregate disclosure and fully revealing worker reports, a low type with $X_i=-1$ can claim the high-type state with $X_i=1$. Preferring its own report would require $\theta\geq1/(1+\gamma)=5/7$ for $\theta>0$, which fails throughout that low-state interval. State-dependent delivery changes these incentives, allowing workers to reveal their private states to AI even though some of that information is concealed from their coworker.

This example combines concealment with additional aggregate disclosure. If deterministic delivery depends only on the public signal and source report, workers can send the corresponding coarser report directly, reproducing the outcome under unchanged forwarding (Appendix~\ref{subsec:proof-full-disclosure-replacement}). Here AI instead uses private aggregate information to decide what to conceal.
\FloatBarrier

Figure~\ref{fig:exchange-nonaligned-reporting-delivery} illustrates a special case at $\gamma=0.9$ and $q_1=q_2=0.5$. The candidate initially pools approximately $[0.35,1.48]$, inducing mean $0.91$. Inside this interval, workers send message 1 for $X_i=-1$ and message 2 for $X_i=1$, regardless of type. AI delivers those messages unchanged below approximately $\theta=0.91$, but delivers messages 3 and 4, respectively, above it.

No worker information is concealed in this case. Delivered messages 1 and 3 identify original report 1 and hence $X_i=-1$; messages 2 and 4 identify original report 2 and hence $X_i=1$. Receiving 1 or 2 additionally indicates the lower aggregate interval, while receiving 3 or 4 indicates the upper interval. The resulting aggregate posterior means are approximately $0.63$ and $1.20$, and the type belief remains $0.5$.

The assistant uses its aggregate information after workers have reported. Forwarding the factual report unchanged and then indicating whether $\theta$ is below or above approximately $0.91$ gives recipients exactly the same information. Both implementations require four delivered labels: two factual reports crossed with two aggregate regions. Releasing that aggregate information before reporting would change the workers' incentives. The candidate therefore illustrates the value of separating elicitation from subsequent project disclosure. It provides no evidence of a gain from concealing worker information.

\begin{figure}[!htbp]
 \centering
 \includegraphics[width=\textwidth]{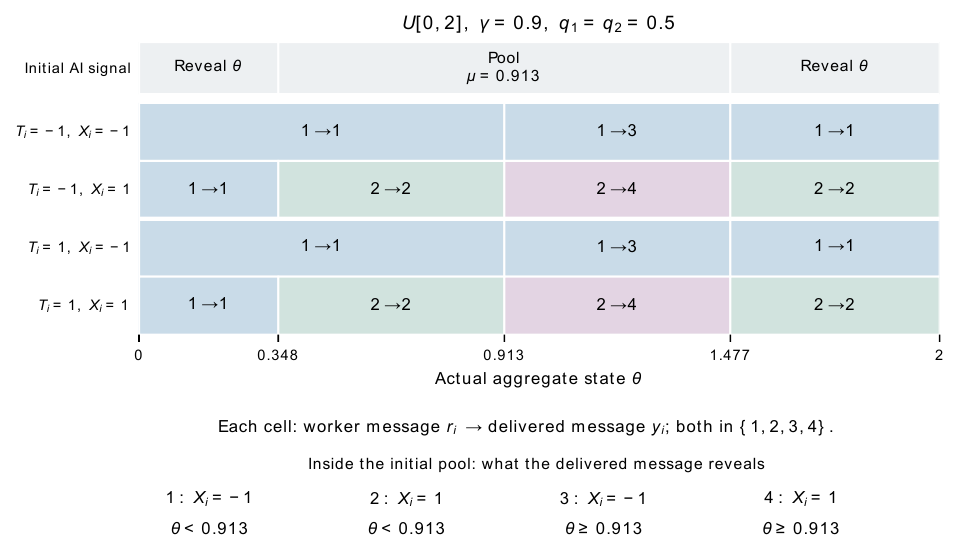}
 \caption{Factual reporting followed by aggregate disclosure}\label{fig:exchange-nonaligned-reporting-delivery}
\par\vspace{8pt}\noindent
\begin{minipage}{\textwidth}\footnotesize
Parameters: $\theta\sim U[0,2]$, $\gamma=0.9$, and $q_1=q_2=0.5$. Within the initial pool, workers use two labels to report $X_i$. The four delivered labels identify both that factual report and an aggregate-state region. Unchanged forwarding followed by the same aggregate signal is informationally equivalent. Each interval shows private states, sent messages, and received messages, with only used messages displayed. Blue delivery arrows and shaded boxes mark replacements; intervals are not drawn to scale.
\end{minipage}
\end{figure}

\FloatBarrier
\subsubsection{Two-period learning}

Figure~\ref{fig:exchange-nonaligned-dynamic-simulation} reports two-period learning and the firm-loss comparison under $U[0,2]$. At $\gamma=0.4$, expected type uncertainty under commitment falls from $0.5$ initially to $0.1$ after period-1 reports and $0.06$ after period-2 reports. Full revelation and high-type revelation identify the worker's type in 80\% of period-1 cases. The remaining 20\% use direction-only reporting and leave type uncertain. Myopic first-period reporting instead leaves expected type uncertainty at $0.4$.

The right panel compares discounted firm losses at each pair of aggregate states. At $\gamma=0.4$, commitment reduces ex-ante firm loss by about $1.2\%$ relative to the myopic benchmark. The gain is larger than in the aligned example, but varies across both aggregate states. Blue and gray cells show that the lower expected loss combines improvements at some state pairs with losses at others. As in the aligned case, the difference can reflect both the future value of learning and the effect of committed disclosure on first-period reporting incentives.

At $\gamma=0.8$ and $0.9$, some period-2 reports reveal type without revealing $X_i$. Because fresh $X_i$ remains equally likely to have either sign, such reports can induce the same actions and losses as babbling. Lower type uncertainty in the left panel therefore need not imply better performance.

\begin{figure}[!htbp]
 \centering
 \includegraphics[width=\textwidth]{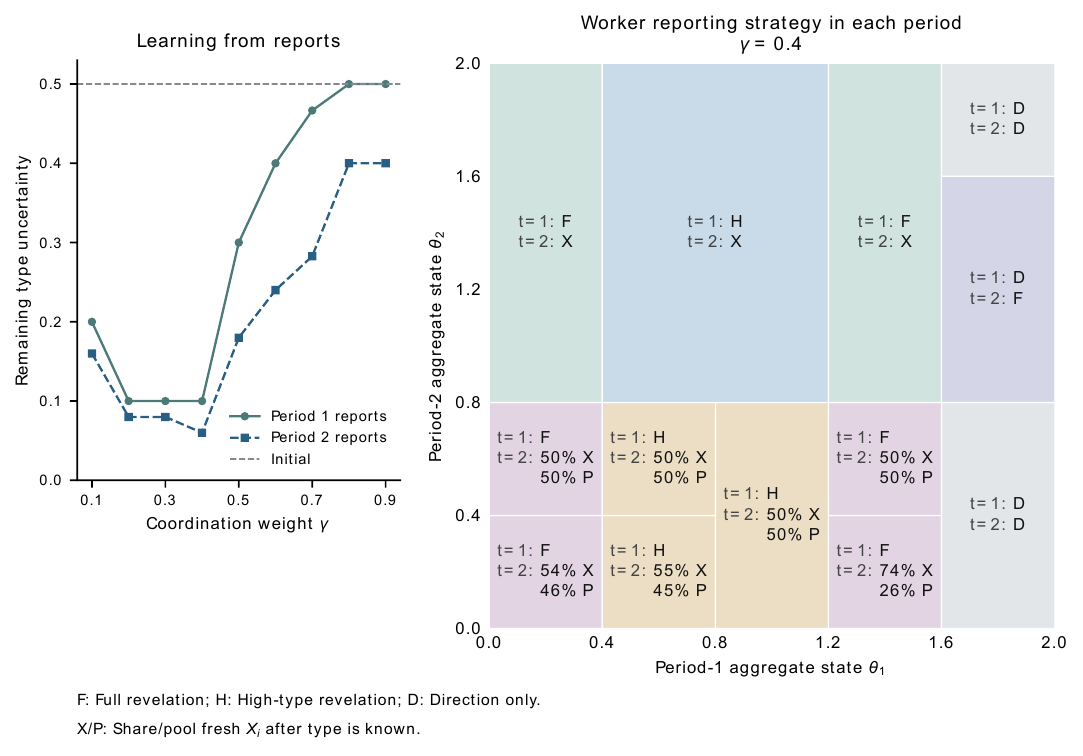}
 \caption{Learning and firm gains from dynamic commitment without alignment}\label{fig:exchange-nonaligned-dynamic-simulation}
\par\vspace{8pt}\noindent
\begin{minipage}{\textwidth}\footnotesize
Parameters: $\theta_t\sim U[0,2]$, $q^0=(0.5,0.5)$, and $\delta=0.8$. Left: type uncertainty after each period's reports under commitment. Right: expected myopic firm loss minus commitment loss, conditional on $(\theta_1,\theta_2)$ at $\gamma=0.4$. Workers are forward-looking in both policies. Blue denotes lower loss under commitment, gray higher loss, and white zero. The color scale is shared with Figure~\ref{fig:exchange-aligned-dynamic-simulation}; cells average private information, report histories, and states within the displayed intervals.
\end{minipage}
\end{figure}

\FloatBarrier

\subsection{Initial policies across aggregate supports}

Figure~\ref{fig:exchange-simulation-alignment-comparison} compares the initial signals at $\gamma=0.9$. Under alignment, full initial disclosure supports the factual reporting preferred by the firm. On the broader support, pooling changes reporting incentives and subsequent delivery can restore some aggregate context. The latter candidate preserves factual reports, as formalized in Appendix~\ref{subsec:exchange-timing-proof}.

\begin{figure}[!htbp]
 \centering
 \includegraphics[width=\textwidth]{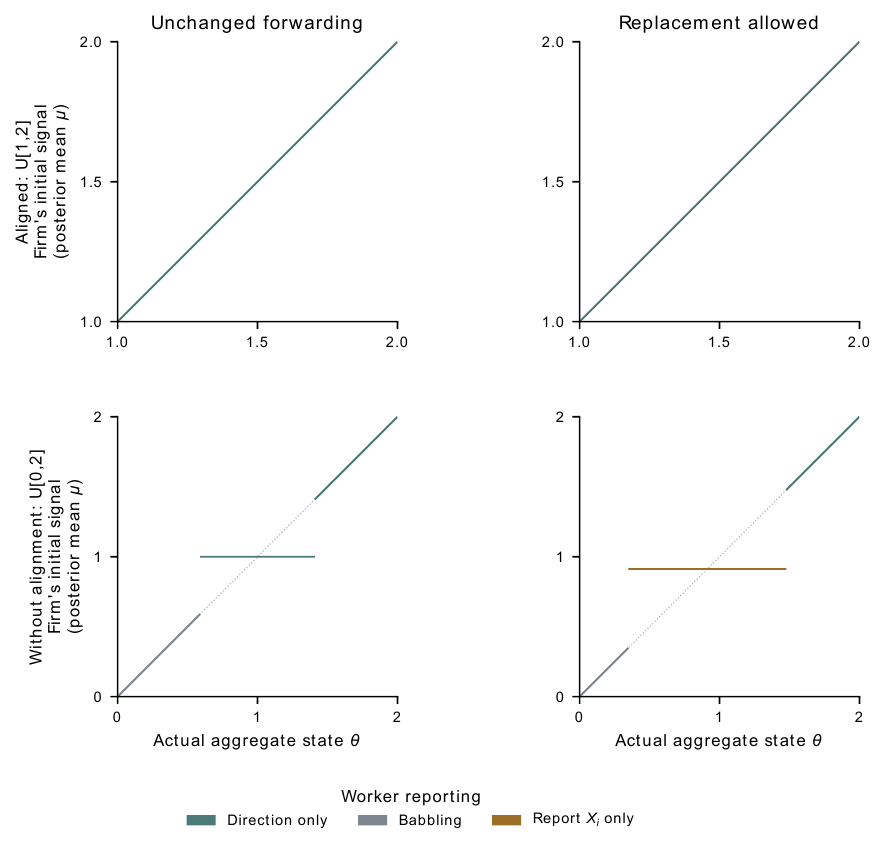}
 \caption{Initial signals and reporting across the two aggregate supports}\label{fig:exchange-simulation-alignment-comparison}
\par\vspace{8pt}\noindent
\begin{minipage}{\textwidth}\footnotesize
Parameters: $\gamma=0.9$ and $q_1=q_2=0.5$. Rows compare aggregate supports; columns compare unchanged forwarding with selective-forwarding candidates. Signals are plotted at their posterior means. Each finite bin displays the signal assigned to its largest portion, with near-diagonal values connected. Line styles and markers identify worker reports before delivery; light gray dashed lines separate reporting regions. Within-bin subdivisions are omitted only from the display.
\end{minipage}
\end{figure}
\FloatBarrier

\begingroup
\setcounter{figure}{0}\renewcommand{\thefigure}{F.\arabic{figure}}\renewcommand{\theHfigure}{appendix.F.\arabic{figure}}
\section{Delayed project disclosure and deterministic implementation}\label{app:exchange-timing}

\subsection{Forwarding reports and project information}\label{subsec:exchange-timing-proof}

\begin{proposition}[Equivalent delivery of reports and context]\label{prop:exchange-context-equivalence}
Fix an initial public signal and deterministic forwarding rules. Identify duplicate input labels that produce the same delivery at every aggregate state, and let $r_i$ index the remaining distinct report alternatives. Suppose $g_i(\theta,s,r_i)=e_i(r_i,C(\theta,s))$, where $e_i$ is one-to-one on attainable report--context pairs, including pairs induced by reporting deviations. Receiving the delivered label is then equivalent to receiving the source report and the public context $C$ after reporting. With the pair encoded in the same message space, workers have identical information, reporting incentives, operating actions, and losses under the two representations.
\end{proposition}

\proofheading{Proof of Proposition~\ref{prop:exchange-context-equivalence}.}
The inverse of $e_i$ recovers the distinct source report and context from the delivered label. Conversely, the pair uniquely determines that label. Every recipient's information is thus the same under the two representations. The initial signal, private observations, and report timing are unchanged. Apply the same recoding to beliefs and operating strategies after every report, including a deviation. Duplicate input labels produce the same delivery distribution and introduce no additional reporting incentive. Each reporting strategy therefore induces the same distribution of actions and losses. Encoding the pair uses the same delivered-message capacity.
\proofend

In the four-label example, use labels 1 and 2 for the two factual reports. After the lower project signal deliver 1 or 2; after the upper signal deliver 3 or 4. Unused input labels can duplicate the factual reports, so they introduce no distinct deviations. This implements the prescribed factual equilibrium with two report alternatives and four delivered labels.

\subsection{Factual reporting with delayed context}\label{subsec:exchange-delayed-binary}

\begin{proposition}[Reporting incentives and a binary disclosure rule]\label{prop:exchange-delayed-context}
Let $q_1=q_2=1/2$, $0<\gamma<1$, $h>0$, $\beta\geq0$, and let the nonnegative aggregate state have positive mean $\mu$ and finite second moment. The initial signal is uninformative. Workers report their local signs before a committed public signal $C$, and each recipient receives the source report and $C$. Put $M=\mathbb E[\theta\mid C]$. Unused report labels duplicate the two sign reports. Factual reporting is an equilibrium if and only if $h\mathbb E[M^2]\geq\beta\mu$. Its operating actions and loss increase from a false-sign report by type $T$ are
\begin{align}
 a_i&=\frac{hM}{1+\gamma}(X_i+\gamma X_j)+(1-\gamma)\beta T_iX_i,
 \label{eq:exchange-delayed-actions}\\
 D(T)&=\frac{4\gamma^2(1-\gamma)h}{1+\gamma}
       \{h\mathbb E[M^2]+\beta T\mu\}.
 \label{eq:exchange-delayed-margin}
\end{align}
If $C$ is instead released before reports, factual reporting requires $hM\geq\beta$ at every positive realized posterior mean. At a zero posterior mean the false-sign report has zero effect.

For $\theta\sim U[\theta_L,\theta_H]$, put $D=\theta_H-\theta_L>0$ and $\mu=(\theta_L+\theta_H)/2$. Among deterministic binary public signals with the prescribed factual-reporting equilibrium, a midpoint cutoff maximizes the low-type reporting margin and minimizes firm loss. Such a binary signal supports factual reporting if and only if
\begin{equation}
 h\left(\mu^2+\frac{D^2}{16}\right)\geq\beta\mu.
 \label{eq:exchange-binary-feasibility}
\end{equation}

\end{proposition}

\proofheading{Proof of Proposition~\ref{prop:exchange-delayed-context}.}
Conditional on $C$ and factual reports, the coworker's type remains balanced. The best responses give~\eqref{eq:exchange-delayed-actions}. If worker $i$ reports $r\in\{-1,1\}$, their coworker's action is
\[
 a_j(r)=(1-\gamma)(hM+\beta T_j)X_j
       +\frac{\gamma^2hMX_j+\gamma hMr}{1+\gamma}.
\]
The deviator optimizes with their actual type and local condition, obtaining
\[
 a_i(r)=(1-\gamma)(hM+\beta T_i)X_i
       +\frac{\gamma^2hMr+\gamma hMX_j}{1+\gamma}.
\]
The deviator's report-dependent loss conditional on $C$, after averaging the independent coworker information and re-optimizing their own action, is
\[
 \gamma(1-\gamma)
 \left\{(hM+\beta T_i)X_i-\frac{\gamma hM}{1+\gamma}r\right\}^2.
\]
The difference between $r=-X_i$ and $r=X_i$ is
$4\gamma^2(1-\gamma)hM(hM+\beta T_i)/(1+\gamma)$.
Reports precede $C$, so expectations yield~\eqref{eq:exchange-delayed-margin}. The low type has the binding condition; duplicate labels exhaust the remaining deviations. If $C$ arrives first, the same calculation holds signal by signal, establishing the early-disclosure condition.

Direct integration of the primitive losses gives firm and average worker losses
\begin{align}
 L&=(1-\gamma)h^2\mathbb E[\theta^2]
       -A_\gamma h^2\mathbb E[M^2]
       +(1-\gamma)^2(1+\gamma)\beta^2,\label{eq:exchange-delayed-firm-loss}\\
 \overline\ell&=(1-\gamma)h^2\mathbb E[\theta^2]
       -A_\gamma h^2\mathbb E[M^2]
       +\gamma(1-\gamma)(2-\gamma)\beta^2,
 \label{eq:exchange-delayed-worker-loss}
\end{align}
where $A_\gamma=(1-\gamma)(1+\gamma^2)/(1+\gamma)^2>0$. Thus more variance in the eventual posterior mean improves both losses and the binding reporting margin within this equilibrium family.

For a binary partition with probability $p$ on one cell, posterior-mean variance is $p(1-p)(m_1-m_0)^2$. At fixed $p$, a cell comprising one endpoint interval maximizes the distance between cell means: among uniform-measure subsets of a given size, the upper interval has the largest integral of $\theta$ and the lower interval the smallest. The endpoint partition has $|m_1-m_0|=D/2$, hence variance at most $p(1-p)D^2/4\leq D^2/16$. The midpoint attains this bound, so it maximizes $\mathbb E[M^2]=\mu^2+\operatorname{Var}(M)$. This proves the optimality and feasibility claims for the stated family.
\proofend

For $\theta\sim U[0,2]$ and $h=\beta=1$, midpoint disclosure gives $M\in\{0.5,1.5\}$ with equal probabilities. The low-type false-report loss is $\gamma^2(1-\gamma)/(1+\gamma)>0$ when context is delayed. Under early disclosure at $M=0.5$, it is $-\gamma^2(1-\gamma)/(1+\gamma)<0$. The two policies provide the same eventual aggregate information and use the same four delivered labels, but differ in when workers can condition their reports on that information.

\begin{figure}[htbp]
\centering
\includegraphics[width=0.64\textwidth]{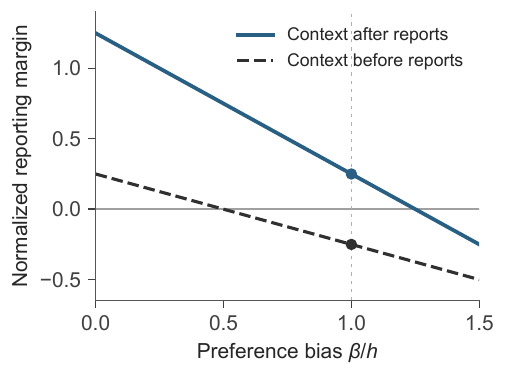}
\caption{Low-type incentives under early and delayed project context}\label{fig:exchange-timing-margin}
\par\vspace{8pt}\noindent\begin{minipage}{\textwidth}\footnotesize
The prior is $U[0,2]$, $h=1$, and type priors are balanced. The early-disclosure margin is conditional on the lower posterior mean 0.5; the delayed margin averages both signals. Margins are divided by $4\gamma^2(1-\gamma)/(1+\gamma)$. Negative values indicate a profitable false-sign deviation from factual reporting. The vertical guide marks $\beta=1$.
\end{minipage}
\end{figure}

\subsection{A deterministic implementation of finite-bin allocations}\label{subsec:exchange-pure-implementation}

Consider a uniform prior partitioned into intervals $I_b=[l_b,u_b]$ with midpoint $m_b$ and length $d_b$. Suppose a finite allocation assigns fractions $\alpha_{bk}\geq0$, $\sum_k\alpha_{bk}=1$, of bin $b$ to signal $k$. Set $c_{b0}=0$ and $c_{bk}=(d_b/2)\sum_{j\leq k}\alpha_{bj}$. Assign signal $k$ to
\[
 [m_b-c_{bk},m_b-c_{b,k-1}]
 \;\cup\;
 [m_b+c_{b,k-1},m_b+c_{bk}]
\]
within that bin, ignoring measure-zero shared endpoints. Each assigned set has length $d_b\alpha_{bk}$ and conditional mean $m_b$. The sets partition the bin, so the resulting signal is a deterministic function of the aggregate state. Its probabilities and posterior means equal those of the allocation.

For unchanged forwarding, reporting incentives depend on the aggregate posterior mean. Conditional operating actions also depend on that mean and workers' reports. Expanding firm loss leaves the state second-moment term $(1-\gamma)h^2\mathbb E[\theta^2]$, which is independent of the disclosure policy; the remaining state terms use the signal probabilities and means. The deterministic construction therefore preserves reporting incentives and aggregate expected loss when the objective includes the continuous prior's actual second moment.

\subsection{Numerical details for the timing comparison}\label{subsec:exchange-timing-numerical}

At $\gamma=0.9$ and balanced type priors, full initial disclosure under firm-preferred pure reporting has expected firm loss $0.0928$ and average worker loss $0.1623$. Enumeration of the fifteen partitions of the four private states gives babbling below approximately $\theta=0.735$, high-type revelation between that cutoff and 1, and factual reporting above 1. Independence makes firm loss additive in the two reporting partitions, and the workers have identical priors, so a symmetric firm-preferred pair exists. Changes at endpoints have zero probability. Conditional losses are quadratic between the thresholds, so their expectations under $U[0,2]$ can be integrated exactly.

A centered initial pool at posterior mean 1 with half-width $0.411771533$ and full disclosure elsewhere has firm loss $0.0883$ and average worker loss $0.1683$. It lowers firm loss by $4.88\%$ relative to full initial disclosure. This pool is a comparison policy, not the unrestricted optimum. 

The timing candidate initially pools $[0.3481,1.4773]$, then discloses the midpoint indicator after factual reports. The smallest false-sign report loss is positive, approximately $2.5\times10^{-6}$. Outside the pool, the policy fully reveals the state and uses babbling below the pool and factual reporting above it. Expected firm and average worker losses are approximately $0.0866$ and $0.1666$. Relative to the centered unchanged-forwarding policy, firm loss falls by about $1.97\%$. Cutoffs are reported to four decimals because the reporting margin is small. Neither policy is claimed to be globally optimal. 

\endgroup
\end{APPENDICES}
\fi
\end{bibunit}

\begin{thebibliography}{42}
\providecommand{\natexlab}[1]{#1}
\providecommand{\url}[1]{\texttt{#1}}
\providecommand{\urlprefix}{URL }

\bibitem[{Alonso et~al.(2008)Alonso, Dessein, \protect\BIBand{}
  Matouschek}]{alonso2008when}
Alonso R, Dessein W, Matouschek N (2008) When does coordination require
  centralization? \emph{American Economic Review} 98(1):145--179.

\bibitem[{Ambrus et~al.(2013)Ambrus, Azevedo, \protect\BIBand{}
  Kamada}]{ambrus2013hierarchical}
Ambrus A, Azevedo EM, Kamada Y (2013) Hierarchical cheap talk.
  \emph{Theoretical Economics} 8(1):233--261.

\bibitem[{Arieli et~al.(2023)Arieli, Geffner, \protect\BIBand{}
  Tennenholtz}]{arieli2023mediated}
Arieli I, Geffner I, Tennenholtz M (2023) Mediated cheap talk design.
  \emph{Proceedings of the AAAI Conference on Artificial Intelligence},
  volume~37, 5456--5463.

\bibitem[{Arieli et~al.(2024)Arieli, Geffner, \protect\BIBand{}
  Tennenholtz}]{arieli2024optimal}
Arieli I, Geffner I, Tennenholtz M (2024) Optimal information design in
  sender-receiver cheap talk interactions. \emph{arXiv preprint
  arXiv:2401.03671} .

\bibitem[{Atakan et~al.(2020)Atakan, Ko{\c{c}}kesen, \protect\BIBand{}
  Kubilay}]{atakan2020starting}
Atakan A, Ko{\c{c}}kesen L, Kubilay E (2020) Starting small to communicate.
  \emph{Games and Economic Behavior} 121:265--296.

\bibitem[{Athey et~al.(2020)Athey, Bryan, \protect\BIBand{}
  Gans}]{athey2020the}
Athey SC, Bryan KA, Gans JS (2020) The allocation of decision authority to
  human and artificial intelligence. \emph{AEA Papers and Proceedings}
  110:80--84.

\bibitem[{Avinadav et~al.(2025)Avinadav, Chernonog, \protect\BIBand{}
  Shamir}]{avinadav2025information}
Avinadav T, Chernonog T, Shamir N (2025) Information provision from a platform
  to competing sellers: The role of strategic ambiguity. \emph{Manufacturing \&
  Service Operations Management} 27(1):269--286.

\bibitem[{Balakrishnan et~al.(2026)Balakrishnan, Ferreira, \protect\BIBand{}
  Tong}]{balakrishnan2026human}
Balakrishnan M, Ferreira KJ, Tong J (2026) Human-algorithm collaboration with
  private information: Na{\"i}ve advice-weighting behavior and mitigation.
  \emph{Management Science} 72(1):265--284.

\bibitem[{Bastani et~al.(2026)Bastani, Bastani, \protect\BIBand{}
  Sinchaisri}]{bastani2026improving}
Bastani H, Bastani O, Sinchaisri WP (2026) Improving human sequential decision
  making with reinforcement learning. \emph{Management Science} 72(1):733--755.

\bibitem[{Bastani \protect\BIBand{} Cachon(2026)}]{bastani2026the}
Bastani H, Cachon GP (2026) The human-{AI} contracting paradox, working paper.

\bibitem[{Bergemann \protect\BIBand{} Morris(2019)}]{bergemann2019information}
Bergemann D, Morris S (2019) Information design: A unified perspective.
  \emph{Journal of Economic Literature} 57(1):44--95.

\bibitem[{Blume et~al.(2007)Blume, Board, \protect\BIBand{}
  Kawamura}]{blume2007noisy}
Blume A, Board OJ, Kawamura K (2007) Noisy talk. \emph{Theoretical Economics}
  2(4):395--440.

\bibitem[{Bonatti et~al.(2026)Bonatti, Dahleh, \protect\BIBand{}
  Horel}]{bonatti2026information}
Bonatti A, Dahleh MA, Horel T (2026) Information design with elicitation and
  strategic coordination. \emph{arXiv preprint arXiv:2302.12223} .

\bibitem[{Boyac{\i} et~al.(2024)Boyac{\i}, Canyakmaz, \protect\BIBand{}
  De~V{\'e}ricourt}]{boyaci2024human}
Boyac{\i} T, Canyakmaz C, De~V{\'e}ricourt F (2024) Human and machine: The
  impact of machine input on decision making under cognitive limitations.
  \emph{Management Science} 70(2):1258--1275.

\bibitem[{Calv{\'o}-Armengol et~al.(2015)Calv{\'o}-Armengol, De~Mart{\'\i},
  \protect\BIBand{} Prat}]{calvo2015communication}
Calv{\'o}-Armengol A, De~Mart{\'\i} J, Prat A (2015) Communication and
  influence. \emph{Theoretical Economics} 10(2):649--690.

\bibitem[{Chen(2012)}]{chen2012value}
Chen Y (2012) Value of public information in sender--receiver games.
  \emph{Economics Letters} 114(3):343--345.

\bibitem[{Corrao \protect\BIBand{} Dai(2023)}]{corrao2023the}
Corrao R, Dai Y (2023) The bounds of mediated communication. \emph{arXiv
  preprint arXiv:2303.06244} .

\bibitem[{Crawford \protect\BIBand{} Sobel(1982)}]{crawford1982strategic}
Crawford VP, Sobel J (1982) Strategic information transmission.
  \emph{Econometrica} 50(6):1431--1451.

\bibitem[{Dai \protect\BIBand{} Taylor(2026)}]{dai2026designing}
Dai T, Taylor T (2026) Designing enterprise {AI} systems: Hallucination,
  creativity, and moral hazard, working paper.

\bibitem[{Dell'Acqua et~al.(2025)Dell'Acqua, Ayoubi, Lifshitz, Sadun, Mollick,
  Mollick, Han, Goldman, Nair, Taub, \protect\BIBand{}
  Lakhani}]{dellacqua2025the}
Dell'Acqua F, Ayoubi C, Lifshitz H, Sadun R, Mollick E, Mollick L, Han Y,
  Goldman J, Nair H, Taub S, Lakhani K (2025) The cybernetic teammate: A field
  experiment on generative {AI} reshaping teamwork and expertise. Working Paper
  33641, National Bureau of Economic Research.

\bibitem[{Dessein(2002)}]{dessein2002authority}
Dessein W (2002) Authority and communication in organizations. \emph{The Review
  of Economic Studies} 69(4):811--838.

\bibitem[{Dreyfuss \protect\BIBand{} Hoong(2026)}]{dreyfuss2026calibrated}
Dreyfuss B, Hoong R (2026) Calibrated coarsening: Designing information for
  {AI}-assisted decisions, working paper.

\bibitem[{Dworczak \protect\BIBand{} Martini(2019)}]{dworczak2019the}
Dworczak P, Martini G (2019) The simple economics of optimal persuasion.
  \emph{Journal of Political Economy} 127(5):1993--2048.

\bibitem[{Galeotti et~al.(2013)Galeotti, Ghiglino, \protect\BIBand{}
  Squintani}]{galeotti2013strategic}
Galeotti A, Ghiglino C, Squintani F (2013) Strategic information transmission
  networks. \emph{Journal of Economic Theory} 148(5):1751--1769.

\bibitem[{Goltsman et~al.(2009)Goltsman, H{\"o}rner, Pavlov, \protect\BIBand{}
  Squintani}]{goltsman2009mediation}
Goltsman M, H{\"o}rner J, Pavlov G, Squintani F (2009) Mediation, arbitration
  and negotiation. \emph{Journal of Economic Theory} 144(4):1397--1420.

\bibitem[{Hagenbach \protect\BIBand{} Koessler(2010)}]{hagenbach2010strategic}
Hagenbach J, Koessler F (2010) Strategic communication networks. \emph{The
  Review of Economic Studies} 77(3):1072--1099.

\bibitem[{Heller \protect\BIBand{} Kuzmics(2024)}]{heller2024communication}
Heller Y, Kuzmics C (2024) Communication, renegotiation and coordination with
  private values. \emph{Games and Economic Behavior} 143:51--76.

\bibitem[{Ibrahim et~al.(2021)Ibrahim, Kim, \protect\BIBand{}
  Tong}]{ibrahim2021eliciting}
Ibrahim R, Kim SH, Tong J (2021) Eliciting human judgment for prediction
  algorithms. \emph{Management Science} 67(4):2314--2325.

\bibitem[{Ivanov(2010{\natexlab{a}})}]{ivanov2010communication}
Ivanov M (2010{\natexlab{a}}) Communication via a strategic mediator.
  \emph{Journal of Economic Theory} 145(2):869--884.

\bibitem[{Ivanov(2010{\natexlab{b}})}]{ivanov2010informational}
Ivanov M (2010{\natexlab{b}}) Informational control and organizational design.
  \emph{Journal of Economic Theory} 145(2):721--751.

\bibitem[{Ivanov(2016)}]{ivanov2016dynamic}
Ivanov M (2016) Dynamic learning and strategic communication.
  \emph{International Journal of Game Theory} 45(3):627--653.

\bibitem[{Kamenica \protect\BIBand{} Gentzkow(2011)}]{kamenica2011bayesian}
Kamenica E, Gentzkow M (2011) Bayesian persuasion. \emph{American Economic
  Review} 101(6):2590--2615.

\bibitem[{Kamenica \protect\BIBand{} Lin(2024)}]{kamenica2024commitment}
Kamenica E, Lin X (2024) Commitment and randomization in communication.
  \emph{arXiv preprint arXiv:2410.17503} .

\bibitem[{Keppler et~al.(2026)Keppler, Sinchaisri, \protect\BIBand{}
  Snyder}]{keppler2026teacher}
Keppler S, Sinchaisri WP, Snyder C (2026) The teacher's knapsack: Generative
  {AI} and teacher productivity. \emph{Available at SSRN 4924786} .

\bibitem[{Kreutzkamp \protect\BIBand{} Lou(2025)}]{kreutzkamp2025persuasion}
Kreutzkamp S, Lou Y (2025) Persuasion without ex-post commitment. \emph{Journal
  of Economic Theory} 228:106058.

\bibitem[{Li \protect\BIBand{} Madar{\'a}sz(2008)}]{li2008when}
Li M, Madar{\'a}sz K (2008) When mandatory disclosure hurts: Expert advice and
  conflicting interests. \emph{Journal of Economic Theory} 139(1):47--74.

\bibitem[{Lin et~al.(2024)Lin, Tomlin, Andreas, \protect\BIBand{}
  Eisner}]{lin2024decision}
Lin J, Tomlin N, Andreas J, Eisner J (2024) Decision-oriented dialogue for
  human-{AI} collaboration. \emph{Transactions of the Association for
  Computational Linguistics} 12:892--911.

\bibitem[{Liu \protect\BIBand{} Sinchaisri(2026)}]{liu2026strategic}
Liu Y, Sinchaisri WP (2026) Strategic advice in the age of personal {AI}.
  \emph{arXiv preprint arXiv:2603.02055} .

\bibitem[{Morris \protect\BIBand{} Shin(2002)}]{morris2002social}
Morris S, Shin HS (2002) Social value of public information. \emph{American
  Economic Review} 92(5):1521--1534.

\bibitem[{{\"O}zer et~al.(2018){\"O}zer, Subramanian, \protect\BIBand{}
  Wang}]{ozer2018information}
{\"O}zer {\"O}, Subramanian U, Wang Y (2018) Information sharing, advice
  provision, or delegation: What leads to higher trust and trustworthiness?
  \emph{Management Science} 64(1):474--493.

\bibitem[{Rantakari(2008)}]{rantakari2008governing}
Rantakari H (2008) Governing adaptation. \emph{The Review of Economic Studies}
  75(4):1257--1285.

\bibitem[{Tamura(2025)}]{tamura2025information}
Tamura W (2025) Information design for adaptive organizations. \emph{arXiv
  preprint arXiv:2501.12669} .

\end{thebibliography}
\end{document}